\documentclass[trackchanges, twocolumn, twocolappendix]{aastex701}

\usepackage{subfigure}
\usepackage{booktabs}
\usepackage{xcolor} 
\usepackage{amsmath}
\usepackage{ulem}
\usepackage{xspace}

\shorttitle{Stellar Tidal Tracks}
\shortauthors{He \& Jiang}

\begin{document}

\title{Tidal Evolution of Satellite Dwarf Galaxies. I. Tidal Tracks of Stellar Mass and Structure}

\author[orcid=0000-0003-0303-4188,sname='He']{Feihong He}
\altaffiliation{Kavli Institute for Astronomy and Astrophysics}
\affiliation{Kavli Institute for Astronomy and Astrophysics, Peking University, Beijing, 100871}
\email[show]{feihong.he@pku.edu.cn}  
\author[orcid=0000-0001-6115-0633,sname='Jiang']{Fangzhou Jiang} 
\altaffiliation{Kavli Institute for Astronomy and Astrophysics}
\affiliation{Kavli Institute for Astronomy and Astrophysics, Peking University, Beijing, 100871}
\email[show]{fangzhou.jiang@pku.edu.cn}
\correspondingauthor{Fangzhou Jiang}


\begin{abstract}

Satellite dwarf galaxies are reshaped by the tidal field of their host, a process that may account for part of the structural diversity of the observed dwarf population. We study how the stellar component of a satellite responds to tidal stripping using a suite of $480$ controlled $N$-body simulations of satellite galaxies consisting of both dark matter (DM) and stars orbiting a Milky-Way-mass host. The simulated dwarfs span a broad range of DM halo structural and stellar-distribution parameters, including halo concentration, inner density slope, stellar-to-halo mass ratio, compactness of the stellar distribution with respect to the halo, as well as the S\'ersic index of the stellar profiles. We present a comprehensive library of tidal tracks, namely, the evolution of the stellar mass, half-mass radius, mean surface density, and line-of-sight velocity dispersion, as functions of the bound mass fraction of the hosting DM subhalo, and provide simple and accurate fitting functions. We also identify the dependence of tidal tracks on key factors in the initial structural conditions. Notably, stellar compactness determines the duration over which the stellar mass remains relatively intact while the surrounding DM is being stripped. The S\'ersic index and inner DM slope, in turn, govern the late-stage stellar mass-loss slope and the contraction or expansion of the remnant. Our simulations and empirical models provide a framework for modelling the stellar evolution of satellites across a broad range of progenitor structures, useful for semi-analytical models of satellite galaxy evolution.

\end{abstract}



\section{Introduction} \label{sec:intro}
 
In the standard $\Lambda$ cold dark matter ($\Lambda$CDM) cosmology, structure grows hierarchically: small dark matter (DM) haloes collapse first and subsequently merge into larger systems, leaving behind a wealth of substructure, subhaloes orbiting within the virialised regions of their hosts \citep{white1978, white1991}. Galaxies form at the centres of haloes, and satellite galaxies are hosted by subhaloes, so that the halo--subhalo hierarchy maps directly onto the central--satellite galaxy hierarchy. While $\Lambda$CDM is remarkably successful on large scales, on the scale of dwarf and satellite galaxies it faces persistent challenges, including the missing-satellites problem \citep{klypin1999, moore1999}, the core--cusp problem \citep{flores1994, Moore1994, deBlok2010}, the too-big-to-fail problem \citep{BoylanKolchin2011}, and the diversity of dwarf-galaxy rotation curves \citep{Oman2015} \citep[see][for reviews]{Bullock2017,Sales2022}. Common to all of these challenges is that they hinge on the internal structure of dwarf galaxies and of the DM haloes that host them, making dwarf structure a key observable for testing the nature of DM.
 
The inner density profile of a dwarf's DM halo encodes information about both baryonic processes and the nature of DM. Collisionless CDM predicts cuspy haloes that are well described by the Navarro--Frenk--White (NFW) profile \citep{Navarro1996a, NFW}, whereas baryonic feedback can substantially modify the central density structure \citep[e.g.,][]{Pontzen2012}. This motivates parametrising halo density profile with more degrees of freedom than NFW. A popular choice is the Dekel--Zhao (DZ) profile \citep{Zhao1996, Dekel2017, Freundlich2020, Jiang2021}, which has been shown to be able to describe baryon-influenced halo profiles at the expense of only one more degree of freedom, with good analytical properties.
 
Satellite systems do not preserve their initial structure: once accreted, subhaloes are subject to tidal stripping and tidal heating within the host potential, which systematically reshape their density profiles \citep{Gnedin1999,ZentnerBullock2003,Hayashi2003,Kravtsov2004,Penarrubia2008,Penarrubia2010,Green2019}. A key result from simulations is that the structural evolution of a subhalo, characterised by its maximum circular velocity $v_{\rm max}$ and the radius $r_{\rm max}$ at which this maximum occurs, is governed almost entirely by the fraction of mass that remains bound, and is largely insensitive to the specific orbit along which the mass was lost. The structural properties therefore trace well-defined ``tidal tracks'' parametrised by the bound mass fraction \citep{Penarrubia2008, Penarrubia2010, Green2019}. 
Crucially, the tidal response itself depends on the inner DM slope: cuspy subhaloes are more resistant to tidal disruption and can evolve towards long-lived bound remnants \citep{Penarrubia2010,Errani2021,Errani2022}. Any attempt to infer DM physics from the structure of observed satellites must therefore account for how tides have reprocessed that structure \citep{Pullen2014, BensonDu2022}.
 
The same concept of structural response to tides extends to the stellar component embedded in the subhalo. In a simple picture of galaxy-halo connection, the satellite galaxy resides deep within the subhalo potential, shielded by the extended DM envelope, so that tidal stripping removes the outer DM while the stellar structure remains nearly intact in its pre-infall state. However, this stage is only temporary: as tidal evolution progresses, the stellar mass, size, and velocity dispersion of the embedded galaxy are also modified \citep{Penarrubia2008, Penarrubia2010, Errani2022}. In the cosmological hydrodynamical simulation COLIBRE \citep{Schaye2026}, \citet{He2026} showed statistically that the stellar mass loss of satellites is a well-defined function of their total bound mass loss, and that the disruption of a subhalo is accompanied by the disruption of the inhabitant galaxy. Tidal evolution is therefore not merely a DM phenomenon: it reshapes the properties of satellite galaxies as well. This could therefore be a natural mechanism to render the morphological diversity of satellite galaxies.

Observations provide a direct motivation for quantifying this stellar response. Dwarf galaxies span a broad range of sizes and surface brightnesses at fixed stellar mass, including ultra-diffuse galaxies \citep[UDGs;][]{vanDokkum2015a,Koda2015}, while some systems have unusually low inferred DM contents \citep{vanDokkum2018b,vanDokkum2019}. Tidal evolution has been proposed to contribute to these properties \citep{Carleton2019,Jiang2019,Ogiya2022,Moreno2022}. Establishing which changes tides can produce requires a systematic description of the stellar tidal tracks across the initial structural parameter space, which is the focus of this paper.
 
Cosmological hydrodynamical and $N$-body simulations are not ideally suited for this task. First, the systems of interest with special structural properties (UDGs, UCDs, DM-deficient galaxies) populate tails of galaxy distributions, so that in a high-resolution simulation that could resolve them, the volume is usually too limited to contain a large statistical sample of them. Besides, their formation conditions cannot be varied in a controlled manner. On the other hand, in large-volume cosmological runs, subhaloes could suffer from artificial disruption driven by limited resolution, discreteness noise, and inadequate force softening, which contaminates precisely the heavily stripped regime that matters most here \citep{vdB2018a, vdB2018b, Errani2020, He2025}. High-resolution idealised simulations have therefore become an ideal alternative approach, in which an $N$-body satellite is evolved within an analytic host potential \citep{Penarrubia2008, Penarrubia2010, Drakos2017, Ogiya2019, Drakos2020, Errani2021, Drakos2022, Du2024}. It is precisely this class of controlled experiments that established the orbit-insensitivity of tidal tracks, that a subhalo's structure is set primarily by its initial structural parameters, with orbital parameters playing at most a second-order role \citep{Penarrubia2008, Green2019, Errani2021}. This nature of tidal tracks also permits a simulation design that scans structural parameters along a single representative orbit, as adopted in this work.
 
Existing idealised studies of the stellar tidal evolution, however, share two important limitations. In the standard approach \citep{Penarrubia2008, Errani2015, Errani2022}, the stellar component is represented by massless tracer particles embedded within an NFW subhalo. First, the massless approximation is valid only if the system is DM-dominated, appropriate for faint and ultra-faint dwarfs, but not for bright dwarfs, compact systems, or the late, deeply stripped stages in which the stars come to dominate the central potential. In such systems the stellar gravity could substantially modify the total potential and hence the dynamical state of both the DM and the stars themselves -- this is particularly relevant for DM-deficient galaxies, which, by definition, feature baryon dominance. Second, dwarf galaxies exhibit substantial structural diversity even before infall, as suggested by observations of field dwarfs. Their stellar distributions span a range of sizes and surface densities \citep{Higgs2021}, while their inferred inner DM distributions also exhibit considerable diversity \citep{Oman2015,Read2016,Read2019}. A single fixed NFW halo combined with one adopted stellar profile therefore cannot capture the range of possible initial satellite structures. What is needed is a suite of self-consistent two-component dwarfs covering the joint structural parameter space of halo and galaxy.
 
In this work, we systematically investigate the tidal evolution of satellite dwarf galaxies using a suite of 480 idealised $N$-body simulations. Each satellite is realised as a self-consistent two-component system of DM and stars with equal particle masses \citep[to suppress spurious mass segregation;][]{Ludlow2019, Ludlow2021, Ludlow2023}. 
The DM halo follows a truncated DZ profile, whose free inner slope spans the range from feedback induced cores to NFW-like or adiabatic-contraction-induced cusps. The stellar component follows a deprojected S\'ersic profile \citep{Sersic1963}. We scan the subhalo concentration and inner slope, the stellar-to-halo mass ratio, the stellar compactness $r_{\rm eff}/r_{\rm vir}$, and the S\'ersic index, and quantify the resulting evolution of the stellar mass, half-mass radius, mean surface density, and line-of-sight velocity dispersion as functions of the bound mass that remains. We also provide simple fitting functions for each track and relate their parameters to the initial structure. 

A companion paper (Paper~II) will apply this framework to the tidal evolution of UDGs and DM-deficient dwarf galaxies (DMDGs), and characterize the conditions under which these properties can arise together. 

The remainder of this paper is organised as follows. In Section~\ref{sec:setup} we describe the simulation setup, initial conditions, parameter space, and the variable used to parametrise the tidal tracks. Section~\ref{sec:results} presents the tracks of stellar mass, size, surface density, and velocity dispersion. Section~\ref{sec:param_dependence} relates the fitted parameters to the initial structure. Section~\ref{sec:discussion} discusses the limitations of our models, and Section~\ref{sec:conclusion} summarises our main findings. The numerical convergence of our results is examined in Appendix~\ref{sec:convergence}.

\section{Simulation Setup}\label{sec:setup}


\subsection{Host halo setup}

We model the host halo as a static, spherical NFW potential \citep{NFW}. A distinct halo is defined as a spherical region with an enclosed mean density of $200$ times the critical density, $\rho_{\rm crit}$, of the Universe, at redshift 0. We adopt cosmological parameters of $\Omega_m=0.3$, $\Omega_{\Lambda}=0.7$, and $H_0=70\,\rm{km\,s^{-1}\,Mpc^{-1}}$, and set the host virial mass to $M_{\rm vir}=10^{12}\,\rm{M}_{\odot}$, approximately the Milky-Way mass. With these definitions, the virial radius of the host halo is $R_{\rm vir}=206\,\rm{kpc}$. The host concentration is $c_h\equiv R_{\rm vir}/R_s=10$, where $R_s$ is the scale radius of the host.

Throughout this paper, capital and lowercase symbols distinguish host and satellite quantities, respectively: for example, $M_{\rm vir}$ and $R_{\rm vir}$ refer to the host virial mass and virial radius, respectively, while $m_{\rm vir}$ and $r_{\rm vir}$ represent those of a subhalo.

\subsection{Initial condition realisation}

The particle realisations are generated with the action-based dynamics library \textsc{Agama} \citep{Vasiliev2019}. We specify the density profiles of the two components, DM and stars, $\rho_{\rm DM}(r)$ and $\rho_{\star}(r)$, and compute the total gravitational potential, $\Phi_{\rm tot}=\Phi_{\rm DM}+\Phi_{\star}$. \textsc{Agama} then obtains the spherically symmetric distribution function of each component in the combined potential through Eddington inversion \citep{Eddington1916}. These distribution functions are isotropic, $f=f(E)$, depending on the binding energy alone. We do not consider anisotropic models of the form $f(E,L)$, so that both the DM and the stars start with isotropic velocity distributions. Particles are sampled from these distribution functions in energy and phase space, producing a self-consistent equilibrium realisation with $\rho_{\rm tot}=\rho_{\rm DM}+\rho_{\star}$. Thus, even when two dwarf systems share the same DM density profile, a different stellar density profile changes the total potential and therefore the equilibrium DM distribution function.

To suppress spurious numerical heating driven by mass segregation, in which massive DM particles artificially transfer kinetic energy to lighter stellar particles through two-body encounters \citep{Ludlow2019, Ludlow2021, Ludlow2023}, we adopt an equal particle mass of $m_{\rm p}=10^{4}\rm{M}_{\odot}$ for both the DM and stellar components. 

\subsection{DM profile}\label{sec:dm_profile}

For the DM component, we follow \citet{Jiang2021} and adopt a truncated Dekel--Zhao (DZ) profile \citep{Zhao1996, Dekel2017, Freundlich2020}. The untruncated DZ density profile is
\begin{equation}
    \begin{aligned}
    \rho_{\rm DZ}(r)&=\frac{\rho_0}{x^{\alpha}\left(1+x^{1/\beta}\right)^{\beta(\gamma-\alpha)}},\\
    \beta&=2,\qquad \gamma=3+\beta^{-1}=3.5,
    \end{aligned}
    \label{eq:DZ}
\end{equation}
with $x\equiv r/r_s$ and $r_s=r_{\rm vir}/c$. The exponent $\beta=2$ is fixed throughout and is unrelated to the size-response index introduced in Section~\ref{sec:size_track}. We impose a smooth outer truncation,
\begin{equation}
    \rho_{\rm DM}(r)=\rho_{\rm DZ}(r)
    \exp\!\left[-\left(\frac{r}{r_{\rm cut}}\right)^{\eta}\right],
    \label{eq:DZ_trunc}
\end{equation}
where $r_{\rm cut}=2r_{\rm vir}$ and $\eta=2$. The normalisation $\rho_0$ is set so that the DM mass enclosed by $r_{\rm vir}$ equals the target $m_{\rm vir}$, corresponding to an average enclosed density of $200\rho_{\rm crit}$.

Relative to the NFW profile, the DZ profile provides independent freedom in the inner slope and concentration while retaining a simple analytic functional form. This flexibility is useful for representing the range of halo structures produced by different assembly histories, baryonic feedback, or DM physics. Its asymptotic outer slope is $-3.5$. The exponential cutoff suppresses the extended outer envelope used in the particle realisation.

Rather than working with $(\alpha,c)$, we use two parameters that have a more interpretable meaning following \citet{Jiang2021}. The first is the conventional concentration
\begin{equation}
    c_2\equiv\frac{r_{\rm vir}}{r_{-2}}
    =\left(\frac{1.5}{2-\alpha}\right)^2 c,
    \label{eq:c2_2_c}
\end{equation}
where $r_{-2}$ is the radius at which the logarithmic slope is $-2$. The logarithmic density slope of a DZ profile is given by
\begin{equation}
    s_{\rm DZ}(r)\equiv-\frac{{\rm d}\ln\rho_{\rm DZ}}{{\rm d}\ln r}
    =\frac{\alpha+3.5\sqrt{x}}{1+\sqrt{x}},
    \label{eq:DZ_slope}
\end{equation}
and we define the inner-slope parameter at one per cent of the virial radius as
\begin{equation}
    s_{001}\equiv s_{\rm DZ}(0.01r_{\rm vir})
    =\frac{\alpha+0.35\sqrt{c}}{1+0.1\sqrt{c}},
    \label{eq:s1_2_alpha}
\end{equation}
and use it as the second parameter.
Because $r_{\rm cut}=2r_{\rm vir}$, the exponential truncation has a negligible effect at $0.01r_{\rm vir}$. For each target pair $(s_{001},c_2)$, we solve Eqs.~\ref{eq:c2_2_c} and~\ref{eq:s1_2_alpha} for $(\alpha,c)$.

Although the untruncated DZ profile has a finite total mass because its outer slope is steeper than $-3$, the convergence is slow. The adopted cutoff therefore suppresses an unnecessarily extended outer envelope while remaining sufficiently gradual to preserve the equilibrium distribution function. In an isotropic system, particles currently inside $r_{\rm vir}$ can have apocentres beyond $r_{\rm vir}$ and particles outside can subsequently return. An excessively sharp cutoff would distort this orbital population. With $r_{\rm cut}=2r_{\rm vir}$ and $\eta=2$, the total initial subhalo mass is typically $m_{\rm sub}\simeq1.2$--$1.5\,m_{\rm vir}$ across our $(s_{001},c_2)$ grid.

In practice, we first generate a trial truncated profile, measure its enclosed mass within $r_{\rm vir}$, and rescale the density normalisation so that this enclosed mass is exactly $m_{\rm vir}$. This procedure fixes the virial mass while retaining the intended DZ shape and outer truncation.

\subsection{Stellar profile}\label{sec:stellar_profile}

Our satellites are intended to represent dwarf spheroidal galaxies, and we model the stellar component as a three-dimensional, spherically symmetric system. For simplicity, we neglect flattening, triaxiality, and any rotating disc, so that the stellar body is fully specified by a radial density profile. Its projected surface-density profile follows the S\'ersic law \citep{Sersic1963},
\begin{equation}
    \Sigma_{\star}(R) = \Sigma_0 \exp \left[ -b_{n} \left(R/r_{\rm eff}\right)^{1/n}\right],
    \label{eq:sersic}
\end{equation}
where $R$ is the projected radius, $n \equiv n_{\rm sersic}$ is the S\'ersic index, and $r_{\rm eff}$ is the projected stellar half-mass radius, enclosing half of the total stellar mass in projection. 
The dimensionless constant $b_{n}$ is fixed by this definition through $\gamma(2n, b_{n}) = \tfrac{1}{2}\,\Gamma(2n)$, where $\Gamma$ and $\gamma$ are the complete and lower incomplete gamma functions. The normalisation $\Sigma_0$ is set so that the integrated stellar mass equals the target value $m_{\star}$. The S\'ersic law provides an accurate description of the surface brightness profiles of observed dwarf galaxies \citep{Sersic1963}, and its index $n$ varies from nearly exponential profiles ($n \simeq 1$), commonly found in low-surface-brightness dwarf galaxies \citep{Venhola2022}, to more centrally concentrated profiles ($n > 1$), characteristic of brighter dwarf ellipticals \citep{GrahamGuzman2003}, making it a natural single-parameter family for modelling the stellar component.

Since our initial conditions require a three-dimensional density field rather than a projected profile, we deproject Eq.~\ref{eq:sersic}. For a spherically symmetric system the projected profile is the same along every line of sight, and the three-dimensional stellar density $\rho_{\star}(r)$ can be recovered directly from $\Sigma_{\star}(R)$ through the inverse Abel transform,
\begin{equation}
    \rho_{\star}(r) = -\frac{1}{\pi} \int_{r}^{\infty} \frac{\mathrm{d}\Sigma_{\star}}{\mathrm{d}R} \, \frac{\mathrm{d}R}{\sqrt{R^2 - r^2}},
    \label{eq:abel}
\end{equation}
where $r$ is the three-dimensional radius. 
We evaluate Eq.~\ref{eq:abel} numerically to construct $\rho_{\star}(r)$, which is then supplied to \textsc{Agama} together with the DM profile to compute the total potential and the corresponding self-consistent stellar distribution function, as described in Section~\ref{sec:setup}. Throughout the tidal evolution, we characterise the galaxy size by $r_{\rm eff}$, measured at each snapshot from the bound stellar particles projected along the $z$-axis of the simulation box. Since all orbits lie in the $x$-$y$ plane, this projection views the satellite perpendicular to its orbital plane. 


\subsection{Parameter grid}

\begin{table}[!t]
\centering
\caption{Initial structural parameter grid of the simulated satellite galaxies.\label{tab:parameter}}
\small
\begin{tabular}{rl}
\toprule
Properties & Values \\
\midrule
$m_{\rm vir}$ & $10^{10}\rm{M}_{\odot}$ \\
$c_{2}$ & 5, 10 \\
$s_{001}$ & 0.6, 1.0, 1.2, 1.4, 1.8 \\
$m_{\star}/m_{\rm vir}$ & 0.001, 0.01, 0.05 \\
$r_{\rm eff}/r_{\rm vir}$ & 0.01, 0.02, 0.04, 0.1 \\
$n_{\rm sersic}$ & 0.5, 1.0, 2.5, 4.0 \\
\bottomrule
\end{tabular}
\par\smallskip
\begin{minipage}{\columnwidth}
\footnotesize\textit{Note.} The virial mass is fixed at $m_{\rm vir}=10^{10}\,\rm{M}_{\odot}$ for all dwarfs. The Cartesian product of the remaining structural choices gives $2\times5\times3\times4\times4=480$ two-component satellite dwarfs.
\end{minipage}
\end{table}

In this work, all simulated DM subhaloes are assigned a virial mass of $m_{\rm vir}=10^{10}\,\rm{M}_{\odot}$. As noted previously, $m_{\rm vir}$  denotes the enclosed DM mass within $r_{\rm vir}$. Since particles are sampled out to radii beyond $r_{\rm vir}$, the total bound mass $m_{\rm bound}$ of each subhalo is slightly larger than $m_{\rm vir}$. This mass scale is chosen because it lies in the dwarf-galaxy regime where baryonic feedback is expected to produce substantial diversity in the inner DM density profiles \citep{Lazar2020,Freundlich2020}. We do not consider more massive subhaloes ($m_{\rm vir}=10^{10.5-11}\rm{M}_{\odot}$) because at those masses, dynamical friction becomes non-negligible for satellites evolving within a Milky-Way-mass host. Since we model the host potential as a fixed, analytic background, dynamical friction is inherently absent, and we wish to avoid this inconsistency. A static host, however, does not eliminate all orbital evolution: material stripped from the satellite can torque the subhalo and cause a slow secular loss of orbital angular momentum and energy \citep{Miller2020, Du2024, Katayama2024}, a process known as ``dynamical self-friction". This effect is generally weaker than classical dynamical friction in the host halo, but it can become most noticeable during strong stripping episodes near pericentre.

The subhalo concentration is set to $c_{2}\in\{5,10\}$. The inner slope parameter $s_{001}$ is varied over $\{0.6, 1.0, 1.2, 1.4, 1.8\}$, spanning the range commonly discussed in the context of the core--cusp problem. For typical concentrations ($c_{2}=10$), an NFW profile has  $s_{001}=1.2$ (whereas the asymptotic logarithmic slope at $r=0$ is -1). Conversely, $s_{001}=0.6$ is the shallowest and most strongly cored configuration considered in our study. More extremely cored profiles are not considered, as such subhaloes undergo tidal disruption too rapidly for the tidal tracks to be meaningfully studied.

The stellar mass fraction $m_{\star}/m_{\rm vir}$ is varied from $0.001$ to $0.05$. The lower bound of 0.001 reflects the typical stellar-to-halo mass ratio at this halo mass. The upper bound, $m_\star/m_{\rm vir}=0.05$, allows for satellites that continue to form stars after infall \citep{Wetzel2013}, and that this star formation is further enhanced when ram pressure compresses the gas and triggers a starburst \citep{BekkiCouch2003, Kapferer2009, Vulcani2018}, elevating the stellar mass fraction.

The compactness parameter $r_{\rm eff}/r_{\rm vir}$ (sometimes referred to as stellar segregation in the literature  \citep{Errani2022, Errani2024}) quantifies the ratio of the dwarf galaxy size to that of its DM subhalo. Typical values for galaxies range from 0.01 to 0.1, with smaller values indicating more compact stellar distributions and larger values more diffuse ones. For our adopted virial mass $m_{\rm vir}=10^{10}\,\rm{M}_{\odot}$, the corresponding virial radius is $r_{\rm vir}=44.4\,\rm{kpc}$, so the half-mass radius spans $r_{\rm eff} \approx 0.44$--$4.4\,\rm{kpc}$.

For dwarf galaxies, the S\'ersic index $n_{\rm sersic}$ is typically close to or below unity. The S\'ersic index of early-type galaxies decreases steadily towards lower luminosities, and most dwarf galaxies observed in the Virgo Cluster have $n_{\rm sersic}\lesssim1$ \citep{Ferrarese2020}. Nevertheless, we scan a broader and more extreme range, $n_{\rm sersic}\in\{0.5,1,2.5,4\}$, in order to explore a wider variety of stellar configurations. For reference, $n_{\rm sersic}=1$ corresponds to an exponential profile, $n_{\rm sersic}=2.5$ is commonly adopted as the dividing value between disc-like and spheroid-like light profiles \citep{Shen2003}, and $n_{\rm sersic}=4$ corresponds to the de Vaucouleurs profile of giant elliptical galaxies \citep{deVaucouleurs1948}. The case $n_{\rm sersic}=0.5$, for which the projected profile is Gaussian, represents stellar bodies with a flat, cored centre.

The full parameter space is summarised in Table~\ref{tab:parameter}. Combining all parameter values yields a total of $2\times5\times3\times4\times4=480$ simulations.


\subsection{N-body integration}\label{sec:nbody}

\begin{table}[!t]
\centering
\caption{Orbital pericentres and apocentres adopted for each pair of DM structural parameters.\label{tab:orbit}}
\small
\begin{tabular}{rlll}
\toprule
$c_2$ & $s_{001}$ & $R_{\rm peri}$ & $R_{\rm apo}$ \\
\midrule
5 & 0.6 & 50\,kpc & 100\,kpc \\
5 & 1.0 & 50\,kpc & 100\,kpc\\
5 & 1.2 & 35\,kpc & 100\,kpc\\
5 & 1.4 & 10\,kpc & 100\,kpc\\
5 & 1.8 & 5\,kpc & 100\,kpc\\
10 & 0.6 & 35\,kpc & 100\,kpc\\
10 & 1.0 & 20\,kpc & 100\,kpc\\
10 & 1.2 & 10\,kpc & 100\,kpc\\
10 & 1.4 & 5\,kpc & 100\,kpc\\
10 & 1.8 & 5\,kpc & 100\,kpc\\
\bottomrule
\end{tabular}
\par\smallskip
\begin{minipage}{\columnwidth}
\footnotesize\textit{Note.} All dwarfs start at a common apocentre of $100\,\rm{kpc}$. The pericentres were selected from preliminary DM-only runs so that each subhalo is stripped deeply over $20\,\rm{Gyr}$.
\end{minipage}
\end{table}

\begin{figure*}
    \centering
    \begin{subfigure}
        \centering
        \includegraphics[scale=0.4]{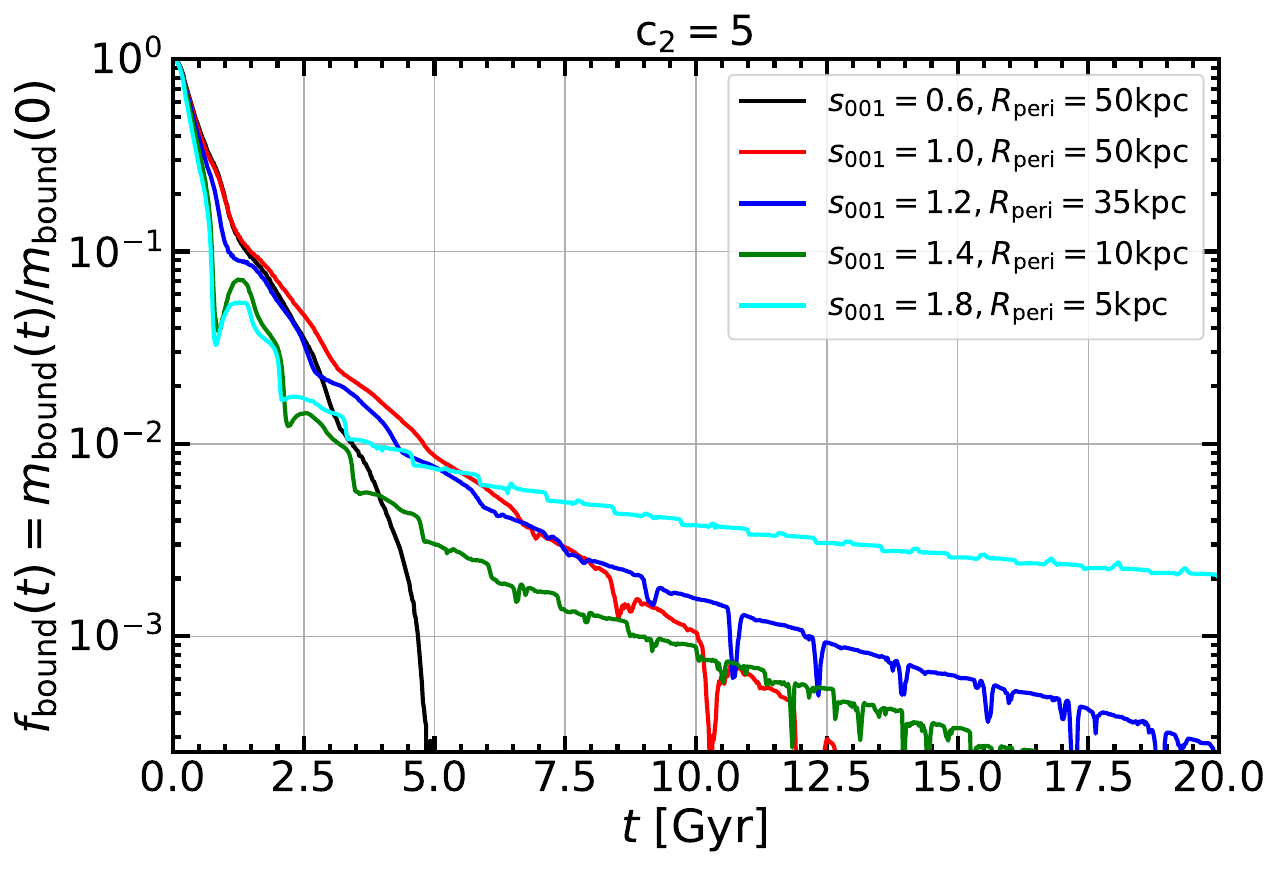}
    \end{subfigure}
    \centering
    \hfill
    \begin{subfigure}
        \centering
        \includegraphics[scale=0.4]{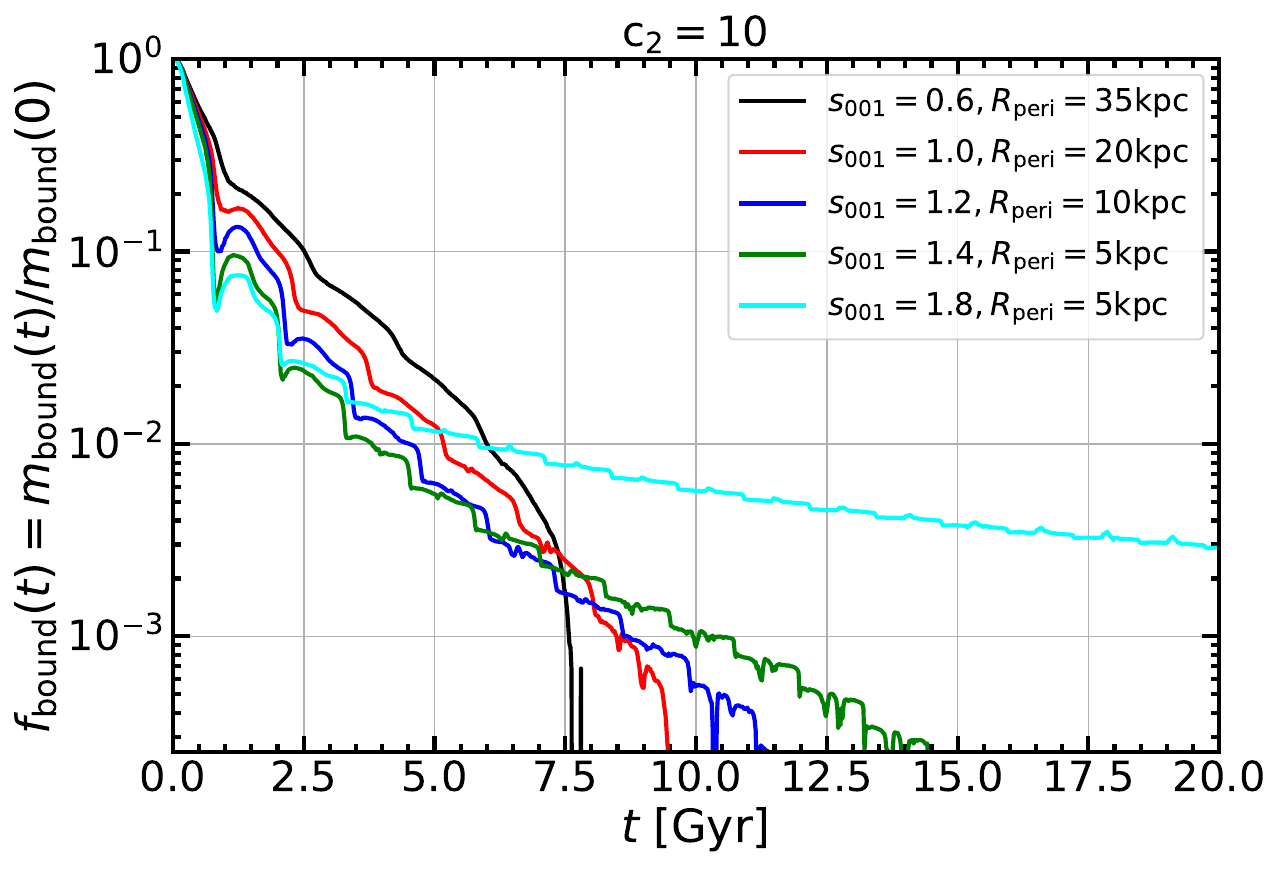}
    \end{subfigure}
    \caption{Bound-mass evolution in the preliminary DM-only orbit-calibration runs. The left and right panels show $c_2=5$ and $c_2=10$, respectively, with colours denoting different inner slopes $s_{001}$. The pericentre for each curve is listed in the legend and in Table~\ref{tab:orbit}. These runs are used to choose orbits that reach the strongly stripped regime.}
    \label{fig:preliminary_dmo}
\end{figure*}

\begin{figure}
    \centering
    \includegraphics[width=\linewidth]{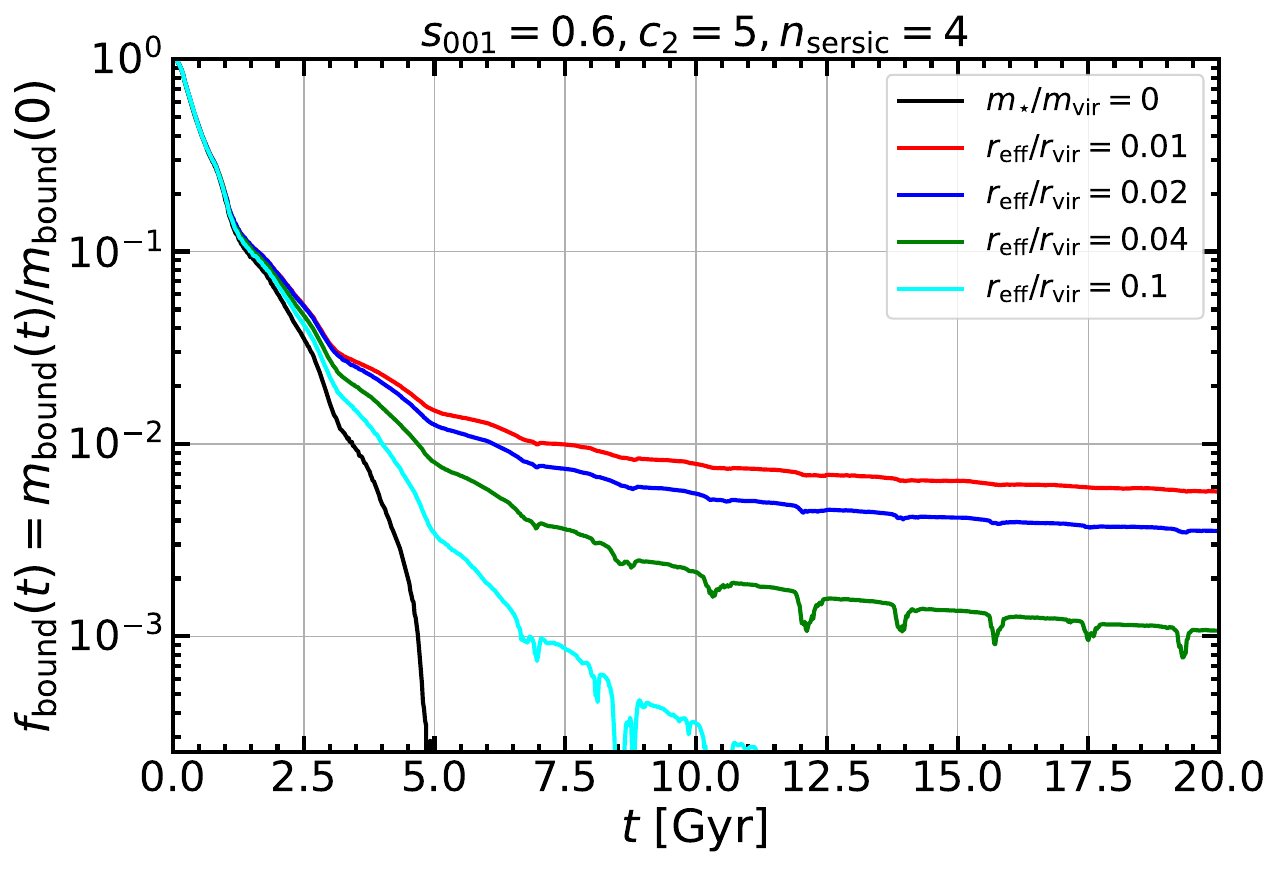}
    \caption{Effect of a stellar component on the total bound-mass evolution for a representative cored subhalo. Coloured curves show otherwise identical two-component satellites with different stellar compactness $r_{\rm eff}/r_{\rm vir}=0.01$--$0.1$, while the black curve is the corresponding DM-only subhalo. The stellar component modifies the central potential and can substantially delay complete disruption.}
    \label{fig:masslosscurve}
\end{figure}

For a subhalo with $m_{\rm vir}=10^{10}\,M_{\odot}$, our adopted particle mass $m_{\rm p}=10^{4}\rm{M}_{\odot}$ corresponds to $N=10^6$ particles within the virial radius. This resolution balances the need to survey a large and diverse parameter space against the requirement for adequate numerical resolution. \citet{vdB2018a} and \citet{vdB2018b} demonstrated that insufficient resolution in $N$-body simulations of subhaloes can lead to artificial disruption driven by discreteness noise or inadequate gravitational softening. We follow the criteria established by \citet{vdB2018b}, who showed that a combination of $N=10^6$ particles and a softening length $\varepsilon=0.003r_{-2}$ is sufficient to prevent numerical artefacts from affecting the tidal evolution of the subhalo.

The simulated dwarf galaxies are initially placed at orbital apocentres with tangential velocity $v_{\rm t}$, and all orbits lie in the $x$-$y$ plane of the simulation box. We integrate their orbital motion and tidal evolution in the static host potential with \textsc{Gadget-4} \citep{gadget4}. Each run is evolved for $20\,\rm{Gyr}$ with snapshots output every $0.02\,\rm{Gyr}$. 
At each snapshot, \textsc{hbt+}\xspace \citep{Han2012, Han2018} iteratively identifies the gravitationally self-bound remnant, including both DM and stellar particles. 

From this bound particle set we measure the component masses, the projected stellar half-mass radius, and the stellar velocity dispersion. The half-mass radius $r_{\rm eff}$ is the radius that encloses half of the bound stellar mass when the particles are projected along the $z$-axis, i.e.\ perpendicular to the orbital plane. The line-of-sight velocity dispersion $\sigma_{\rm los}$ is measured separately for projections along the $x$-, $y$-, and $z$-axes. For each projection, we take the position and velocity of the most bound particle as the centre of the galaxy, select the bound stellar particles within $r_{\rm eff}$, and compute the mass-weighted dispersion of their line-of-sight velocities. $\sigma_{\rm los}$ is the average of the three values. Unless otherwise stated, $m_{\star}$ and $m_{\rm DM}$ denote the bound stellar and DM masses, the total satellite bound mass is $m_{\rm bound}=m_{\star}+m_{\rm DM}$, and a subscript 0 denotes the initial value. We refer to $f_{\rm bound}\equiv m_{\rm bound}/m_{\rm bound,0}$ as the bound mass fraction and use it to describe the mass-loss histories.

Given that the scan over the structural parameter space of dwarf galaxies and their host subhaloes already produces a large number of simulation runs, simultaneously varying the orbital parameters for each structural configuration would incur a prohibitive computational cost. Previous studies \citep{Penarrubia2010, Errani2021, Errani2022} have shown that the evolutionary tracks of subhalo structural properties are primarily governed by the initial structural parameters, with the choice of orbital parameters playing at most a secondary role. 
Therefore, for each structural parameter combination, we run the simulation on a fixed initial orbit that is chosen in a way described below. We test these choices in Appendix~\ref{sec:orbit_convergence}, where representative dwarfs are rerun on orbits with different pericentres.

Because the stars are initially embedded more deeply than most of the DM, the orbit must yield the removal of a substantial fraction of the outer subhalo before the stellar body responds. We therefore opt for relatively low-energy orbits that drive the satellite into the strongly stripped regime. A single deeply plunging orbit cannot be used for every subhalo, however, because cored subhaloes would then be disrupted rapidly before their structural evolution could be adequately sampled.

We calibrate the orbit case by case for each of the ten $(s_{001},c_2)$ combinations using DM-only simulations. The adopted pericentres and common apocentre are listed in Table~\ref{tab:orbit}, and the resulting mass-loss histories are shown in Fig.~\ref{fig:preliminary_dmo}. The pericentres are chosen such that, if the remnant survives, the system reaches $f_{\rm bound}\sim10^{-3.5}$--$10^{-2.5}$ by the end of $20\,\rm{Gyr}$. 

Also evident from Fig.~\ref{fig:preliminary_dmo} is the strong dependence of tidal survival on the inner halo structure: cuspy dwarfs approach a long-lived bound remnant, whereas the most weakly bound cored dwarfs, e.g., with $(s_{001},c_2)=(0.6,5)$, are disrupted within $\sim10\,\rm{Gyr}$ even on a comparatively mild (high pericentre) orbit.

The DM-only calibration does not necessarily capture the mass-loss history once stellar mass is included. Fig.~\ref{fig:masslosscurve} compares a representative cored DM-only configuration with otherwise similar two-component satellites that differ in $r_{\rm eff}/r_{\rm vir}$. The stellar component deepens the central potential and delays disruption relative to the DM-only case, with the detailed response depending on stellar compactness. This comparison motivates treating Fig.~\ref{fig:preliminary_dmo} as an orbit-selection benchmark while measuring the actual tidal tracks from the full two-component simulations. For a fixed $(s_{001},c_2)$, all stellar configurations $(m_{\star}/m_{\rm vir},r_{\rm eff}/r_{\rm vir},n_{\rm sersic})$ are evolved on the same calibrated orbit.

\subsection{Input variable of the tidal tracks}\label{sec:xbound}

Tidal tracks are conventionally expressed as functions of the bound mass fraction $f_{\rm bound}$ or $m_{\rm max}/m_{\rm max,0}$ \citep{Penarrubia2008, Penarrubia2010, Green2019}. In this work we instead parametrise them by the scaled bound mass
\begin{equation}
    x_{\rm bound}\equiv\frac{m_{\rm bound}}{m_{\rm max,0}}=x_{\rm bound,0}\,f_{\rm bound},
    \label{eq:xbound}
\end{equation}
where $m_{\rm max,0}$ is the mass enclosed within $r_{\rm max,0}$, the radius at which the circular velocity of the initial satellite peaks, and $x_{\rm bound,0}\equiv m_{\rm bound,0}/m_{\rm max,0}$ is the initial value of $x_{\rm bound}$. Across our parameter grid, $m_{\rm bound,0}$ exceeds $m_{\rm max,0}$ by a factor of about 3 to 5, depending mainly on the halo concentration. 
In contrast to $f_{\rm bound}$, the scaled bound mass therefore exceeds unity at early times and falls below unity only once the satellite has lost more mass than was initially contained within $r_{\rm max,0}$.

Two considerations motivate this choice. First, $m_{\rm bound,0}$ is not a uniquely defined property of the initial satellite, because it depends on how the outer density profile is truncated. Different truncation schemes can change $m_{\rm bound,0}$ substantially (the cutoff adopted here gives $m_{\rm sub}\simeq1.2$ to $1.5\,m_{\rm vir}$, see Section~\ref{sec:dm_profile}), whereas the loosely bound outer envelope is removed early in the tidal evolution and has little effect on the stripping of the inner structure that follows. The mass within $r_{\rm max,0}$ is instead set by this inner structure. \citet{Du2025} showed that tidal tracks expressed in terms of $m_{\rm bound}/m_{\rm max,0}$ are insensitive to the initial truncation, whereas those expressed in terms of $f_{\rm bound}$ are not \citep[see also][]{Errani2022}, and we follow their normalisation. Second, one of our aims is to follow satellites into the DM-deficient regime. Tidal tracks built on the DM structure, such as those of $v_{\rm max}$, $r_{\rm max}$, and $m_{\rm max}$, lose their meaning once most of the DM has been removed, and are undefined for a remnant without DM. We therefore use the total bound mass $m_{\rm bound}=m_{\rm DM}+m_{\star}$ in the numerator of Eq.~\ref{eq:xbound}, while $m_{\rm max,0}$ only needs to be evaluated in the initial conditions. This keeps $x_{\rm bound}$ well defined along the entire evolution from a DM-dominated dwarf to a DM-deficient remnant.



\section{Stellar Tidal Tracks}\label{sec:results}

We describe the stellar response through tracks of mass, size, mean surface density, and velocity dispersion as functions of $x_{\rm bound}$. Their empirical forms and parameter meanings are compiled in Table~\ref{tab:tracks} at the end of this section.  Section~\ref{sec:param_dependence} examines how the fitted parameters depend on the initial dwarf structure.

\subsection{Tidal track for stellar mass}\label{sec:mass_track}

We first quantify the stellar mass evolution by the remaining stellar mass fraction, $f_{\star}\equiv m_{\star}/m_{\star,0}$. \citet{He2026} showed in the cosmological hydrodynamical simulation \textsc{COLIBRE} \citep{Schaye2026} that stellar mass loss is tightly related to the total bound mass loss of satellites. Here we use controlled simulations to determine how this relation depends on the initial structure of the dwarf. Following the functional form introduced by \citet{He2026} and \citet{Wan2026}, we write
\begin{equation}
    f_{\star}=1-\exp\!\left[-\ln 2\left(\frac{x_{\rm bound}}{f_{1/2}}\right)^b\right],
    \label{eq:fstar}
\end{equation}
where $f_{1/2}$ and $b$ are two free parameters, which we refer to as the half-mass stripping scale and the stripping sharpness, respectively. By construction, $f_{\star}=1/2$ when $x_{\rm bound}=f_{1/2}$, so $f_{1/2}$ is the characteristic scaled bound mass fraction at which half of the initial stellar mass has been stripped -- it measures how deeply the satellite must be stripped before substantial stellar mass loss occurs, and marks the onset of efficient stellar stripping. The stripping sharpness $b$ controls how abrupt this transition is and sets the late-time logarithmic slope. For $x_{\rm bound}\gg f_{1/2}$, $f_{\star}\simeq1$, whereas for $x_{\rm bound}\ll f_{1/2}$,
\begin{equation}
    f_{\star}\simeq \ln 2\left(\frac{x_{\rm bound}}{f_{1/2}}\right)^b.
    \label{eq:fstar_asymptotic}
\end{equation}
Eq.~\ref{eq:fstar} therefore captures both the early phase when stars are shielded by subhalo from tidal stripping and the late phase when the stripping of stellar mass and DM mass are coupled in a compact form.

\begin{figure}
    \centering
    \begin{subfigure}
        \centering
        \includegraphics[width=\linewidth]{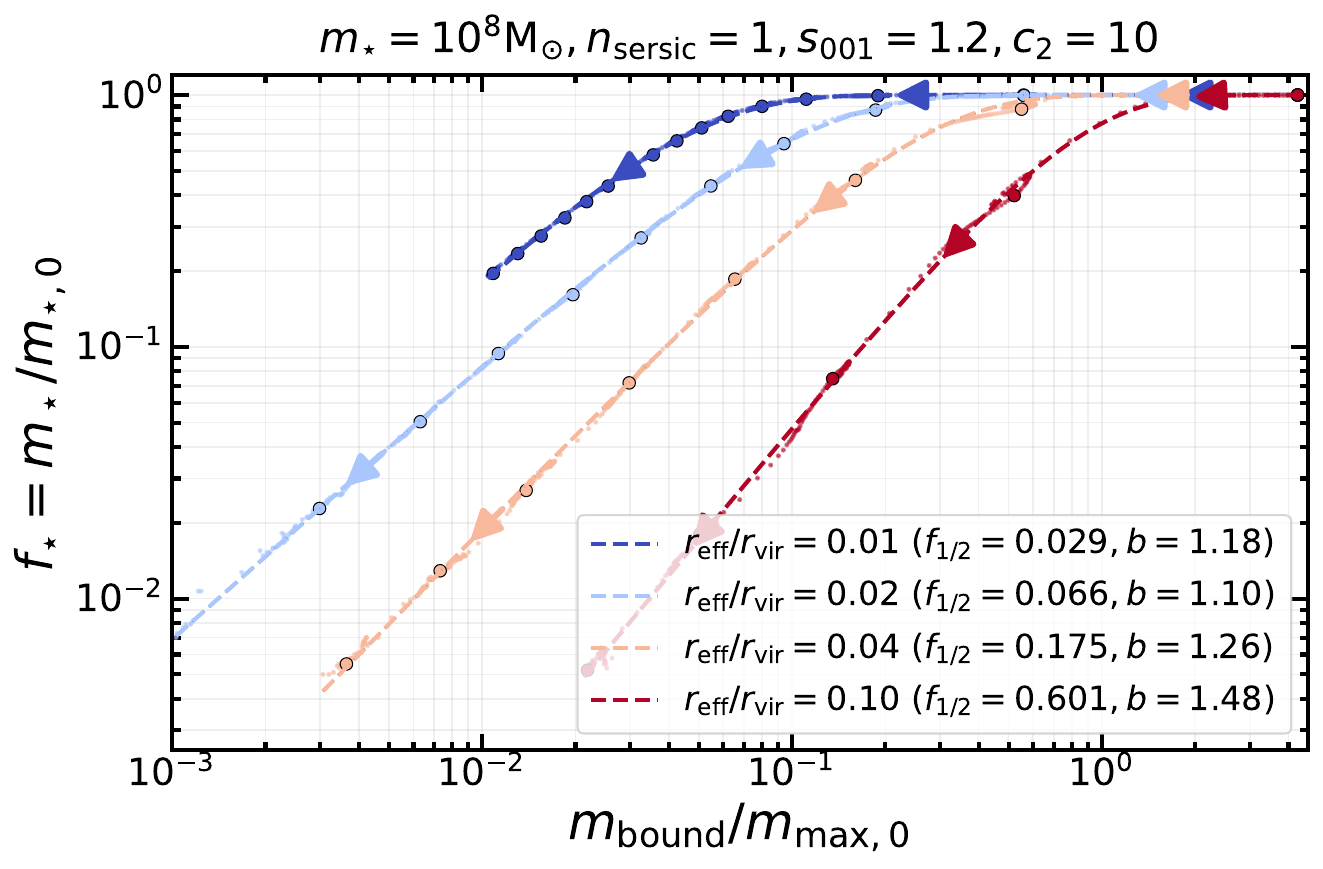}
    \end{subfigure}
    \centering
    \hfill
    \begin{subfigure}
        \centering
        \includegraphics[width=\linewidth]{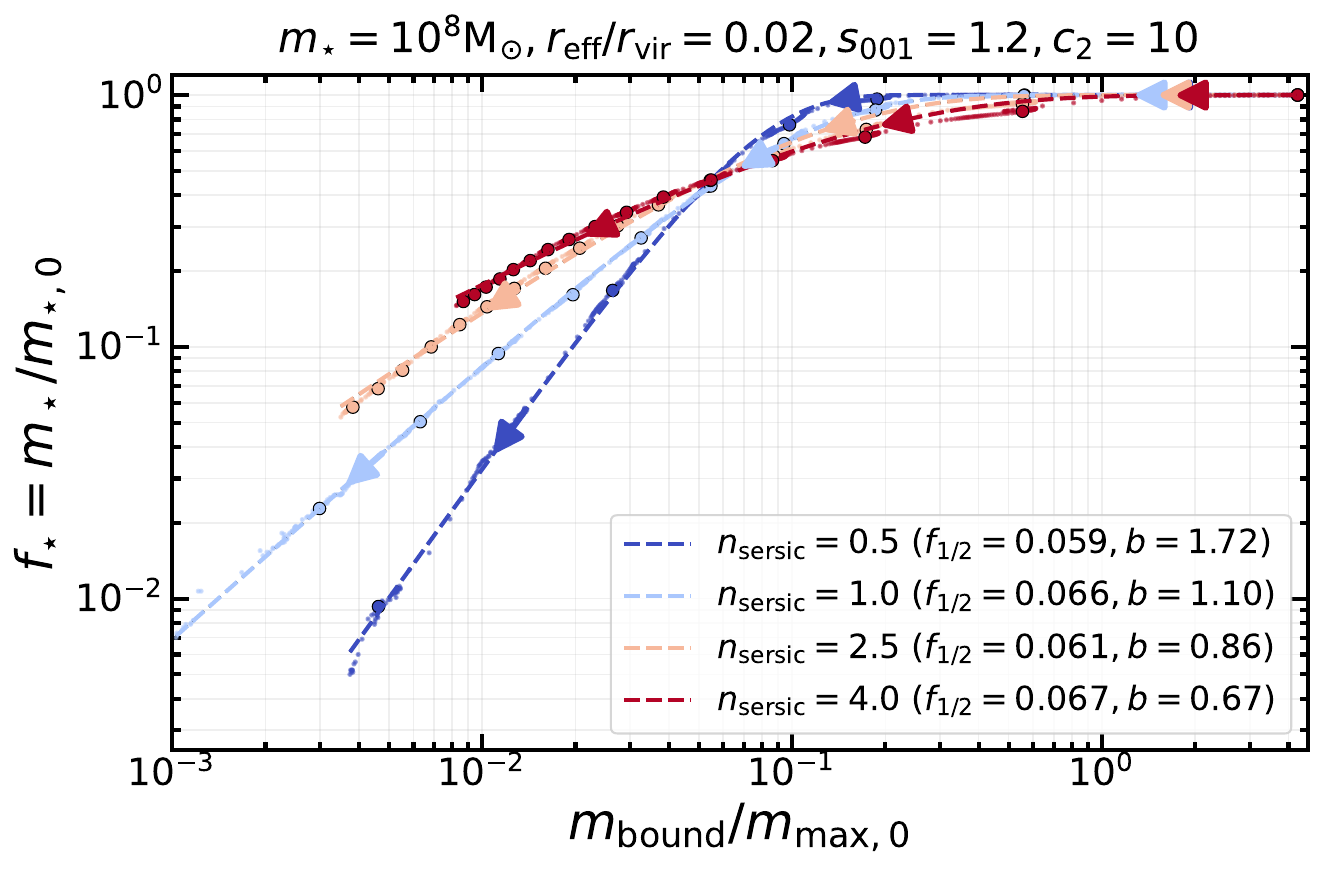}
    \end{subfigure}
    \caption{Tidal track for stellar mass. The vertical axis shows the remaining stellar mass fraction $f_{\star}=m_{\star}/m_{\star,0}$ and the horizontal axis shows the scaled bound mass $x_{\rm bound}=m_{\rm bound}/m_{\rm max,0}$ (Eq.~\ref{eq:xbound}), which starts at $x_{\rm bound,0}\simeq4.4$ for these dwarfs. Filled circles mark apocentre snapshots and dashed curves show fits of Eq.~\ref{eq:fstar}. The fitted half-mass stripping scale $f_{1/2}$ and stripping sharpness $b$ are listed in the legends. Arrows indicate the direction of evolution. \textit{Top:} $r_{\rm eff}/r_{\rm vir}$ is varied from $0.01$ to $0.1$ at fixed $m_{\star}=10^{8}\,\rm{M}_{\odot}$, $c_2=10$, $s_{001}=1.2$, and $n_{\rm sersic}=1$. \textit{Bottom:} $n_{\rm sersic}$ is varied from $0.5$ to $4$ at fixed $r_{\rm eff}/r_{\rm vir}=0.02$ and the other fiducial parameters. Stellar compactness mainly sets $f_{1/2}$, so that more compact dwarfs lose half of their stars only after more extensive stripping, whereas the S\'ersic index mainly sets $b$, which decreases with increasing $n_{\rm sersic}$.}
    \label{fig:mass_track}
\end{figure}

Fig.~\ref{fig:mass_track} illustrates these dependences, where we define a fiducial case with $m_{\star}=10^{8}\,\rm{M}_{\odot}$, $c_2=10$, $s_{001}=1.2$, $n_{\rm sersic}=1$, and $r_{\rm eff}/r_{\rm vir}=0.02$, and vary one stellar structural parameter at a time. The top panel varies $r_{\rm eff}/r_{\rm vir}$ and shows that stellar compactness primarily shifts the half-mass stripping scale, from $f_{1/2}=0.029$ at $r_{\rm eff}/r_{\rm vir}=0.01$ to $0.60$ at $r_{\rm eff}/r_{\rm vir}=0.1$: more compact stellar distributions are more deeply embedded and must be stripped to a smaller $x_{\rm bound}$ before losing half of their stars. The bottom panel varies the S\'ersic index and shows that $n_{\rm sersic}$ primarily changes the stripping sharpness, from $b=1.72$ at $n_{\rm sersic}=0.5$ to $0.67$ at $n_{\rm sersic}=4$, while $f_{1/2}$ remains close to $0.06$. Low-$n_{\rm sersic}$ systems therefore show a steeper late-stage decline in $f_{\star}$, while high-$n_{\rm sersic}$ systems retain their stars more efficiently once the subhalo has been heavily stripped.

The stripping sharpness also determines the relative late-stage evolution of the stellar and total bound components. For $b=1$, the remaining stellar fraction becomes proportional to $x_{\rm bound}$, and hence to the total bound mass, in the asymptotic regime. For $b>1$, $f_{\star}$ declines more rapidly than the bound mass, whereas for $b<1$ it declines more slowly. This distinction is central to whether a remnant evolves towards a stellar-poor or a DM-deficient state.

\begin{figure}
    \centering
    \begin{subfigure}
        \centering
        \includegraphics[width=\linewidth]{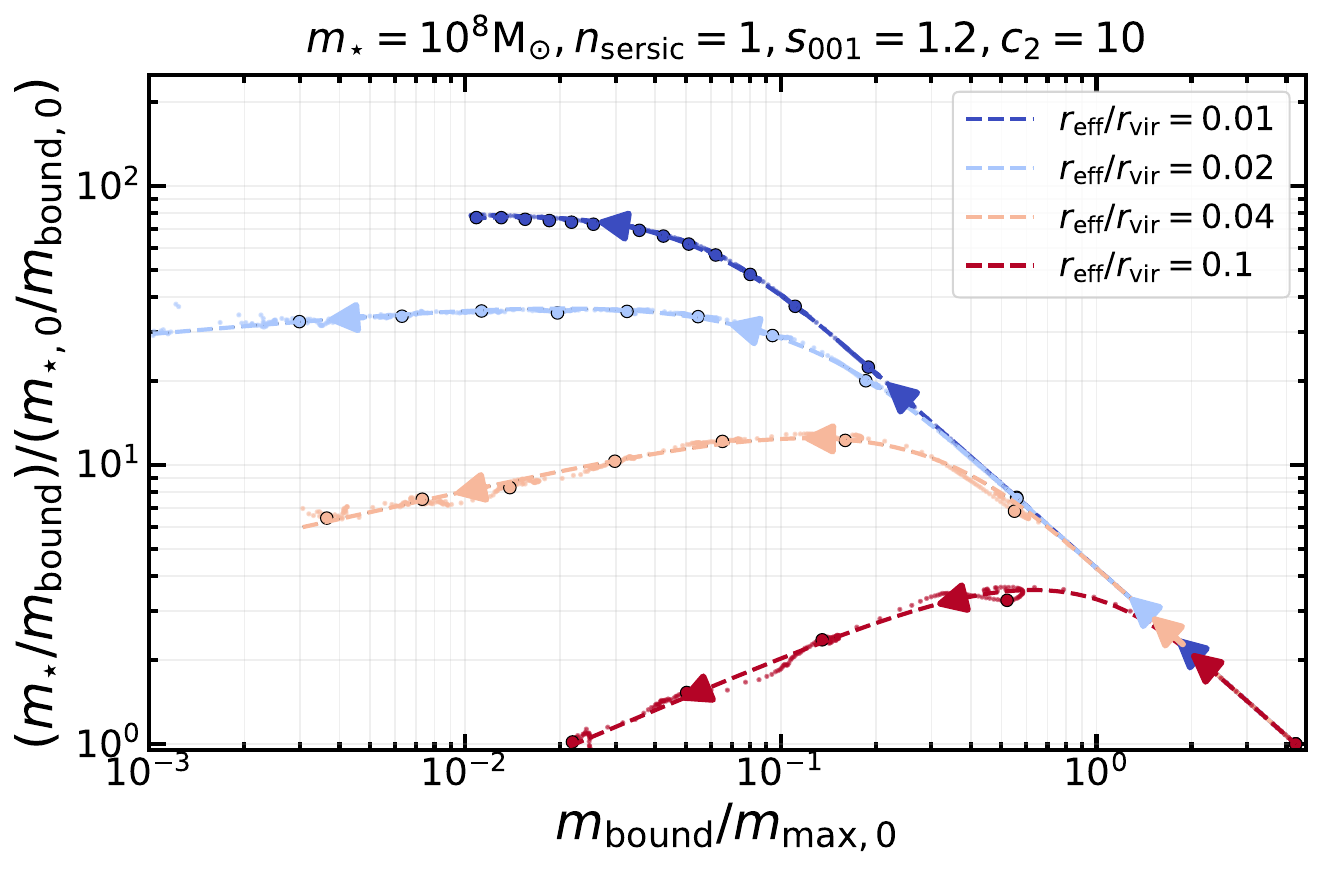}
    \end{subfigure}
    \centering
    \hfill
    \begin{subfigure}
        \centering
        \includegraphics[width=\linewidth]{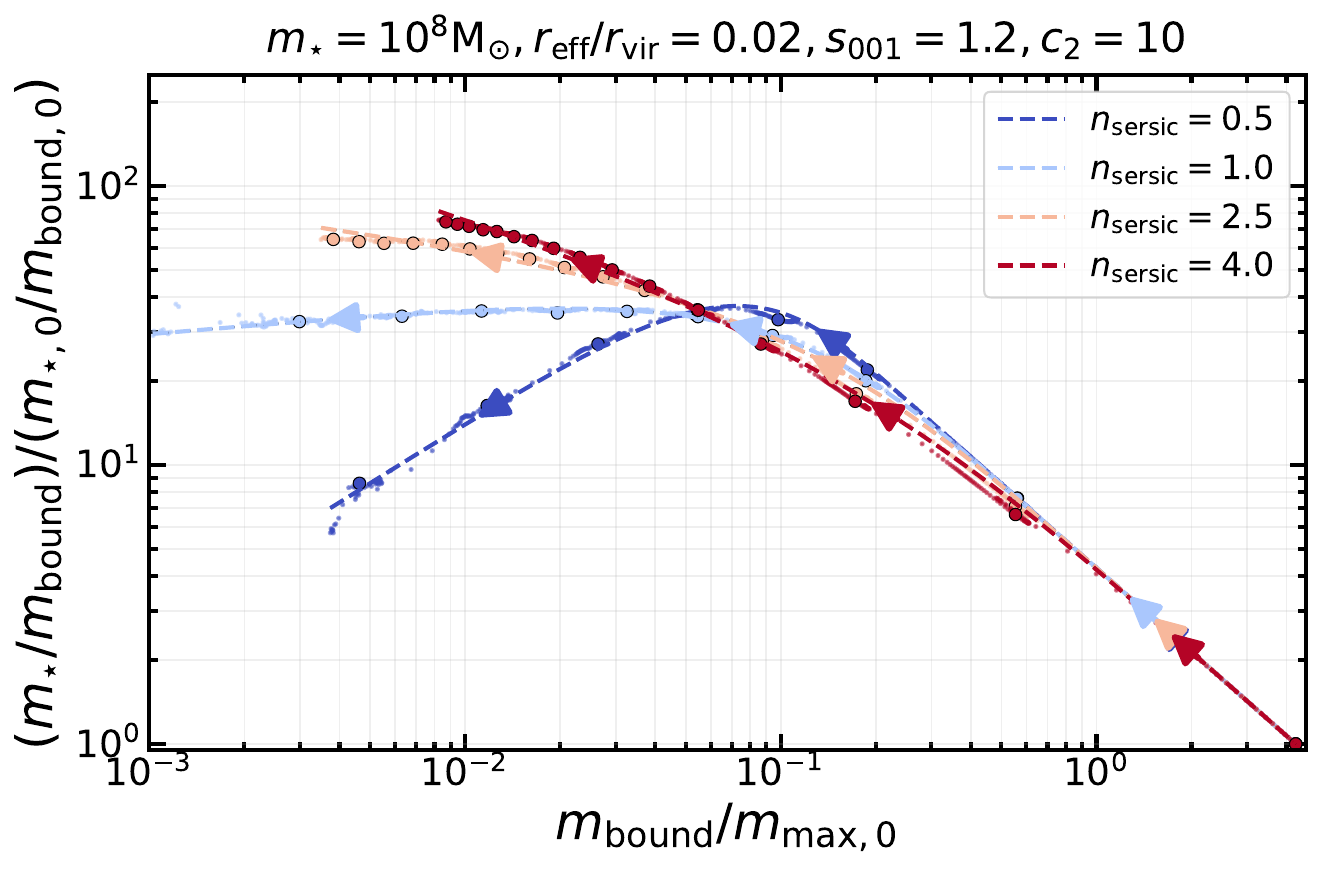}
    \end{subfigure}
    \caption{Evolution of the normalised stellar-to-bound mass ratio $\xi/\xi_0$ as a function of the scaled bound mass $x_{\rm bound}$ for the same examples as Fig.~\ref{fig:mass_track}. Filled circles mark apocentre snapshots, and dashed curves are the prediction of Eq.~\ref{eq:fxi} using the stellar-mass-track fits. Arrows indicate the direction of evolution. In the shielded phase all tracks follow the common relation $\xi/\xi_0\simeq x_{\rm bound,0}/x_{\rm bound}$. 
    \textit{Top:} varying $r_{\rm eff}/r_{\rm vir}$ changes how far the stellar-to-bound mass ratio can rise before efficient stellar stripping begins. \textit{Bottom:} varying $n_{\rm sersic}$ changes the late-time behaviour through the stripping sharpness $b$: $\xi/\xi_0$ keeps rising for $b<1$, approaches a constant for $b\simeq1$, and turns over for $b>1$.}
    \label{fig:xi_track}
\end{figure}

Because the stellar-to-bound mass ratio is used repeatedly below when discussing DM deficiency, we denote it compactly as $\xi\equiv m_{\star}/m_{\rm bound}$. Normalising by its initial value $\xi_0=m_{\star,0}/m_{\rm bound,0}$ and using Eq.~\ref{eq:xbound} gives
\begin{align}
    \frac{\xi}{\xi_0}
    &=\frac{m_{\star}/m_{\rm bound}}{m_{\star,0}/m_{\rm bound,0}}
      =\frac{f_{\star}}{f_{\rm bound}} \nonumber\\
    &=\frac{x_{\rm bound,0}}{x_{\rm bound}}\left\{1-\exp\!\left[-\ln 2\left(\frac{x_{\rm bound}}{f_{1/2}}\right)^b\right]\right\}.
    \label{eq:fxi}
\end{align}
Because $\xi/\xi_0$ is defined relative to the initial state, the change of variable introduces the constant factor $x_{\rm bound,0}$. In the shielded phase, $f_{\star}\simeq1$ and $\xi/\xi_0\simeq x_{\rm bound,0}/x_{\rm bound}$, so all tracks initially rise along a common line of slope $-1$ that starts from $\xi/\xi_0=1$ at $x_{\rm bound}=x_{\rm bound,0}$. Fig.~\ref{fig:xi_track} shows $\xi/\xi_0$ for the same examples as Fig.~\ref{fig:mass_track}. In the deeply stripped limit,
\begin{equation}
    \frac{\xi}{\xi_0}\simeq
    \frac{x_{\rm bound,0}\ln 2}{f_{1/2}^{\,b}}\,x_{\rm bound}^{\,b-1},
    \label{eq:fxi_asymptotic}
\end{equation}
so the sign of $b-1$ determines the asymptotic behavior of the stellar-to-bound mass ratio.

For $b<1$, the ratio in Eq.~\ref{eq:fxi_asymptotic} rises as stripping proceeds, corresponding to preferential removal of DM and the approach to a stellar-dominated remnant. This growth cannot continue indefinitely because $\xi\leq1$. Once the system is effectively a bare stellar remnant, the simple two-component asymptote ceases to apply. For $b\simeq1$, the ratio approaches the constant $x_{\rm bound,0}\ln 2/f_{1/2}$. For $b>1$, the ratio first rises while the outer DM subhalo is removed but declines after stellar stripping becomes sufficiently strong. 

\subsection{Tidal track for satellite galaxy size}\label{sec:size_track}

\begin{figure}
    \centering
    \begin{subfigure}
        \centering
        \includegraphics[width=\linewidth]{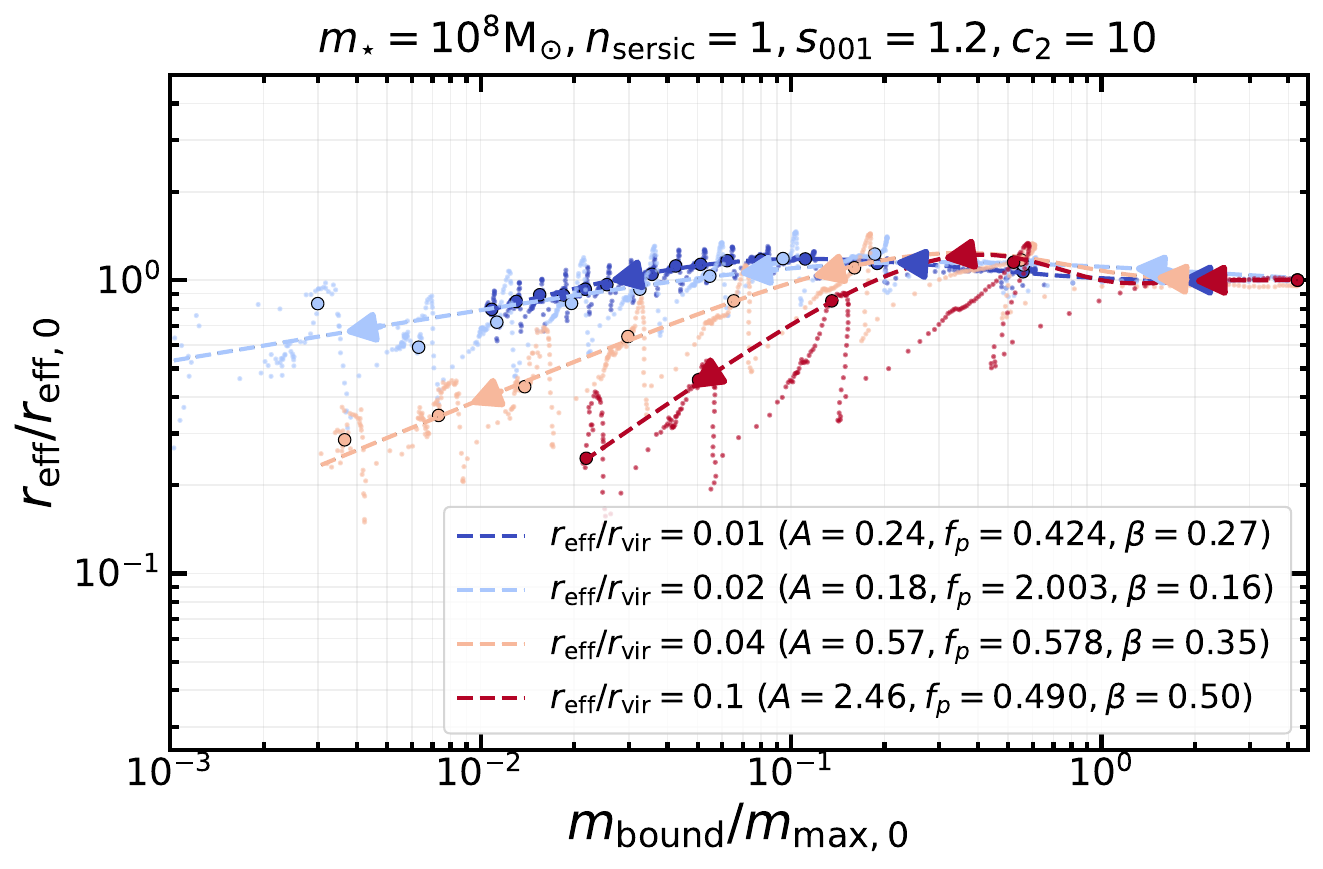}
    \end{subfigure}
    \centering
    \hfill
    \begin{subfigure}
        \centering
        \includegraphics[width=\linewidth]{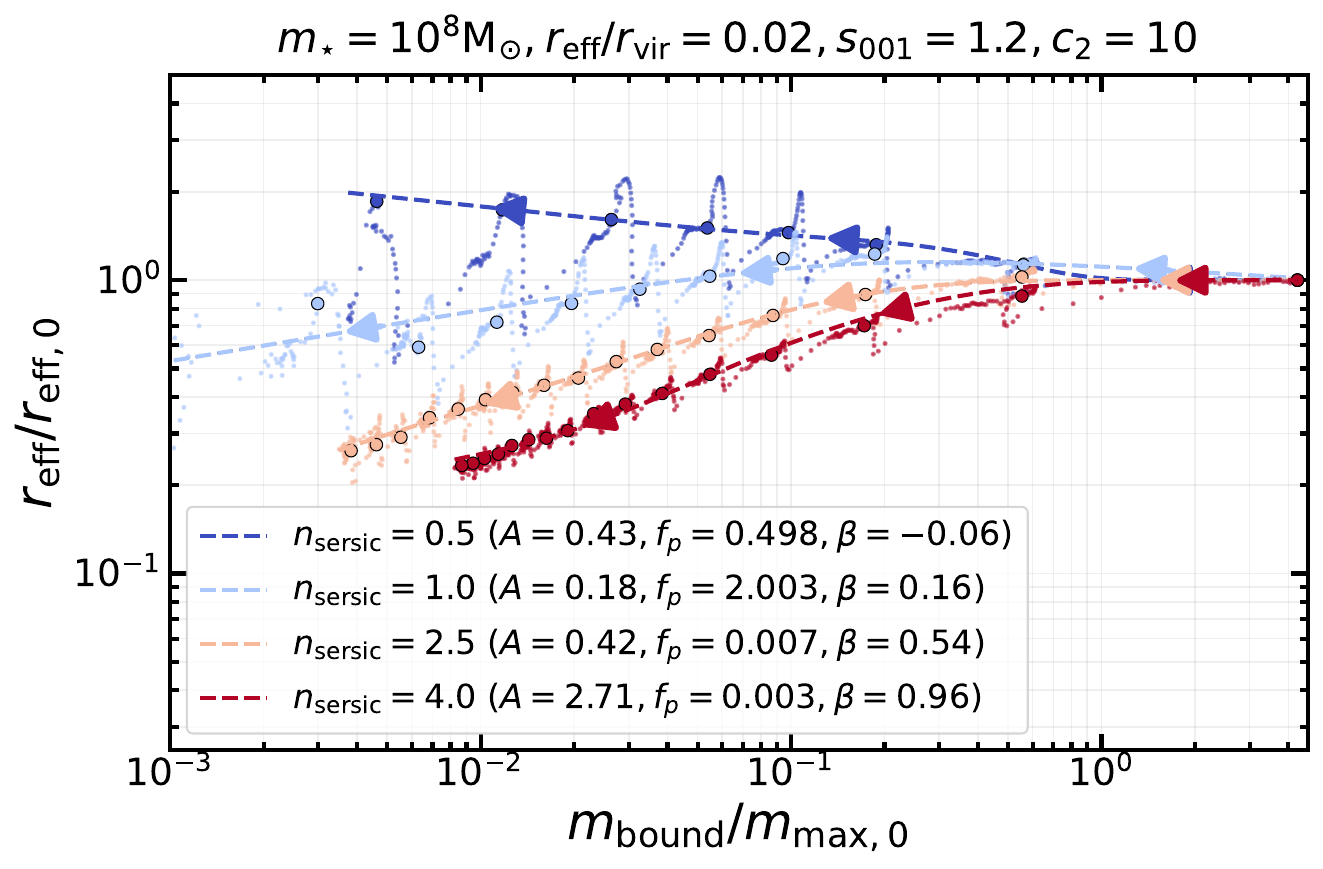}
    \end{subfigure}
    \caption{Tidal track for satellite galaxy size. The vertical axis shows the half-mass radius $r_{\rm eff}$ scaled by its initial value $r_{\rm eff,0}$, and the horizontal axis shows the scaled bound mass $x_{\rm bound}$. Faint dotted lines show the simulation output at every snapshot, filled circles the output at each apocentre, and dashed lines the best fits of Eq.~\ref{eq:freff}. $f_{1/2}$ and $b$ are inherited from the stellar-mass fit, while the fitted puff-up amplitude $A$, puff-up scale $f_{\rm p}$, and size-response index $\beta$ are listed in the legends. Arrows indicate the direction of evolution. \textit{Top panel:} $r_{\rm eff}/r_{\rm vir}$ is varied from $0.01$ (most compact, blue) to $0.1$ (most diffuse, dark red) at fixed $m_{\star}=10^{8}\,\rm{M}_{\odot}$, $c_2=10$, $s_{001}=1.2$, and $n_{\rm sersic}=1$. \textit{Bottom panel:} $n_{\rm sersic}$ is varied from $0.5$ to $4$ at fixed $r_{\rm eff}/r_{\rm vir}=0.02$ and the other fiducial parameters. Beneath the orbital-phase oscillations, the secular trend ranges from strong contraction for the most concentrated stellar profile ($\beta=0.96$ at $n_{\rm sersic}=4$) to net expansion for the shallowest one ($\beta=-0.06$ at $n_{\rm sersic}=0.5$).} 
    \label{fig:size_track}
\end{figure}

We quantify the galaxy size by the projected half-mass radius $r_{\rm eff}$ (Section~\ref{sec:stellar_profile}). Fig.~\ref{fig:size_track} shows the size evolution, $r_{\rm eff}/r_{\rm eff,0}$ as a function of $x_{\rm bound}$, for the same set of satellites presented in Fig.~\ref{fig:mass_track}. In contrast to the smooth behaviour of the mass tracks, the size tidal tracks display distinctly more complicated structure.

First, all simulation runs exhibit periodic oscillations: the size contracts as the satellite approaches pericentre and expands as it moves outward toward apocentre, and this pattern repeats over successive orbits. These oscillations, however, are superimposed on an underlying trend. Consistent with the behaviour of the mass tracks, we find that when $x_{\rm bound}$ lies above the half-mass stripping scale $f_{1/2}$, the half-mass radius remains essentially unchanged, corresponding to the plateau in the stellar mass track. Once $x_{\rm bound}$ falls below $f_{1/2}$ and the satellite begins to lose stellar mass, the half-mass radius generally decreases as the satellite is progressively stripped. This secular contraction is not necessarily monotonic. For the $n_{\rm sersic} = 0.5$ case shown in the bottom panel of Fig.~\ref{fig:size_track}, for example, the satellite undergoes a pronounced ``puffing up'' phase: its half-mass radius grows steadily once $x_{\rm bound}\lesssim0.5$ and reaches nearly twice its initial value by $x_{\rm bound}\sim5\times10^{-3}$.

To summarise, the tidal evolution of the half-mass radius comprises the superposition of three distinct behaviours: (i) a two-phase secular contraction analogous to the mass tidal track, characterised by an initial plateau followed by a decline once $x_{\rm bound} \lesssim f_{1/2}$; (ii) a structure-dependent puffing-up phase whose amplitude and timing depend on the internal properties of the satellite; and (iii) short-timescale oscillations driven by the orbital motion between pericentre and apocentre.

In the present work, we characterise satellite evolution as a function of $x_{\rm bound}$, which measures the remaining bound mass due to tidal stripping, and therefore focus on the first two behaviours, deferring a quantitative treatment of the orbital-phase-driven periodic variations to future work.

\begin{figure}
    \centering
    \begin{subfigure}
        \centering
        \includegraphics[width=\linewidth]{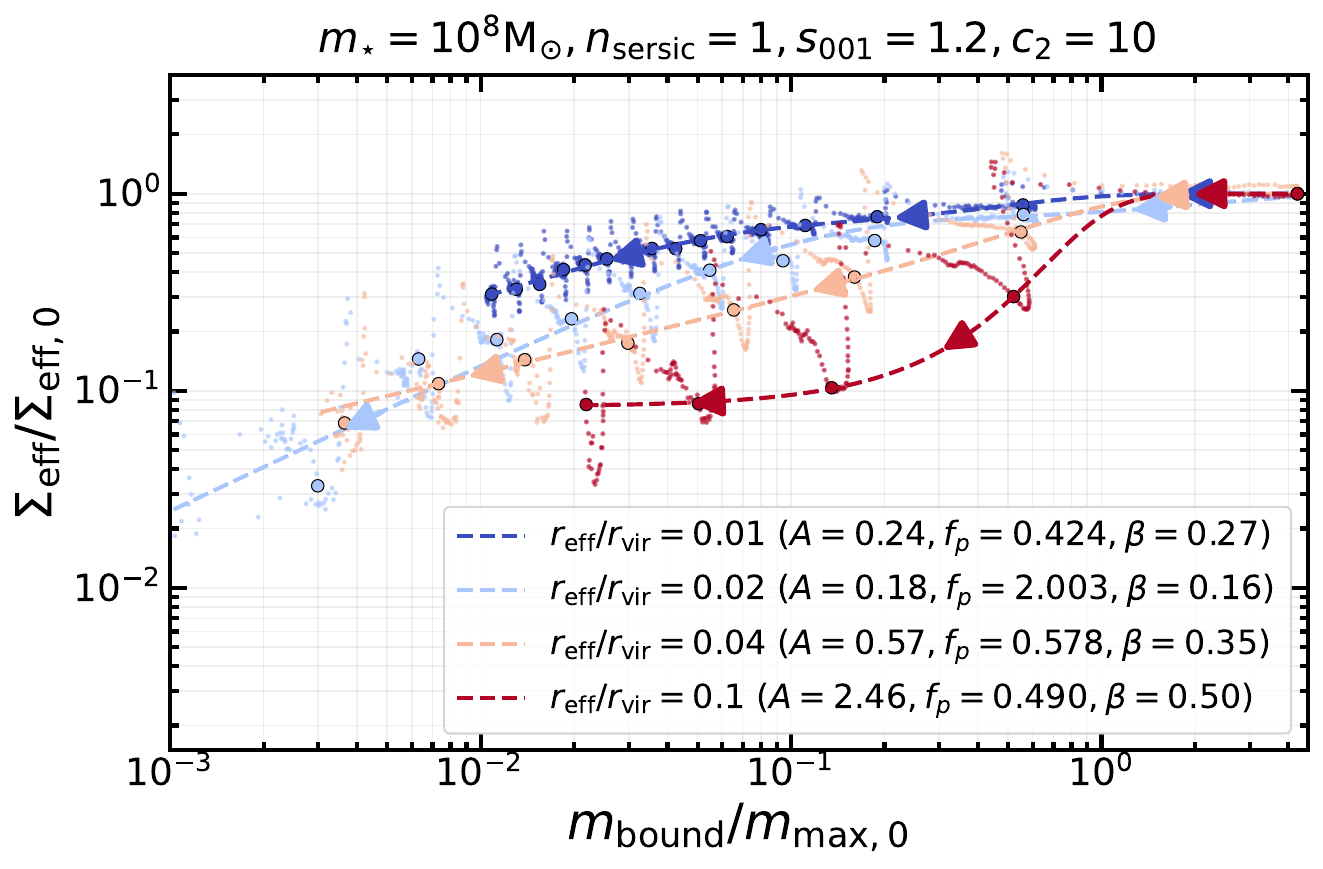}
    \end{subfigure}
    \centering
    \hfill
    \begin{subfigure}
        \centering
        \includegraphics[width=\linewidth]{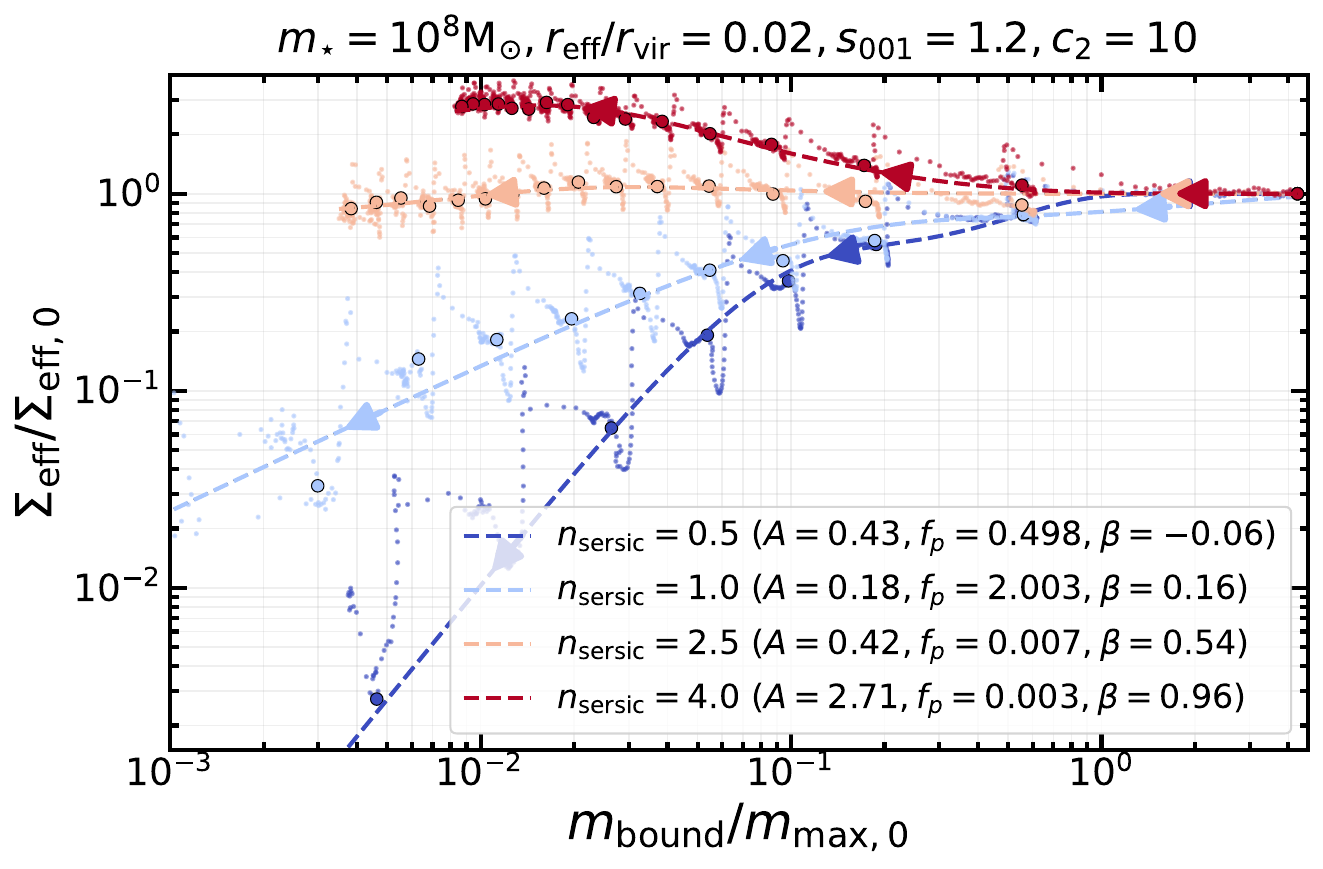}
    \end{subfigure}
    \caption{Tidal track of the mean stellar surface density within the half-mass radius, $\Sigma_{\rm eff}$ (Eq.~\ref{eq:sigma_eff}), scaled by its initial value $\Sigma_{\rm eff,0}$, as a function of the scaled bound mass $x_{\rm bound}$. Faint dotted lines show the simulation output at every snapshot, filled circles the output at each apocentre, and dashed lines the prediction of Eq.~\ref{eq:fSD} from the stellar-mass fit together with the best-fit size-track parameters ($A$, $f_{\rm p}$, $\beta$) listed in the legends. Arrows indicate the direction of evolution. \textit{Top panel:} $r_{\rm eff}/r_{\rm vir}$ is varied from $0.01$ (most compact, blue) to $0.1$ (most diffuse, dark red) with the other fiducial parameters fixed. \textit{Bottom panel:} $n_{\rm sersic}$ is varied from $0.5$ to $4$ with the other fiducial parameters fixed. Satellites with $\beta>0.5$ become denser as they are stripped, whereas those with $\beta<0.5$ fade towards the diffuse regime.}
    \label{fig:SB_track}
\end{figure}
We model the size tidal track using the following functional form:
\begin{equation}
    \frac{r_{\rm eff}}{r_{\rm eff,0}} =\left\{1 + A\exp\!\left[-\left(\frac{x_{\rm bound}}{f_{\rm p}}\right)^b \right] \right\} f_{\star}^{\beta},
    \label{eq:freff}
\end{equation}
where $f_{\star}$ is the remaining stellar mass fraction given by Eq.~\ref{eq:fstar}. The right-hand side decomposes the evolution of $r_{\rm eff}$ into two distinct contributions.

The factor $f_{\star}^{\beta}$ describes the secular contraction associated with stellar mass loss. Its power-law index $\beta$ measures the size response to stellar stripping, and we refer to it hereafter as the size-response index. It quantifies how strongly the half-mass radius shrinks as stars are lost: a larger $\beta$ implies a more pronounced contraction for a given fractional stellar mass loss, while $\beta<0$ implies that the half-mass radius grows as stars are removed.

The first factor describes an \textit{expansion} term, where the puff-up amplitude $A$ sets the strength of the expansion and the puff-up scale $f_{\rm p}$ is a characteristic value of $x_{\rm bound}$. When $x_{\rm bound} \gg f_{\rm p}$, this factor approaches unity and the size evolution is dominated by the contraction term. When $x_{\rm bound} \ll f_{\rm p}$, the expansion term moves towards $1+A$, with the same sharpness parameter $b$ controlling the rapidity of the transition. We emphasise that ``expansion'' here refers to a departure from the secular contraction that would otherwise result from stellar mass loss, rather than a growth of $r_{\rm eff}$ relative to $r_{\rm eff,0}$ in absolute terms. Whether the half-mass radius exceeds its initial value at any point, i.e.\ whether $r_{\rm eff}/r_{\rm eff,0} > 1$, depends on the specific combination of parameters.

Combining Eqs.~\ref{eq:fstar} and~\ref{eq:freff}, our joint tidal tracks of both stellar mass and size evolution include a total of five parameters: the half-mass stripping scale $f_{1/2}$ and stripping sharpness $b$ characterise the stellar mass evolution, while the puff-up amplitude $A$, the puff-up scale $f_{\rm p}$, and the size-response index $\beta$ characterise the size evolution. Fig.~\ref{fig:size_track} shows the size tidal tracks for the same simulations as in Fig.~\ref{fig:mass_track}, with the best-fit parameters listed in the legend of each panel. The dashed curves demonstrate that Eq.~\ref{eq:freff} provides a good description of the underlying secular trend across the parameter space explored, capturing both the two-phase contraction and the expansion phase. The expansion is most prominent for the $n_{\rm sersic} = 0.5$ case in the bottom panel, for which $\beta\simeq0$ and the half-mass radius keeps growing above $r_{\rm eff,0}$ as stripping proceeds. As discussed above, the model does not aim to reproduce the orbital-phase oscillations superimposed on this secular behaviour.

A natural corollary of Eqs.~\ref{eq:fstar} and~\ref{eq:freff} is that the evolution of the mean stellar surface density within the half-mass radius,
\begin{equation}
    \Sigma_{\rm eff}\equiv\frac{m_{\star}(<r_{\rm eff})}{\pi r_{\rm eff}^{2}}=\frac{m_{\star}}{2\pi r_{\rm eff}^{2}},
    \label{eq:sigma_eff}
\end{equation}
can be derived directly:
\begin{align}
    \frac{\Sigma_{\rm eff}}{\Sigma_{\rm eff,0}} &= \frac{f_{\star}}{(r_{\rm eff}/r_{\rm eff,0})^2}\nonumber\\
    &=\left\{1 + A\exp\!\left[-\left(\frac{x_{\rm bound}}{f_{\rm p}}\right)^b \right] \right\}^{-2} f_{\star}^{1-2\beta}.
    \label{eq:fSD}
\end{align}
The squared size dependence in Eq.~\ref{eq:fSD} makes the puff-up factor enter as the inverse square. Fig.~\ref{fig:SB_track} shows the resulting surface density tracks for the same simulations. The size-response index $\beta$ plays a decisive role in setting the direction of surface density evolution under tidal stripping. In the case of $\beta = 0.5$, the exponent $1 - 2\beta$ in the contraction term vanishes, and the surface density remains nearly invariant under tidal evolution (provided $A$ is not too large). For $\beta > 0.5$, the half-mass radius contracts faster than mass is lost, so that the surface density increases with stripping: the tidal remnant becomes progressively more compact, evolving in the direction of compact dwarfs. This behaviour is illustrated by the $n_{\rm sersic} = 4$ case in the bottom panel of Fig.~\ref{fig:SB_track} ($\beta = 0.96$), which reaches $\Sigma_{\rm eff}/\Sigma_{\rm eff,0} \simeq 3$ by the end of the simulation. Conversely, for $\beta < 0.5$, mass is lost faster than the size contracts, so the surface density decreases and the satellite fades towards the diffuse-dwarf regime. This behaviour is exemplified by the $n_{\rm sersic} = 0.5$ case ($\beta = -0.06$), in which $\Sigma_{\rm eff}/\Sigma_{\rm eff,0}$ drops by more than two orders of magnitude over the course of the simulation. Because $\beta$ is slightly negative in this case, the half-mass radius grows rather than shrinks as stars are lost, which further accelerates the decline in surface density.

\subsection{Tidal track for velocity dispersion}\label{sec:sigma_track}

\begin{figure}
    \centering
    \begin{subfigure}
        \centering
        \includegraphics[width=\linewidth]{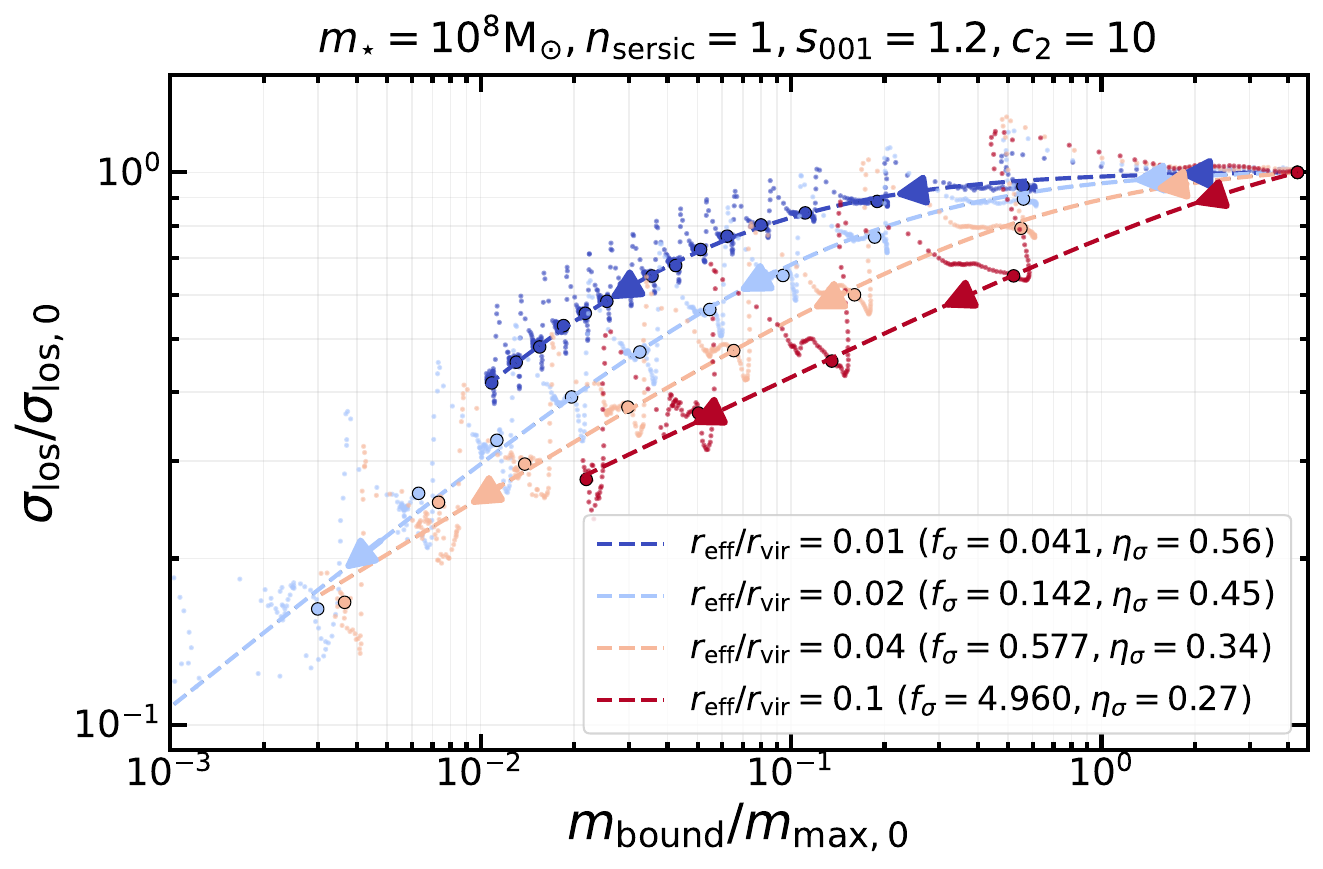}
    \end{subfigure}
    \centering
    \hfill
    \begin{subfigure}
        \centering
        \includegraphics[width=\linewidth]{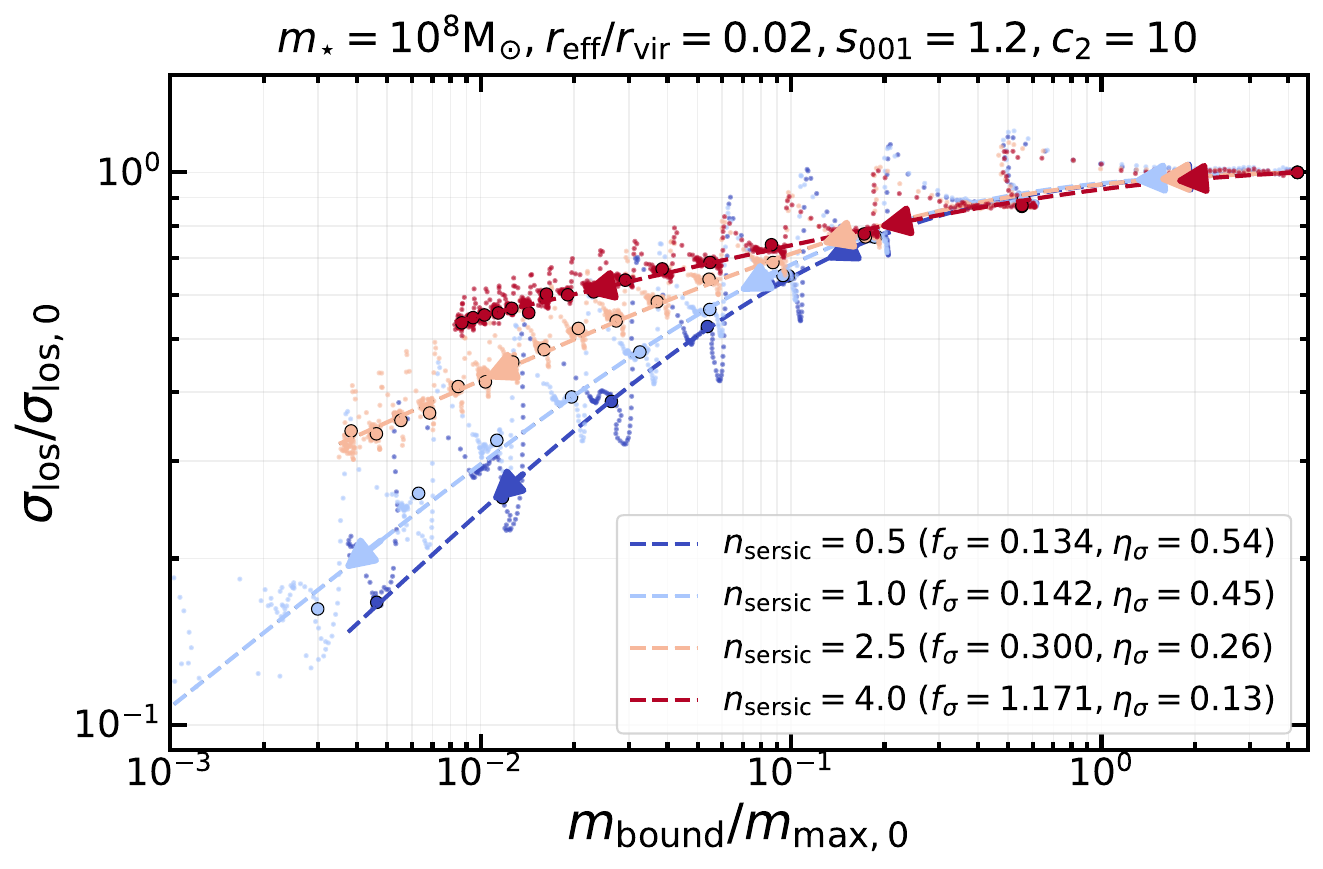}
    \end{subfigure}
    \caption{Line-of-sight velocity dispersion tidal track. The vertical axis shows the dispersion $\sigma_{\rm los}$ of the bound stars within the projected half-mass radius, averaged over three orthogonal projections and scaled by its initial value $\sigma_{\rm los,0}$. The horizontal axis shows the scaled bound mass $x_{\rm bound}$. Faint dotted lines show the simulation output at every snapshot, filled circles the output at each apocentre, and dashed curves the best fits of Eq.~\ref{eq:fsigma}, with the fitted dispersion transition scale $f_{\sigma}$ and decline index $\eta_{\sigma}$ given in the legends. Arrows indicate the direction of evolution. \textit{Top panel:} $r_{\rm eff}/r_{\rm vir}$ is varied from $0.01$ (most compact, blue) to $0.1$ (most diffuse, dark red) at fixed $m_{\star}=10^{8}\,\rm{M}_{\odot}$, $c_2=10$, $s_{001}=1.2$, and $n_{\rm sersic}=1$. \textit{Bottom panel:} $n_{\rm sersic}$ is varied from $0.5$ to $4$ at fixed $r_{\rm eff}/r_{\rm vir}=0.02$ and the other fiducial parameters. The dispersion follows a two-phase track whose onset is set mainly by how deeply the stars are embedded, through $f_{\sigma}$, and whose late-time slope is set mainly by the concentration of the stellar profile, through $\eta_{\sigma}$.}
    \label{fig:sigma_los_track}
\end{figure}

The internal kinematics of a satellite provide a further probe of its tidal evolution, and one that is directly accessible to spectroscopic observations \citep{Penarrubia2008, Penarrubia2010, Errani2022}. We therefore also follow the line-of-sight velocity dispersion $\sigma_{\rm los}$ of the bound stars, measured within the projected half-mass radius and averaged over three orthogonal projections as described in Section~\ref{sec:nbody}. 

Fig.~\ref{fig:sigma_los_track} shows the resulting tracks, $\sigma_{\rm los}/\sigma_{\rm los,0}$ as a function of $x_{\rm bound}$, for the same satellites presented in Fig.~\ref{fig:mass_track}. The behaviour parallels that of the stellar mass track: while stripping is confined to the outer DM envelope, the central kinematics are essentially unaffected, and $\sigma_{\rm los}$ begins to decline only once the stripping front reaches the stellar body. Superimposed on this secular trend are the same orbital-phase oscillations that affect the size track, visible as excursions of the faint dotted lines near each pericentric passage. As for the size track, we therefore fit the apocentre snapshots, at which the remnant is closest to a relaxed state. We describe the secular behaviour with the two-parameter form
\begin{equation}
    \frac{\sigma_{\rm los}}{\sigma_{\rm los,0}}
    =\left[\frac{1+f_{\sigma}/x_{\rm bound,0}}{1+f_{\sigma}/x_{\rm bound}}\right]^{\eta_{\sigma}},
    \label{eq:fsigma}
\end{equation}
where the dispersion transition scale $f_{\sigma}$ marks the transition between the two phases and the dispersion decline index $\eta_{\sigma}$ sets the asymptotic power-law slope of the decline. 
The numerator ensures that $\sigma_{\rm los}=\sigma_{\rm los,0}$ in the initial state, $x_{\rm bound}=x_{\rm bound,0}$. For $f_{\sigma}\ll x_{\rm bound}\leq x_{\rm bound,0}$ the bracket remains close to unity, reproducing the plateau, whereas for $x_{\rm bound}\ll f_{\sigma}$,
\begin{equation}
    \frac{\sigma_{\rm los}}{\sigma_{\rm los,0}}
    \simeq\left(1+\frac{f_{\sigma}}{x_{\rm bound,0}}\right)^{\eta_{\sigma}}
    \left(\frac{x_{\rm bound}}{f_{\sigma}}\right)^{\eta_{\sigma}}.
    \label{eq:fsigma_asymptotic}
\end{equation}
Eqs.~\ref{eq:fsigma} and~\ref{eq:fsigma_asymptotic} therefore mirror the two-phase structure of the stellar mass track, with $f_{\sigma}$ playing the role of $f_{1/2}$ and $\eta_{\sigma}$ that of $b$.

The top panel of Fig.~\ref{fig:sigma_los_track} varies the stellar compactness around the fiducial dwarf. As for the mass track, compactness is the dominant control on the transition scale: $f_{\sigma}$ increases monotonically from $0.041$ at $r_{\rm eff}/r_{\rm vir}=0.01$ to $4.96$ at $r_{\rm eff}/r_{\rm vir}=0.1$. Compact dwarfs therefore preserve their central dispersion until they have been stripped very deeply, whereas the most diffuse system has $f_{\sigma}$ larger than its initial value $x_{\rm bound,0}\simeq4.4$, so that its dispersion declines from the onset of stripping without any plateau. The fitted slope decreases in the opposite sense, from $\eta_{\sigma}=0.56$ to $0.27$, although this ordering should be read with caution because $f_{\sigma}$ and $\eta_{\sigma}$ are covariant when the track samples only a limited range beyond the transition, in the same way as the puff-up parameters discussed in Section~\ref{sec:param_dependence}.

The bottom panel varies the S\'ersic index. Here the transition scale changes by less than an order of magnitude, from $f_{\sigma}=0.134$ at $n_{\rm sersic}=0.5$ to $1.17$ at $n_{\rm sersic}=4$, whereas the slope changes substantially, from $\eta_{\sigma}=0.54$ to $0.13$. Stellar bodies with shallow central profiles therefore lose their central kinematic support rapidly once stripping reaches them, while concentrated profiles retain a nearly constant dispersion deep into the stripped regime. This is the kinematic counterpart of the trends identified for the mass and size tracks, in which high-$n_{\rm sersic}$ dwarfs both retain their stars and contract, and thereby preserve a deep central potential.

\centerwidetable
\begin{deluxetable*}{l p{0.28\textwidth} p{0.44\textwidth}}
\tabletypesize{\footnotesize}
\setlength{\tabcolsep}{3pt}
\tablewidth{0pt}
\tablecaption{Empirical stellar tidal tracks and the physical meanings of their parameters.\label{tab:tracks}}
\tablehead{\colhead{Target quantity} & \colhead{Functional form ($x\equiv x_{\rm bound}=m_{\rm bound}/m_{\rm max,0}$)} & \colhead{Parameters and interpretation}}
\startdata
Stellar mass & $f_{\star}\equiv m_{\star}/m_{\star,0}=1-e^{-\ln2\,\left (\frac{x}{f_{1/2}}\right )^b}$ & $f_{1/2}$ (transition scale): scaled bound mass at which half of the initial stars remain. $b$ (sharpness): late-stage slope, $f_{\star}\propto x^b$. \\[5pt]
Stellar-to-bound mass ratio & $\xi/\xi_0=(x_0/x)f_{\star}(x)$ & $\xi=m_{\star}/m_{\rm bound}$, no additional fit parameters. At late stages, $\xi/\xi_0\propto x^{b-1}$. \\[5pt]
Half-mass radius & $r_{\rm eff}/r_{\rm eff,0}=P(x)f_{\star}^{\beta}$ & 
$P(x)\equiv1+A\exp[-(x/f_{\rm p})^b]$, $A$: asymptotic excess of the puff-up factor above unity. $f_{\rm p}$: its characteristic bound-mass scale. $\beta$: size response to stellar mass loss. \\[5pt]
Mean surface density & $\Sigma_{\rm eff}/\Sigma_{\rm eff,0}=P(x)^{-2}f_{\star}^{1-2\beta}$ & No additional fit parameters. $\beta>1/2$ gives increasing surface density as stars are lost, $\beta<1/2$ gives decreasing surface density. \\[5pt]
Velocity dispersion & $\sigma_{\rm los}/\sigma_{\rm los,0}=[(1+f_{\sigma}/x_0)/(1+f_{\sigma}/x)]^{\eta_{\sigma}}$ & $f_{\sigma}$: transition scale of the dispersion decline. $\eta_{\sigma}$: late-stage slope, $\sigma_{\rm los}\propto x^{\eta_{\sigma}}$. \\
\enddata
\tablecomments{The mass and size tracks share $f_{1/2}$ and $b$. $A$, $f_{\rm p}$, and $\beta$ add three parameters for size, and $f_{\sigma}$ and $\eta_{\sigma}$ add two for dispersion, giving seven in total. The stellar mass-fraction and surface-density tracks are derived from the mass and size tracks, not fitted with independent parameters. The full forms and limitations are discussed in Eqs.~\ref{eq:fstar}, \ref{eq:fxi}, \ref{eq:freff}, \ref{eq:fSD}, and~\ref{eq:fsigma}.}
\end{deluxetable*}

\section{Dependence of Track Parameters on Initial Structure}\label{sec:param_dependence}

\begin{figure}
    \centering
    \includegraphics[width=\linewidth]{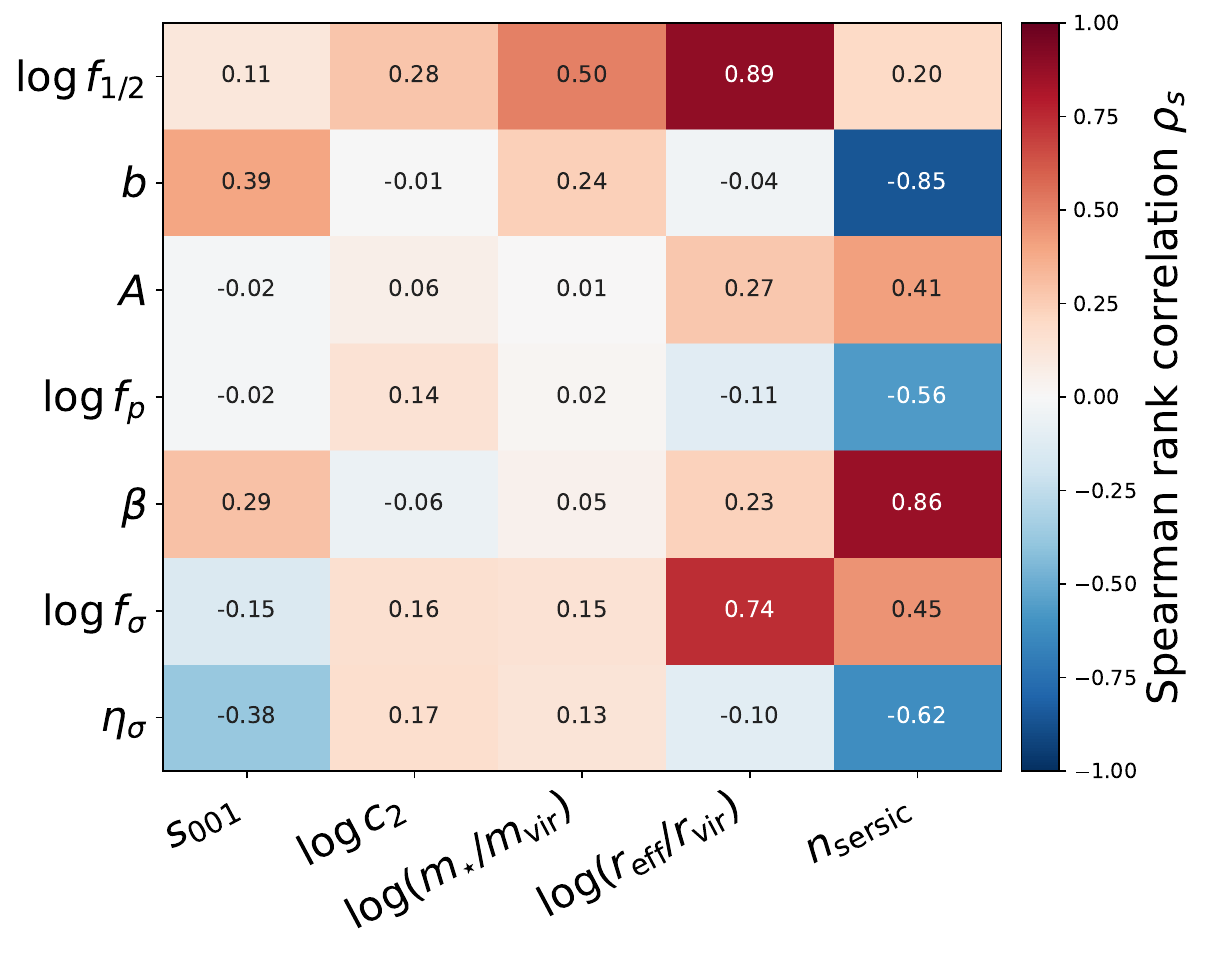}
    \caption{Spearman rank-correlation matrix between the initial structural parameters (columns) and the fitted tidal-track parameters (rows): the half-mass stripping scale $f_{1/2}$ and stripping sharpness $b$ of the stellar mass track, the puff-up amplitude $A$, puff-up scale $f_{\rm p}$, and size-response index $\beta$ of the size track, and the transition scale $f_{\sigma}$ and decline index $\eta_{\sigma}$ of the velocity dispersion track. Each element is the Spearman coefficient $\rho_{\rm S}$, computed using all systems for which both quantities are available. Red and blue denote positive and negative monotonic correlations, respectively, and values close to zero indicate little monotonic association. The strongest associations are those of $f_{1/2}$ and $f_{\sigma}$ with the stellar compactness ($r_{\rm eff}/r_{\rm vir}$), and of $b$, $\beta$, and $\eta_{\sigma}$ with the S\'ersic index.}
    \label{fig:correlation}
\end{figure}

\begin{figure}
    \centering
    \begin{subfigure}
        \centering
        \includegraphics[width=\linewidth]{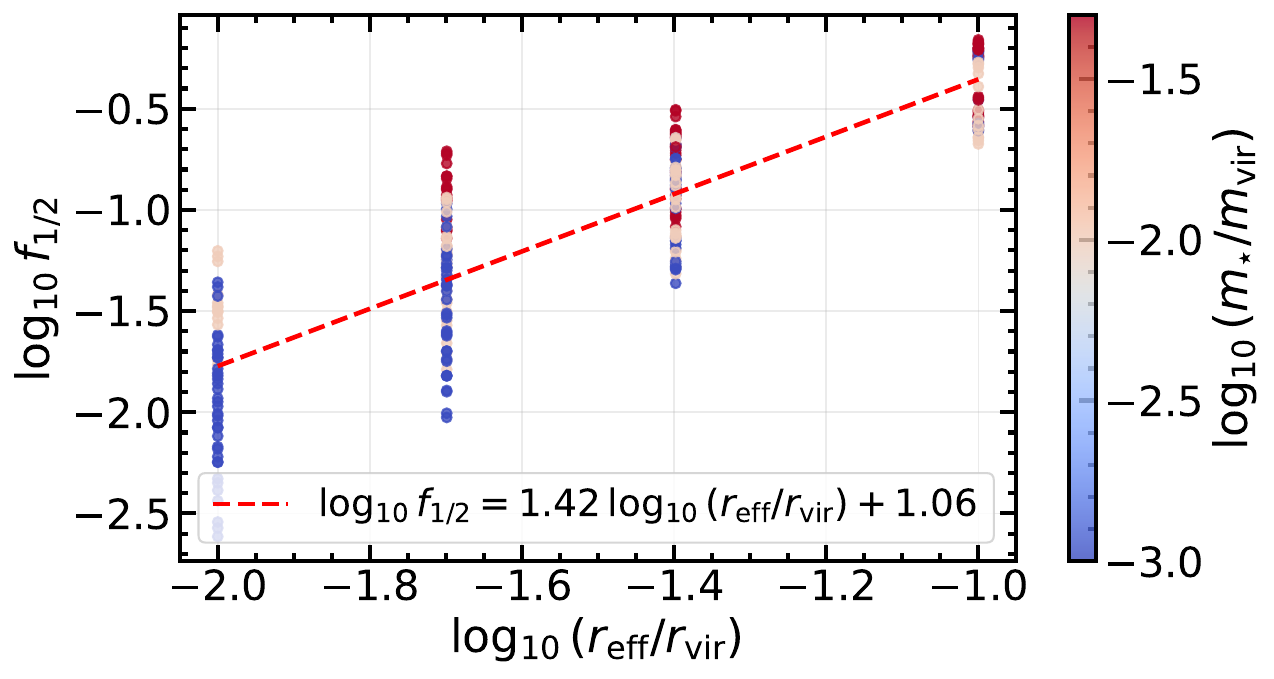}
    \end{subfigure}
    \centering
    \vfill
    \begin{subfigure}
        \centering
        \includegraphics[width=\linewidth]{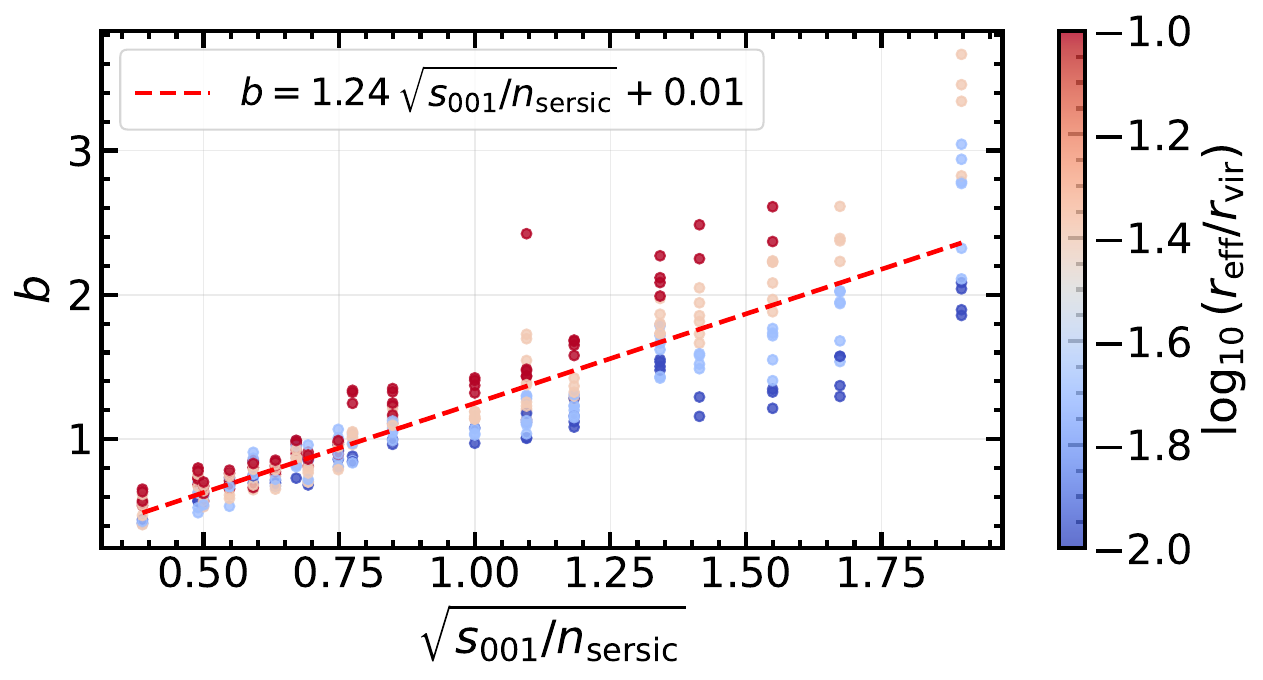}
    \end{subfigure}
    \vfill
    \begin{subfigure}
        \centering
        \includegraphics[width=\linewidth]{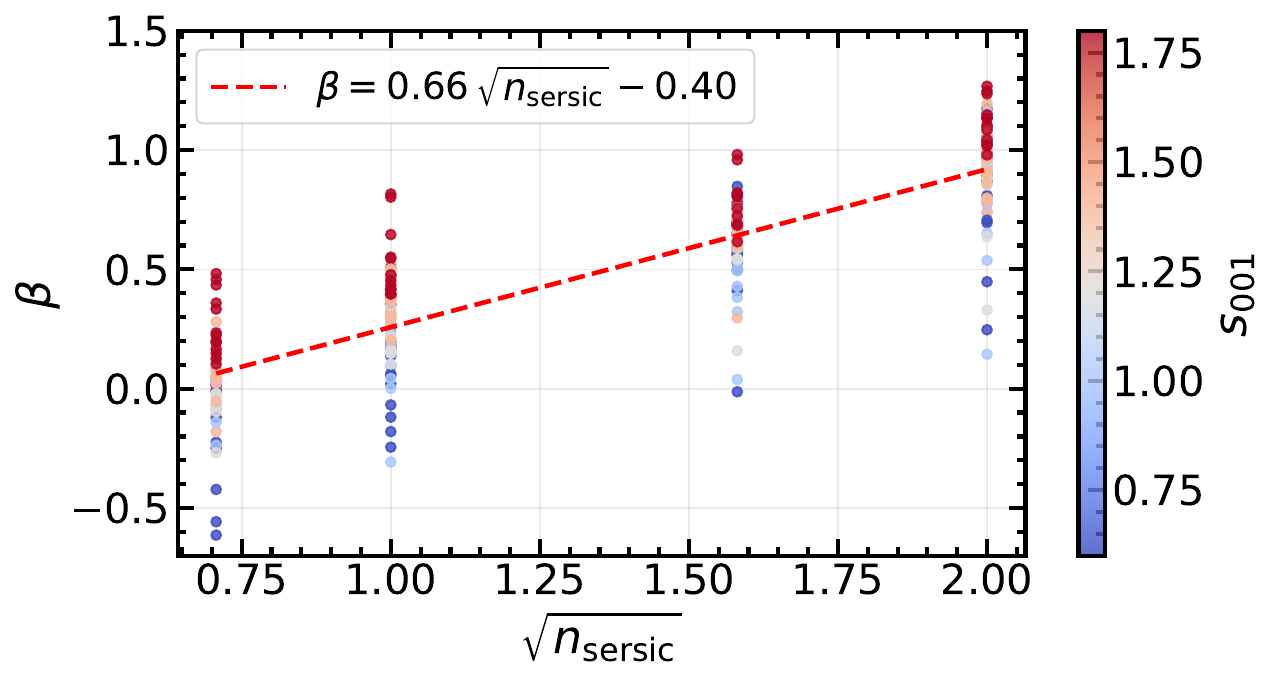}
    \end{subfigure}
    \caption{Dependence of the three best-determined tidal-track parameters on their dominant initial-structure combinations. Each point is one simulation, and dashed lines show the first-order relations of Table~\ref{tab:dependence}. \textit{Top:} half-mass stripping scale, $\log_{10}f_{1/2}$, versus $\log_{10}(r_{\rm eff}/r_{\rm vir})$, colour-coded by $\log_{10}(m_{\star}/m_{\rm vir})$. \textit{Middle:} stripping sharpness $b$ versus $\sqrt{s_{001}/n_{\rm sersic}}$, colour-coded by $\log_{10}(r_{\rm eff}/r_{\rm vir})$. \textit{Bottom:} size-response index $\beta$ versus $\sqrt{n_{\rm sersic}}$, colour-coded by $s_{001}$. The colour gradients at fixed abscissa reveal the leading secondary dependences: a higher stellar mass fraction raises $f_{1/2}$, a more diffuse stellar body raises $b$, and a cuspier halo raises $\beta$.}
    \label{fig:parameter}
\end{figure}

Having established the functional forms of the stellar mass, size, and velocity dispersion tidal tracks (Eqs.~\ref{eq:fstar}, \ref{eq:freff}, and~\ref{eq:fsigma}), we now ask how they are set by the initial dwarf structure, parameterised by $(s_{001},\,c_{2},\,m_{\star}/m_{\rm vir},\,r_{\rm eff}/r_{\rm vir},\,n_{\rm sersic})$. The mass-track parameters are measured from all snapshots, whereas the size, surface-density, and velocity dispersion tracks are fit to apocentre snapshots in order to suppress orbital-phase oscillations and isolate the relaxed secular evolution. The half-mass radius and surface density are fit jointly in logarithmic space, since $\Sigma_{\rm eff}/\Sigma_{\rm eff,0}=f_{\star}/(r_{\rm eff}/r_{\rm eff,0})^2$ and the two observables provide complementary leverage on $\beta$. Our analysis proceeds in two steps. We first use a correlation analysis to identify the structural parameters on which each track parameter depends most strongly, and then derive empirical fits for a few tidal-track parameters that are strongly correlated with initial structural parameters. The resulting relations are collected in Table~\ref{tab:dependence}.

The correlation analysis is based on the Spearman rank-correlation coefficient $\rho_{\rm S}$. For two variables $X$ and $Y$ measured in $N$ simulations, each value is first replaced by its rank, that is, by its position, when the $N$ values of that variable are sorted in increasing order, with tied values assigned their average rank. The Spearman coefficient is then the ordinary correlation coefficient between the two sets of ranks.
A value $\rho_{\rm S}=+1$ ($-1$) means that $Y$ increases (decreases) strictly with $X$, while $\rho_{\rm S}\simeq0$ does not indicate a monotonic trend. 
Correlations among fitted track parameters can also reflect degeneracies in the fitting function itself. To quantify the predictive power of the calibrated relations, we use the coefficient of determination $R^2=1-\sum_i(y_i-\hat{y}_i)^2/\sum_i(y_i-\bar{y})^2$, where $y_i$ is the track parameter fitted to simulation $i$ (or its logarithm, where stated), $\hat{y}_i$ is the value predicted by the relation, and $\bar{y}$ is the sample mean. $R^2=1$ corresponds to a perfect prediction and $R^2=0$ to no improvement over the sample mean.

Several robust trends are highlighted in Fig.~\ref{fig:correlation}. The half-mass stripping scale is dominated by the initial stellar compactness, with $\rho_{\rm S}[\log f_{1/2},\log(r_{\rm eff}/r_{\rm vir})]=0.89$. The next strongest marginal correlations are with $m_{\star}/m_{\rm vir}$ and $c_2$, with $\rho_{\rm S}=0.50$ and $0.28$, respectively. The stripping sharpness $b$ is strongly anticorrelated with $n_{\rm sersic}$ ($\rho_{\rm S}=-0.85$) and positively correlated with $s_{001}$ ($\rho_{\rm S}=0.39$), while the size-response index $\beta$ shows the opposite trend, strongly correlated with $n_{\rm sersic}$ ($\rho_{\rm S}=0.86$) and more weakly with $s_{001}$ ($\rho_{\rm S}=0.29$). The puff-up parameters are less cleanly tied to the initial structure: $A$ correlates most strongly with $n_{\rm sersic}$ ($\rho_{\rm S}=0.41$), whereas $f_{\rm p}$ anticorrelates with it ($\rho_{\rm S}=-0.56$). The velocity dispersion parameters repeat the pattern of the mass track in a weaker form: $f_{\sigma}$ correlates most strongly with the stellar compactness ($\rho_{\rm S}=0.74$) and more weakly with $n_{\rm sersic}$ ($\rho_{\rm S}=0.45$), whereas $\eta_{\sigma}$ anticorrelates with $n_{\rm sersic}$ ($\rho_{\rm S}=-0.62$) and $s_{001}$ ($\rho_{\rm S}=-0.38$).

\begin{deluxetable*}{lll}
\tabletypesize{\footnotesize}
\tablewidth{0pt}
\tablecaption{Calibrated dependence of the tidal-track parameters on the initial structure of the dwarf.\label{tab:dependence}}
\tablehead{
\colhead{Parameter} & \colhead{Relation} & \colhead{$R^2$}
}
\startdata
\cutinhead{First-order dependence}
$f_{1/2}$ & $\log f_{1/2}=1.063+1.418\log(r_{\rm eff}/r_{\rm vir})$ & 0.761 \\
$b$ & $b=0.010+1.239\,(s_{001}/n_{\rm sersic})^{1/2}$ & 0.812 \\
$\beta$ & $\beta=-0.405+0.662\,n_{\rm sersic}^{1/2}$ & 0.679 \\
$f_{\sigma}$ & $\log f_{\sigma}=1.598+1.311\log(r_{\rm eff}/r_{\rm vir})$ & 0.485 \\
$\eta_{\sigma}$ & $\eta_{\sigma}=0.337\,n_{\rm sersic}^{-0.435}$ & 0.341 \\
\cutinhead{Including secondary dependences}
$f_{1/2}$ & $f_{1/2}=1.50\,(r_{\rm eff}/r_{\rm vir})^{1.326}\,c_2^{1.194}\,(m_{\star}/m_{\rm vir})^{0.154}\,s_{001}^{0.751}$ & 0.944 \\
$b$ & $b=1.816\,s_{001}^{0.496}\,n_{\rm sersic}^{-0.489}\,(r_{\rm eff}/r_{\rm vir})^{0.101}\,$ & 0.925 \\
$\beta$ & $\beta=-0.777+0.660\,n_{\rm sersic}^{1/2}+0.303\,s_{001}$ & 0.768 \\
$f_{\sigma}$ & $f_{\sigma}=4.05\,(r_{\rm eff}/r_{\rm vir})^{1.309}\,c_2^{0.710}\,(m_{\star}/m_{\rm vir})^{-0.137}\,n_{\rm sersic}^{0.407}\,s_{001}^{-0.160}$ & 0.594 \\
$\eta_{\sigma}$ & $\eta_{\sigma}=0.361\,n_{\rm sersic}^{-0.433}\,s_{001}^{-0.452}$ & 0.416 \\
\enddata
\tablecomments{Parameters: $f_{1/2}$, half-mass stripping scale, and $b$, stripping sharpness (Eq.~\ref{eq:fstar}), $\beta$, size-response index (Eq.~\ref{eq:freff}), $f_{\sigma}$, dispersion transition scale, and $\eta_{\sigma}$, dispersion decline index (Eq.~\ref{eq:fsigma}). First-order relations use only the dominant structural parameter or combination of parameters. The relations in the lower block include the leading secondary dependence on initial structural parameters. All logarithms are decimal. The coefficient of determination $R^2$ is evaluated in $\log f_{1/2}$, $\log f_{\sigma}$, and $\log\eta_{\sigma}$, in $\log b$ for the multivariate relation of $b$, and in linear units for the first-order relation of $b$ and for $\beta$. The puff-up parameters $A$ and $f_{\rm p}$ are not calibrated because of their strong mutual degeneracy.}
\end{deluxetable*}

The dominant dependence of $f_{1/2}$ on $r_{\rm eff}/r_{\rm vir}$ is physically expected. More compact stellar systems are more deeply embedded within the DM halo and can undergo substantially more DM stripping before half of their stars are removed. A single-variable fit gives $f_{1/2}\simeq11.57\,(r_{\rm eff}/r_{\rm vir})^{1.418}$ with $R^2=0.761$ in $\log f_{1/2}$ (Table~\ref{tab:dependence} and top panel of Fig.~\ref{fig:parameter}). The remaining scatter contains genuine secondary structural information. Allowing all initial structural parameters to enter as a power law raises the coefficient of determination to $R^2=0.944$. 

The half-mass stripping scale can be compared with previous measurements once the difference in normalisation is taken into account. \citet{He2026}, using satellites in the \textsc{COLIBRE} simulations, measured a median half-mass stripping scale of $f_{1/2}\simeq 0.04$, with a typical range of $0.024$--$0.065$. Independently, \citet{Wan2026}, combining JWST observations of dwarf satellites at cosmic noon with models of satellite tidal evolution, found that the observations allow $f_{1/2}\simeq 0.015$--$0.09$, broadly consistent with the \textsc{COLIBRE} results. Both studies express the stellar mass loss in terms of the bound mass fraction $f_{\rm bound}=m_{\rm bound}/m_{\rm bound,0}$, rather than our scaled bound mass $x$. From Eq.~\ref{eq:xbound}, the half-mass scale in their convention corresponds to $f_{1/2}/x_{\rm bound,0}$ in our model. Since $x_{\rm bound,0}$ ranges from approximately 3 to 5, depending on concentration, our values of $f_{1/2}$ are systematically larger by the same factor. The dependence on compactness can nevertheless be compared directly. Both our simulations and the \textsc{COLIBRE} satellites of \citet{He2026} show that $f_{1/2}$ increases with the size of the galaxy relative to its halo, i.e.\ that more extended stellar bodies begin to lose their stars earlier. The logarithmic slope that we find, $\simeq1.4$, is considerably steeper than the slope of about $0.5$ found by \citet{He2026}. Several factors may contribute to this difference. First, our compactness is defined through the projected half-mass radius, whereas \citet{He2026} use the three-dimensional stellar half-mass radius. The ratio between the two depends on the stellar profile, so the two measures are not related by a single constant factor across a population. Second, satellites in cosmological simulations are resolved with far fewer particles, and collisional heating of stellar particles by more massive DM particles can spuriously inflate the sizes of poorly resolved galaxies \citep{Ludlow2021, Ludlow2023}, which compresses the dynamic range in compactness and can bias the fitted slope. Third, \textsc{COLIBRE} is a hydrodynamical simulation in which the gas content of satellites, its removal, and its conversion into stars all modify the potential in which the stars are embedded, whereas our controlled simulations do not model these effects. 

For sharpness $b$, the pairwise correlation matrix makes $n_{\rm sersic}$ appear substantially more important than $s_{001}$. Symbolic regression instead identifies the compact combination $\sqrt{s_{001}/n_{\rm sersic}}$, and a direct calibration gives $b\simeq0.010+1.239\sqrt{s_{001}/n_{\rm sersic}}$ with $R^2=0.812$ (middle panel of Fig.~\ref{fig:parameter}). The fitted intercept is negligible. Forcing the relation through the origin gives $b\simeq1.247\sqrt{s_{001}/n_{\rm sersic}}$ with the same $R^2$ to three significant figures. The difference between the Spearman coefficients of $s_{001}$ and $n_{\rm sersic}$ is therefore not inconsistent with comparable logarithmic sensitivities to the two parameters. Across the simulation grid, $n_{\rm sersic}$ spans a factor of 8 whereas $s_{001}$ spans a factor of 3. So $n_{\rm sersic}$ produces a larger variation in $b$ and hence dominates its rank ordering. Allowing all initial structural parameters to enter gives $R^2=0.925$ in $\log b$ (Table~\ref{tab:dependence}). The remaining exponents are small, demonstrating that the leading dependence is well described by $b\propto\sqrt{s_{001}/n_{\rm sersic}}$ with only a weak correction from the initial stellar compactness.

The size-response index $\beta$ is controlled primarily by the S\'ersic index. Among simple one-variable transformations, the relation $\beta\simeq-0.405+0.662\sqrt{n_{\rm sersic}}$ gives $R^2=0.679$ (bottom panel of Fig.~\ref{fig:parameter}). The leading secondary dependence is on the inner DM slope: adding a term linear in $s_{001}$ increases the coefficient of determination to $R^2=0.768$. Thus almost all of the predictable variation in $\beta$ is captured by $n_{\rm sersic}$ and $s_{001}$, while the other structural parameters provide only minor corrections. Physically, a larger $\beta$ means that the half-mass radius contracts more rapidly for a given amount of stellar mass loss. Its strong increase with $n_{\rm sersic}$ therefore links the initial stellar concentration directly to the compact-versus-diffuse bifurcation of tidally stripped remnants.

The puff-up parameters $A$ and $f_{\rm p}$ describe the same exponential term,
\begin{equation}
    E(x_{\rm bound}) \equiv A\exp\!\left[-\left(\frac{x_{\rm bound}}{f_{\rm p}}\right)^b\right],
    \label{eq:puffup}
\end{equation}
and are therefore intrinsically covariant whenever the track samples only a limited range around the onset of the puff-up. At a fixed pivot $x_{\rm bound}=x$, keeping $E(x)$ approximately constant gives
\begin{equation}
    {\rm d}\ln A \simeq
    -b\left(\frac{x}{f_{\rm p}}\right)^b {\rm d}\ln f_{\rm p},
    \label{eq:Afp_degeneracy}
\end{equation}
so that a larger $f_{\rm p}$ can be compensated by a smaller $A$. This degeneracy is directly reflected in the fitted sample, for which $A$ and $\log f_{\rm p}$ are strongly anticorrelated, $\rho_{\rm S}=-0.76$. The individual structural correlations of $A$ and $f_{\rm p}$ should therefore not be interpreted as independent physical laws: they mix genuine changes in the size response with movement along a fitting degeneracy. 
We therefore do not attempt to provide a universal calibration for $A$ and $f_{\rm p}$ separately. 

The poor determination of the puff-up term has several origins. First, we measure the size tracks at orbital apocentre, but the cored satellites that are most prone to puffing up are also the most readily disrupted, often surviving only a few orbital periods. Their tracks are then sampled by no more than two or three apocentre points, too few to constrain the puff-up parameters. Second, the puff-up is not a monotonic function of the inner DM slope: it appears most clearly in cored satellites with low $n_{\rm sersic}$, yet a non-negligible expansion term persists even for cuspier or higher-$n_{\rm sersic}$ systems, so the effect cannot be predicted from a single structural parameter. Third, the puff-up typically develops only in the heavily stripped regime at very low $x_{\rm bound}$. For a fixed orbit, many satellites are not stripped deeply enough to reach this stage, so the relevant portion of their track is simply not sampled.

The velocity dispersion track shows qualitatively similar but substantially weaker structural predictability. The one-variable calibration of the transition scale on stellar compactness gives $f_{\sigma}\simeq39.6\,(r_{\rm eff}/r_{\rm vir})^{1.311}$ with $R^2=0.485$ in $\log f_{\sigma}$, and allowing all structural parameters raises this only to $R^2=0.594$. The decline index is even less predictable. A single-variable relation in $n_{\rm sersic}$ gives $R^2=0.341$ in $\log\eta_{\sigma}$. Including $s_{001}$ raises this to $0.416$. The inverse dependence of $\eta_{\sigma}$ on both $n_{\rm sersic}$ and $s_{001}$ is physically sensible, since a more concentrated stellar profile or a cuspier inner halo produces a more resilient central potential and hence a shallower late-time decline of the dispersion. The weaker predictability has three origins. First, $f_{\sigma}$ and $\eta_{\sigma}$ are themselves moderately anticorrelated ($\rho_{\rm S}=-0.45$), so part of the scatter reflects movement along a fitting degeneracy rather than a genuine change in the kinematic response. Second, a non-negligible fraction of the fits approach the adopted parameter limits. Third, because the dispersion is measured within $r_{\rm eff}$ from a bound particle set that becomes small in the heavily stripped regime, the late-time portion of the track that determines $\eta_{\sigma}$ is also the noisiest. 



\section{Discussion}\label{sec:discussion}
Our simulations rely on several simplifying assumptions that define the applicability of the tidal tracks presented in this work.

First, the satellites contain no gas. In real dwarf galaxies, gas removal by ram pressure, gas outflows and consumption during post-infall star formation, and feedback-driven outflows can modify the gravitational potential and consequently the tidal response of both stars and DM. Second, both components are initialised with isotropic distribution functions, $f(E)$, and we do not explore anisotropic models of the form $f(E,L)$. \citet{Chiang2025} showed that radially anisotropic subhaloes lose mass considerably faster than isotropic ones and can undergo complete disruption, whereas tangentially anisotropic subhaloes are more resilient. Our tidal tracks should therefore be interpreted as predictions for initially isotropic dwarfs.

Third, the stellar component is represented by a spherical Sérsic spheroid, without flattening or a rotating disc. While this provides a reasonable approximation for dwarf spheroidal galaxies, the tidal response of flattened or rotation-supported dwarfs may differ from the tracks presented here. Finally, our analysis focuses on the evolution of integrated stellar quantities, with the Sérsic index entering as a parameter of the initial stellar distribution. The evolution of the full stellar density profile, including changes in the Sérsic index, together with the evolution of the DM density profile, will be investigated in future work.

\section{Conclusion \& Summary}\label{sec:conclusion}

We have investigated the tidal evolution of the stellar structure of satellite dwarf galaxies using a suite of $480$ idealised $N$-body simulations spanning a broad range of initial structural parameters. Each satellite is modelled as a two-component system, with a Dekel--Zhao DM halo characterised by its concentration $c_{2}$ and inner slope $s_{001}$, and a S\'ersic stellar component characterised by its stellar-to-halo mass ratio $m_{\star}/m_{\rm vir}$, compactness $r_{\rm eff}/r_{\rm vir}$, and S\'ersic index $n_{\rm sersic}$. The initial conditions are generated self-consistently with \textsc{Agama}, and the satellites are evolved for $20\,$Gyr on fixed orbits within a static Milky-Way-mass NFW host potential using \textsc{Gadget-4}, with the bound remnant identified and tracked at each snapshot using \textsc{hbt+}\xspace. Our main results are as follows.

\begin{enumerate}

\item We characterise the evolution of the stellar mass, half-mass radius, mean surface density, and line-of-sight velocity dispersion of each satellite as a function of its scaled bound mass $x_{\rm bound}=m_{\rm bound}/m_{\rm max,0}$, and refer to these relations as \emph{tidal tracks} (Section~\ref{sec:results}). Normalising the bound mass by the initial mass within $r_{\rm max}$ removes the dependence on the arbitrary outer truncation of the initial satellite and keeps the variable well defined for DM-deficient remnants. 

\item The stellar mass track (Eq.~\ref{eq:fstar}) is well described by a two-phase function,
\[
f_{\star} \equiv \frac{m_{\star}}{m_{\star,0}}
= 1-\exp\left[-\ln 2
\left(\frac{x_{\rm bound}}{f_{1/2}}\right)^b\right],
\]
with two parameters: the half-mass stripping scale $f_{1/2}$, which marks the onset of stellar stripping and quantifies how long the stars are shielded by the surrounding DM, and the stripping sharpness $b$, which sets the rate of stellar mass loss relative to the DM in the second phase.

\item The size track (Eq.~\ref{eq:freff}) follows
\[
\frac{r_{\rm eff}}{r_{\rm eff,0}}
=
\left\{
1+A\exp\left[
-\left(\frac{x_{\rm bound}}{f_{\rm p}}\right)^b
\right]
\right\}f_{\star}^{\beta}.
\]
It comprises a secular contraction driven by stellar mass loss and governed by the size-response index $\beta$, an expansion (``puff-up'') term with amplitude $A$ and scale $f_{\rm p}$, and orbital-phase oscillations that we do not model. The sign of $\beta-0.5$ determines the direction of surface-density evolution (Eq.~\ref{eq:fSD}):
\[\frac{\Sigma_{\rm eff}}{\Sigma_{\rm eff,0}} =\left\{1 + A\exp\!\left[-\left(\frac{x_{\rm bound}}{f_{\rm p}}\right)^b \right] \right\}^{-2} f_{\star}^{1-2\beta},\]
satellites with $\beta>0.5$ are compressed towards higher surface densities, while those with $\beta<0.5$ fade towards the diffuse regime.

\item The line-of-sight velocity dispersion (Eq.~\ref{eq:fsigma}) follows a two-phase track
\[\sigma_{\rm los}/\sigma_{\rm los,0}=[(1+f_{\sigma}/x_{\rm bound,0})/(1+f_{\sigma}/x_{\rm bound})]^{\eta_{\sigma}},\]
analogous to that of the stellar mass, with a transition scale $f_{\sigma}$ set mainly by the stellar compactness and a decline index $\eta_{\sigma}$ that decreases with increasing S\'ersic index (Section~\ref{sec:sigma_track}).

\item A correlation analysis across the full simulation suite yields simple, physically interpretable mappings from the initial structure to the track parameters (Section~\ref{sec:param_dependence} and Table~\ref{tab:dependence}). The half-mass stripping scale is governed primarily by the stellar compactness, $f_{1/2}\propto(r_{\rm eff}/r_{\rm vir})^{1.4}$. The stripping sharpness scales approximately as $b\propto\sqrt{s_{001}/n_{\rm sersic}}$. The size-response index $\beta$ increases approximately linearly with $\sqrt{n_{\rm sersic}}$, with a secondary dependence on $s_{001}$. The puff-up parameters $A$ and $f_{\rm p}$ are strongly covariant, and the velocity dispersion parameters are only loosely tied to the initial structure (see Table~\ref{tab:dependence}).

\end{enumerate}

Our results demonstrate that the tidal response of the stellar body depends systematically on its initial structure and can be described by a small set of empirical functions (Table~\ref{tab:tracks}). These functions provide a practical framework for modelling the stellar structural evolution of satellite dwarf galaxies under tidal stripping.

In Paper~II of this series, we will apply these results to investigate the tidal evolution of ultra-diffuse galaxies (UDGs) and dark-matter-deficient galaxies (DMDGs), and explore how different initial structures and tidal tracks can account for the diversity of observed dwarf galaxies.



\software{Agama \citep{Vasiliev2019}, \textsc{Gadget-4} \citep{gadget4}, \textsc{hbt+}\xspace \citep{Han2012,Han2018}, NumPy (\url{https://numpy.org}), SciPy (\url{https://scipy.org}), Astropy (\url{https://www.astropy.org}), Matplotlib (\url{https://matplotlib.org}), gplearn (\url{https://gplearn.readthedocs.io})}

\begin{acknowledgments}
This work is supported by the National Natural Science Foundation of China (NSFC; Grant No. 12473007) and the China Manned Space Program (Grant No. CMS-CSST-2025-A03). The authors thank Xiaolong Du and Jiaxin Han for helpful discussions. The authors also thank Yimeng Tang for helpful discussions from an observational perspective. F.J. thanks Aaron Ludlow for helpful discussions. F.H. thanks Zhaozhou Li for suggestions on constructing the initial conditions for the idealised simulations. F.H. also thanks Guangze Sun for setting up and maintaining the Milkomeda computing server, on which the simulations in this work were performed.
\end{acknowledgments}

\section*{Data Availability}

The simulation data underlying this study are available
from the corresponding author upon reasonable request.

\appendix

\section{Convergence Study}\label{sec:convergence}

\subsection{Resolution convergence}\label{sec:res_convergence}

\begin{figure*}
    \centering
    \includegraphics[width=\linewidth]{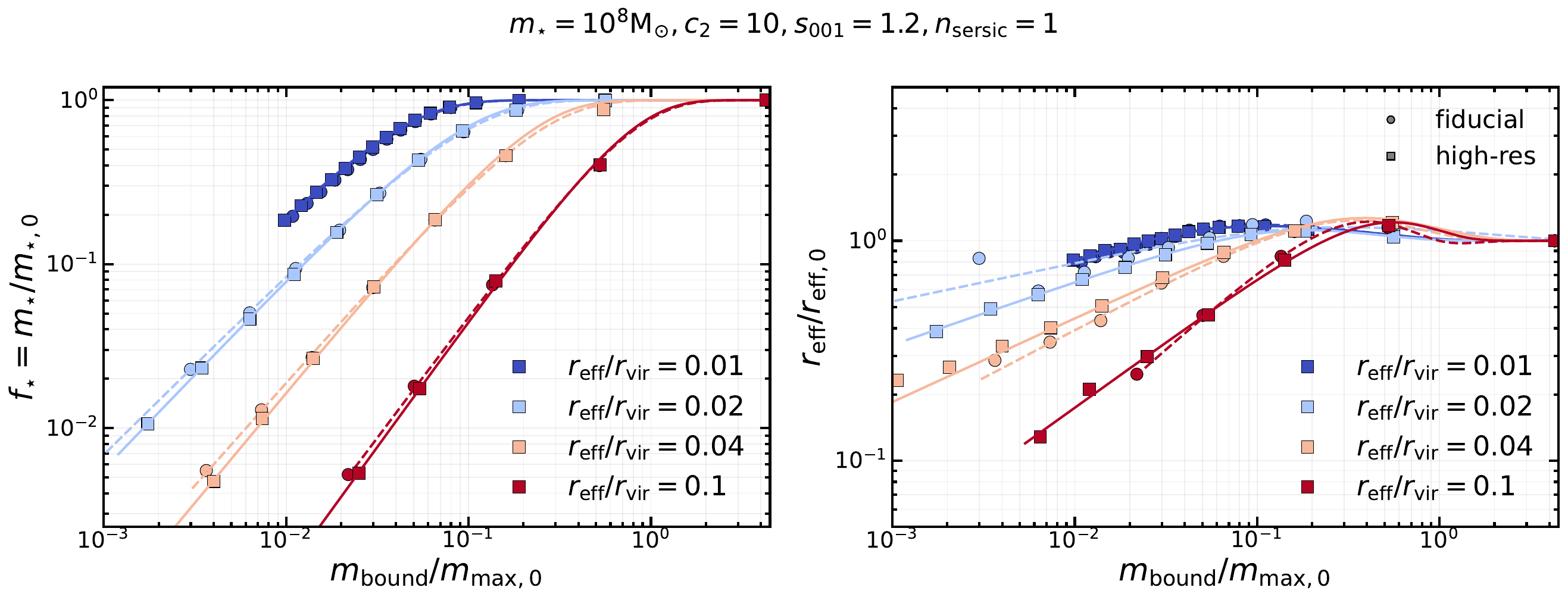}
    \caption{Resolution convergence of the stellar mass and size tidal tracks for different initial stellar compactness. All models have $m_{\star}=10^{8}\,\rm{M}_{\odot}$, $c_2=10$, $s_{001}=1.2$, and $n_{\rm sersic}=1$. Colours denote $r_{\rm eff}/r_{\rm vir}$ from $0.01$ (most compact, blue) to $0.1$ (most diffuse, dark red). The left panel shows the remaining stellar mass fraction $f_{\star}=m_{\star}/m_{\star,0}$ and the right panel the normalised half-mass radius $r_{\rm eff}/r_{\rm eff,0}$, both as functions of the scaled bound mass $x_{\rm bound}$. Filled circles and squares mark the apocentre snapshots of the fiducial ($N=10^{6}$) and high-resolution ($N=10^{7}$) runs, respectively, while dashed and solid curves show the corresponding fits of Eqs.~\ref{eq:fstar} and~\ref{eq:freff}. The two resolutions trace the same tracks over the full range of stellar compactness, with the agreement being closer for the mass track than for the size track.}
    \label{fig:conv_reff}
\end{figure*}

\begin{figure*}
    \centering
    \includegraphics[width=\linewidth]{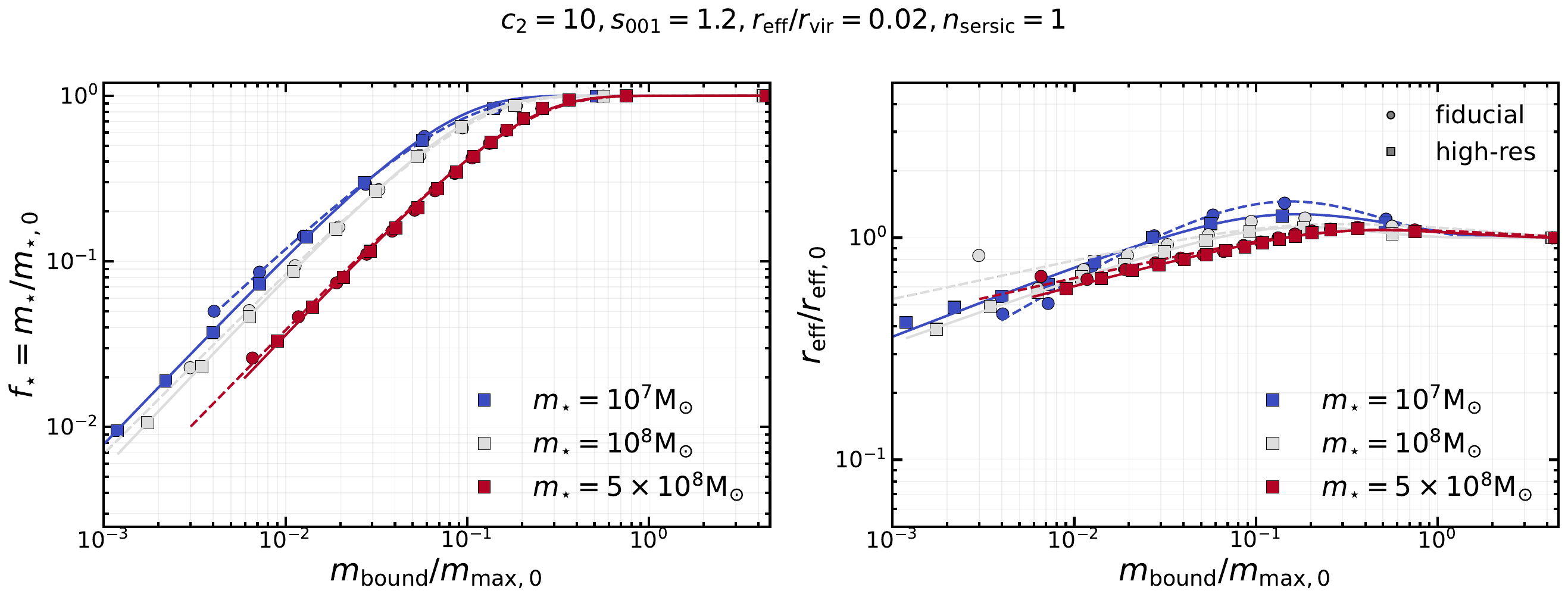}
    \caption{Resolution convergence of the tidal tracks for different initial stellar masses, at fixed $c_2=10$, $s_{001}=1.2$, $r_{\rm eff}/r_{\rm vir}=0.02$, and $n_{\rm sersic}=1$. Colours denote $m_{\star}=10^{7}$, $10^{8}$, and $5\times10^{8}\,\rm{M}_{\odot}$, and the symbols and curves follow the convention of Fig.~\ref{fig:conv_reff}. In the fiducial runs the $m_{\star}=10^{7}\, \rm{M}_{\odot}$ model contains only $N_{\star}\sim10^{3}$ stellar particles, rising to $N_{\star}\sim10^{4}$ at high resolution. Both tracks are recovered consistently even when the stellar component is sampled by as few as $\sim10^{3}$ particles.}
    \label{fig:conv_mstar}
\end{figure*}

\begin{figure*}
    \centering
    \includegraphics[width=\linewidth]{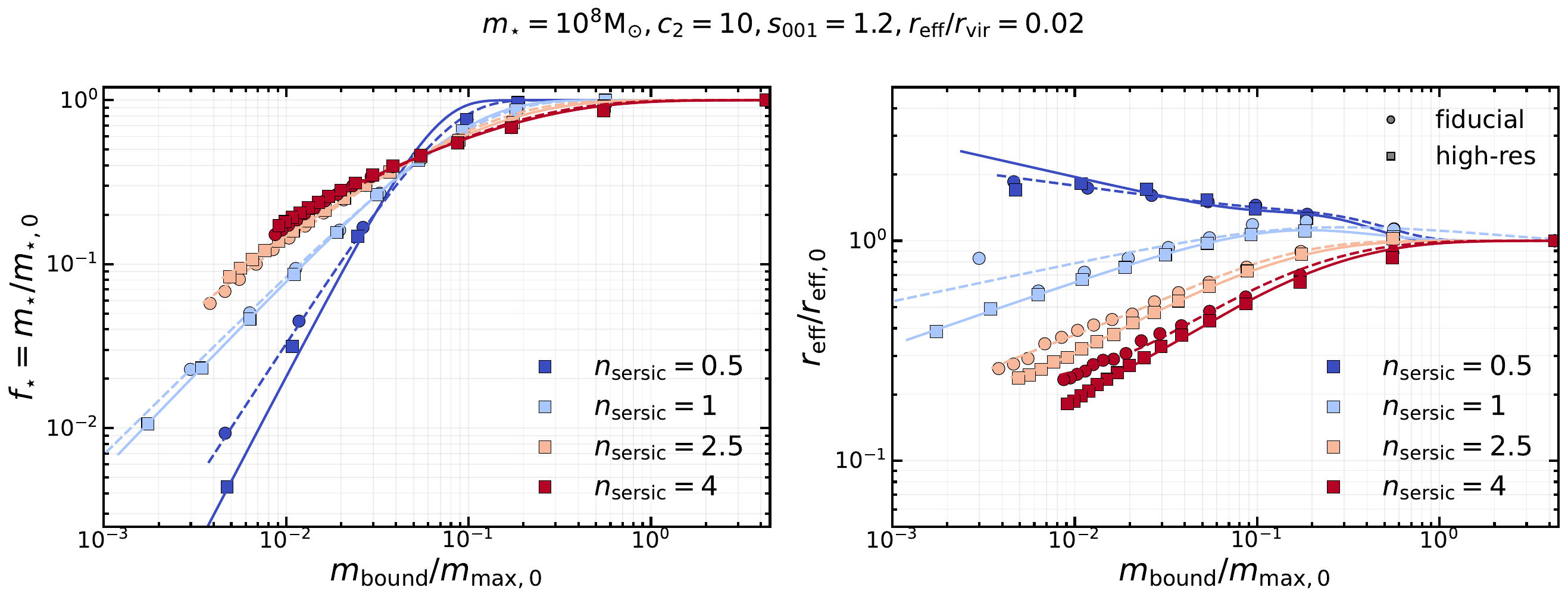}
    \caption{Resolution convergence of the tidal tracks for different stellar S\'ersic indices, at fixed $m_{\star}=10^{8}\,\rm{M}_{\odot}$, $c_2=10$, $s_{001}=1.2$, and $r_{\rm eff}/r_{\rm vir}=0.02$. Colours denote $n_{\rm sersic}=0.5$, $1$, $2.5$, and $4$, and the symbols and curves follow the convention of Fig.~\ref{fig:conv_reff}. The mass and size tracks are converged between the two resolutions over the full range of stellar structures considered.}
    \label{fig:conv_nsersic}
\end{figure*}

\begin{figure*}
    \centering
    \includegraphics[width=\linewidth]{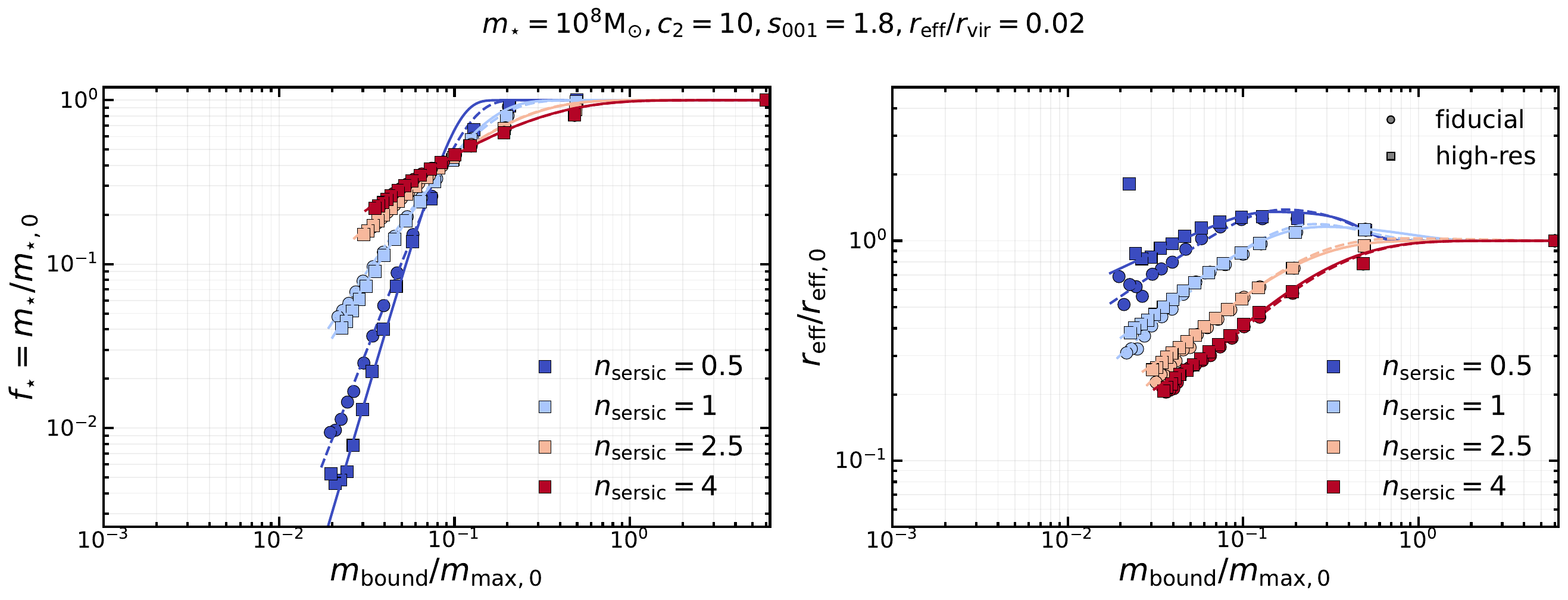}
    \caption{Resolution convergence of the tidal tracks in a cuspy subhalo, with $m_{\star}=10^{8}\,\rm{M}_{\odot}$, $c_2=10$, $s_{001}=1.8$, and $r_{\rm eff}/r_{\rm vir}=0.02$. Colours, symbols, and curves follow the convention of Fig.~\ref{fig:conv_nsersic}. The high-resolution runs reproduce the fiducial tracks at least as well as in the $s_{001}=1.2$ models of Fig.~\ref{fig:conv_nsersic}, the largest residual differences occurring for the most diffuse stellar configuration, $n_{\rm sersic}=0.5$.}
    \label{fig:conv_nsersic_cusp}
\end{figure*}

\begin{figure*}
    \centering
    \includegraphics[width=\linewidth]{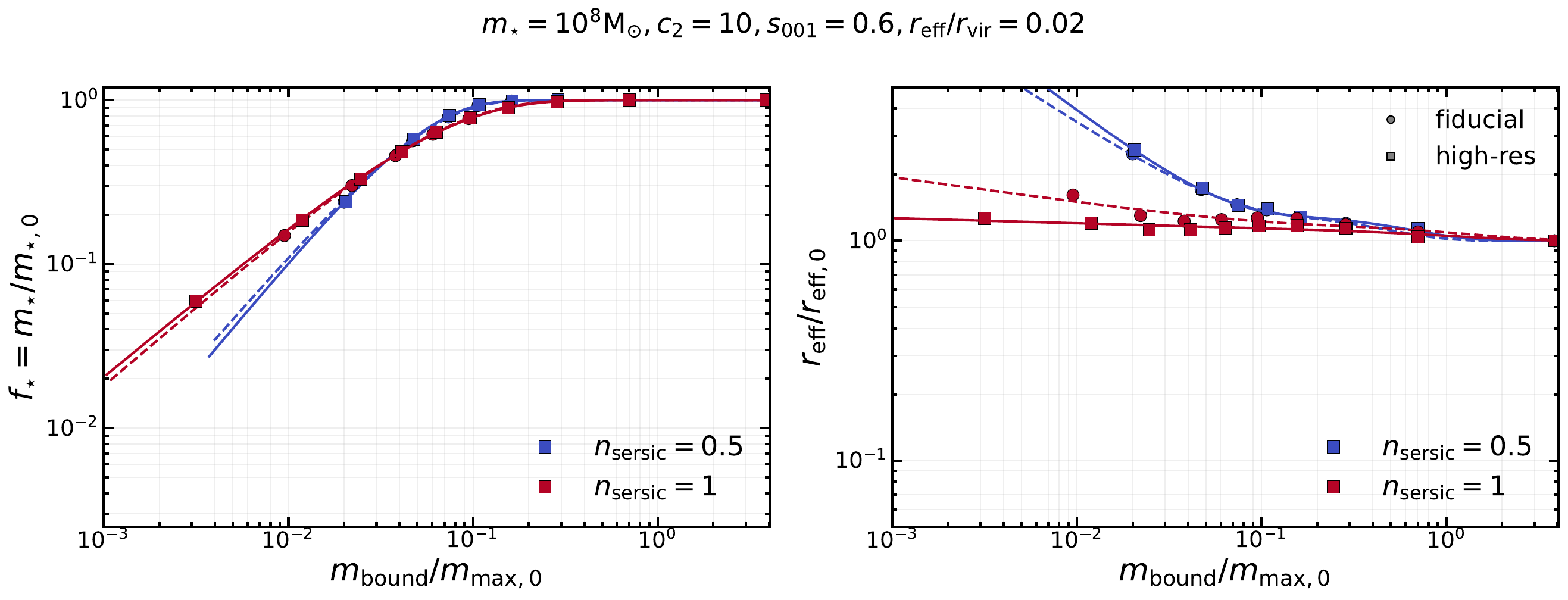}
    \caption{Resolution convergence of the tidal tracks in the most strongly cored subhalo considered here, with $m_{\star}=10^{8}\,\rm{M}_{\odot}$, $c_2=10$, $s_{001}=0.6$, and $r_{\rm eff}/r_{\rm vir}=0.02$. Colours denote $n_{\rm sersic}=0.5$ and $1$, and the symbols and curves follow the convention of Fig.~\ref{fig:conv_reff}. Although the size evolution of these diffuse stellar systems is intrinsically noisier, the tracks remain converged between the two resolutions.}
    \label{fig:conv_nsersic_core}
\end{figure*}

The wide parameter space scanned in this work, together with the finite computational resources available, requires a compromise between numerical resolution and the number of simulations. We therefore repeat a representative subset of the runs presented in Section~\ref{sec:results} at higher resolution. The subset is selected by starting from the typical dwarf models discussed in Section~\ref{sec:results} and extending along the directions in parameter space along which the tidal tracks are expected to be most sensitive to resolution, namely the stellar compactness, the stellar mass, the S\'ersic index, and the inner slope of the DM halo.

The high-resolution initial conditions are constructed with \textsc{Agama} following exactly the procedure of Section~\ref{sec:setup}, the only difference being that the target particle number is raised from $N_{\rm target}=10^{6}$ to $N_{\rm target}=10^{7}$. The gravitational softening is scaled with particle number following \citet{Du2024},
\begin{equation}
    \varepsilon=0.003\,r_{-2}\left(\frac{N}{10^{6}}\right)^{-1/3}.
    \label{eq:softening_scaling}
\end{equation}
The stellar mass and size tidal tracks are then measured with the same procedure as in the fiducial runs.

Figs.~\ref{fig:conv_reff}--\ref{fig:conv_nsersic_core} compare the high-resolution tracks with their fiducial counterparts. In all of these figures, filled circles and squares mark the apocentre snapshots of the fiducial ($N=10^{6}$) and high-resolution ($N=10^{7}$) runs, and dashed and solid curves show the corresponding fits of Eqs.~\ref{eq:fstar} and~\ref{eq:freff}. Across the tested range of structural parameters the two resolutions produce consistent tracks, so the fiducial resolution is adequate for characterising the stellar tidal evolution over the parameter space explored here. The agreement is systematically closer for the stellar mass track than for the size track, as expected given the greater sensitivity of $r_{\rm eff}$ to orbital-phase fluctuations and to the detailed structure of the outer stellar envelope.

Fig.~\ref{fig:conv_reff} varies the initial stellar compactness at fixed $m_{\star}=10^{8}\,\rm{M}_{\odot}$, $c_2=10$, $s_{001}=1.2$, and $n_{\rm sersic}=1$. The mass and size tracks are recovered consistently between $N=10^{6}$ and $N=10^{7}$ over the full range $r_{\rm eff}/r_{\rm vir}=0.01$--$0.1$, with the residual differences in the size track remaining modest.

Fig.~\ref{fig:conv_mstar} tests the dependence on the initial stellar mass at fixed $c_2=10$, $s_{001}=1.2$, $r_{\rm eff}/r_{\rm vir}=0.02$, and $n_{\rm sersic}=1$. This is the most demanding test of the stellar sampling: for $m_{\star}=10^{7}\, \rm{M}_{\odot}$ the fiducial run resolves the stellar component with only $N_{\star}\sim10^{3}$ particles. Increasing the particle number by an order of magnitude, so that $N_{\star}\sim10^{4}$, leaves both tracks unchanged, indicating that the tracks are not limited by discreteness noise in the stellar component even in this regime.

Fig.~\ref{fig:conv_nsersic} varies the S\'ersic index with the remaining parameters held at the fiducial values. Convergence is maintained for both the mass and the size track across the full range of stellar structures considered. Figs.~\ref{fig:conv_nsersic_cusp} and~\ref{fig:conv_nsersic_core} then examine the dependence on the inner halo slope. For the cuspy subhalo with $s_{001}=1.8$ (Fig.~\ref{fig:conv_nsersic_cusp}), the agreement between the two resolutions is comparable to, and in places better than, that of the fiducial $s_{001}=1.2$ models. Within this set, the $n_{\rm sersic}=0.5$ case shows the largest residual differences in both tracks, because such cored stellar configurations are the most diffuse and therefore the most susceptible to tidal heating. Fig.~\ref{fig:conv_nsersic_core} shows the opposite extreme, the most strongly cored subhalo considered here with $s_{001}=0.6$, for which the two resolutions again agree.

\subsection{Orbit convergence}\label{sec:orbit_convergence}

\begin{figure}
    \centering
    \begin{subfigure}
        \centering
        \includegraphics[width=\linewidth]{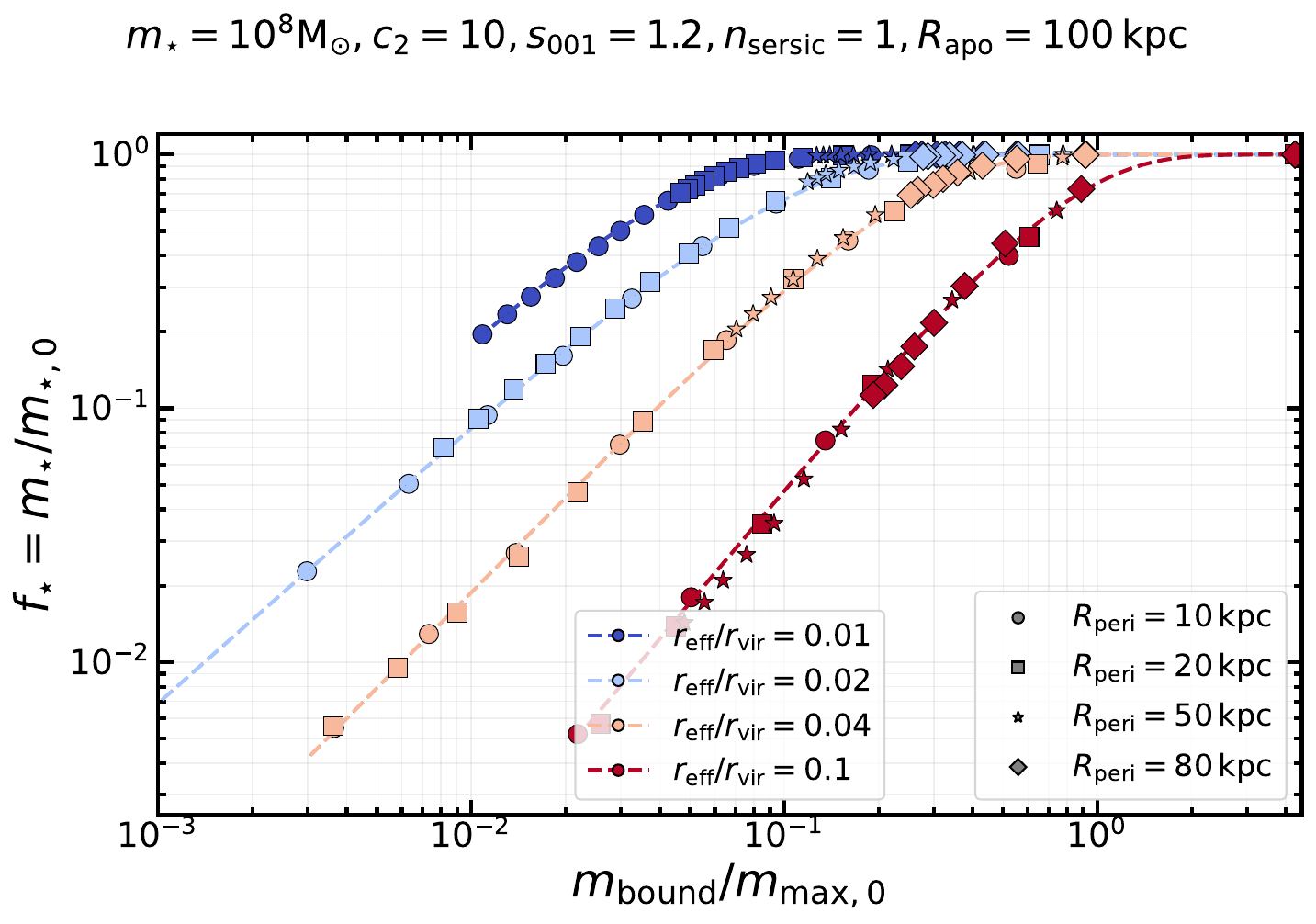}
    \end{subfigure}
    \vfill
    \begin{subfigure}
        \centering
        \includegraphics[width=\linewidth]{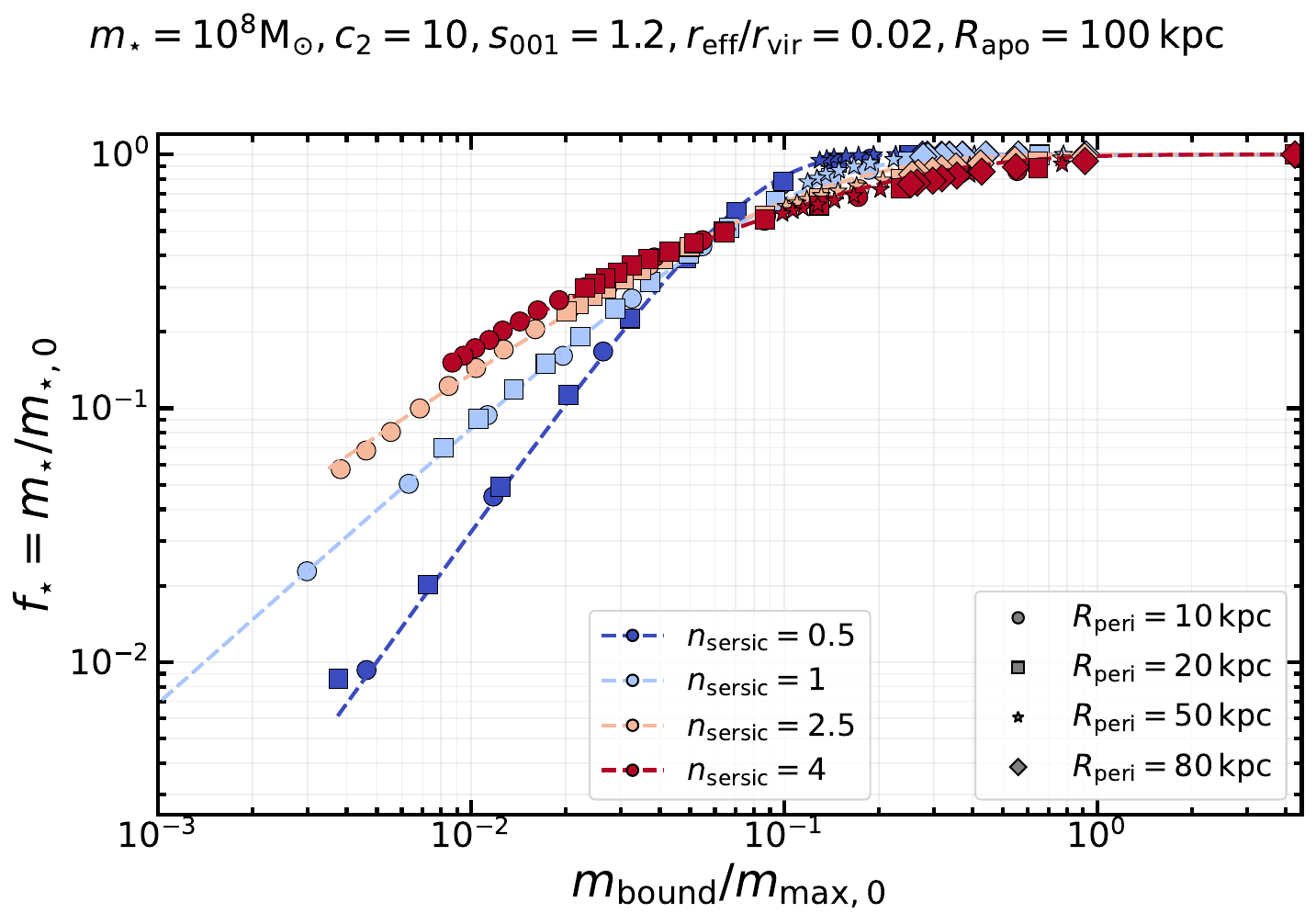}
    \end{subfigure}
    \caption{Orbit dependence of the stellar mass tidal track. The remaining stellar mass fraction $f_{\star}$ is shown as a function of $x_{\rm bound}$ for dwarfs released with $R_{\rm peri}=10$, $20$, $50$, and $80\,{\rm kpc}$ at fixed $R_{\rm apo}=100\,{\rm kpc}$, distinguished by the marker shapes given in the legend. Colours denote the structural parameter that is varied, and the dashed curve of each colour is the fit of Eq.~\ref{eq:fstar} to the corresponding fiducial run. \textit{Top:} $r_{\rm eff}/r_{\rm vir}$ varied at fixed $m_{\star}=10^{8}\,\rm{M}_{\odot}$, $c_2=10$, $s_{001}=1.2$, and $n_{\rm sersic}=1$. \textit{Bottom:} $n_{\rm sersic}$ varied at fixed $r_{\rm eff}/r_{\rm vir}=0.02$ and the same remaining parameters. The stellar mass track is independent of the orbit, which controls only the range of $x_{\rm bound}$ that can be reached within $20\,{\rm Gyr}$.}
    \label{fig:orbit_mass}
\end{figure}

\begin{figure*}
    \centering
    \begin{subfigure}
        \centering
        \includegraphics[width=0.45\linewidth]{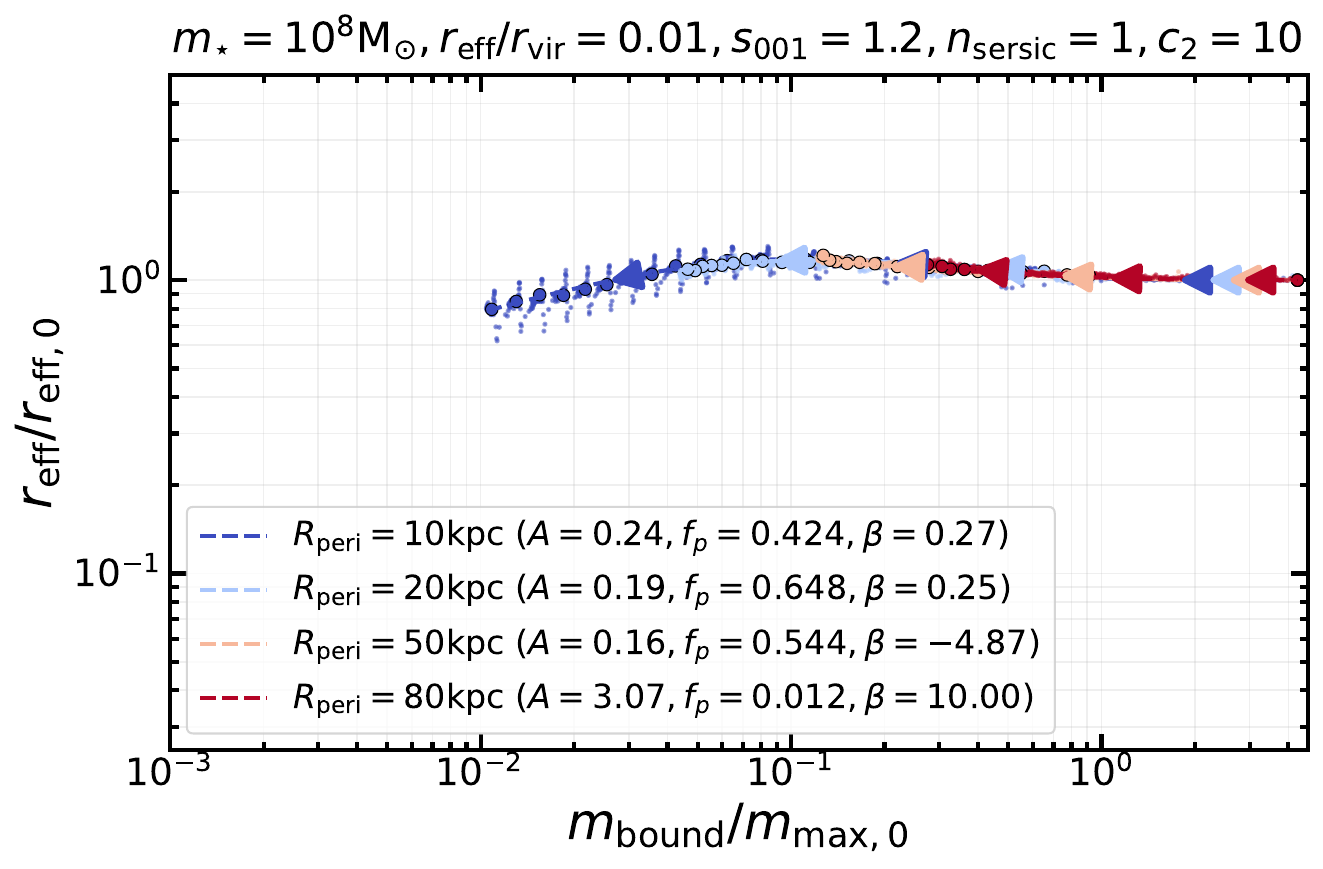}
    \end{subfigure}
    \centering
    \hfill
    \begin{subfigure}
        \centering
        \includegraphics[width=0.45\linewidth]{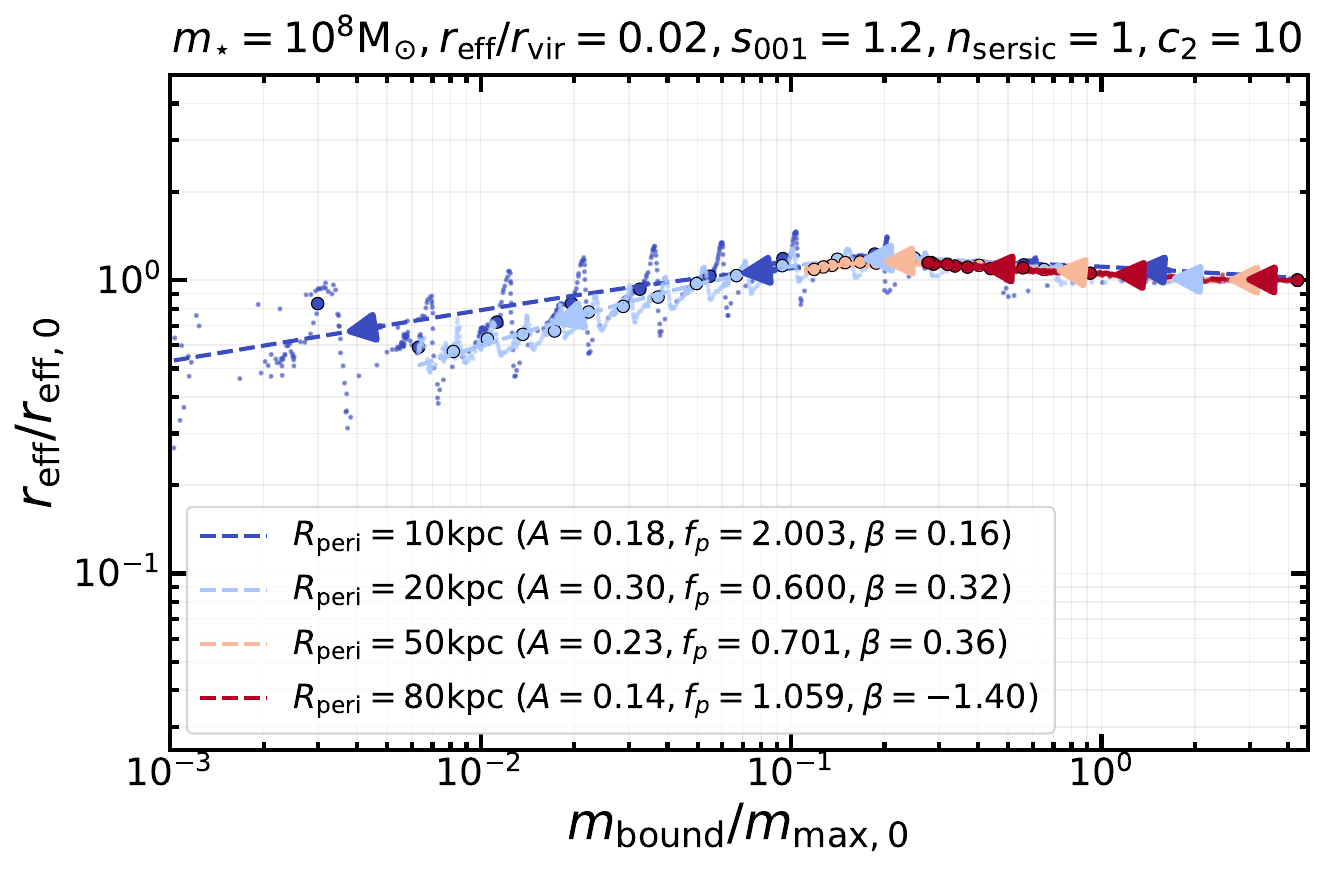}
    \end{subfigure}
    \vfill
    \centering
    \begin{subfigure}
        \centering
        \includegraphics[width=0.45\linewidth]{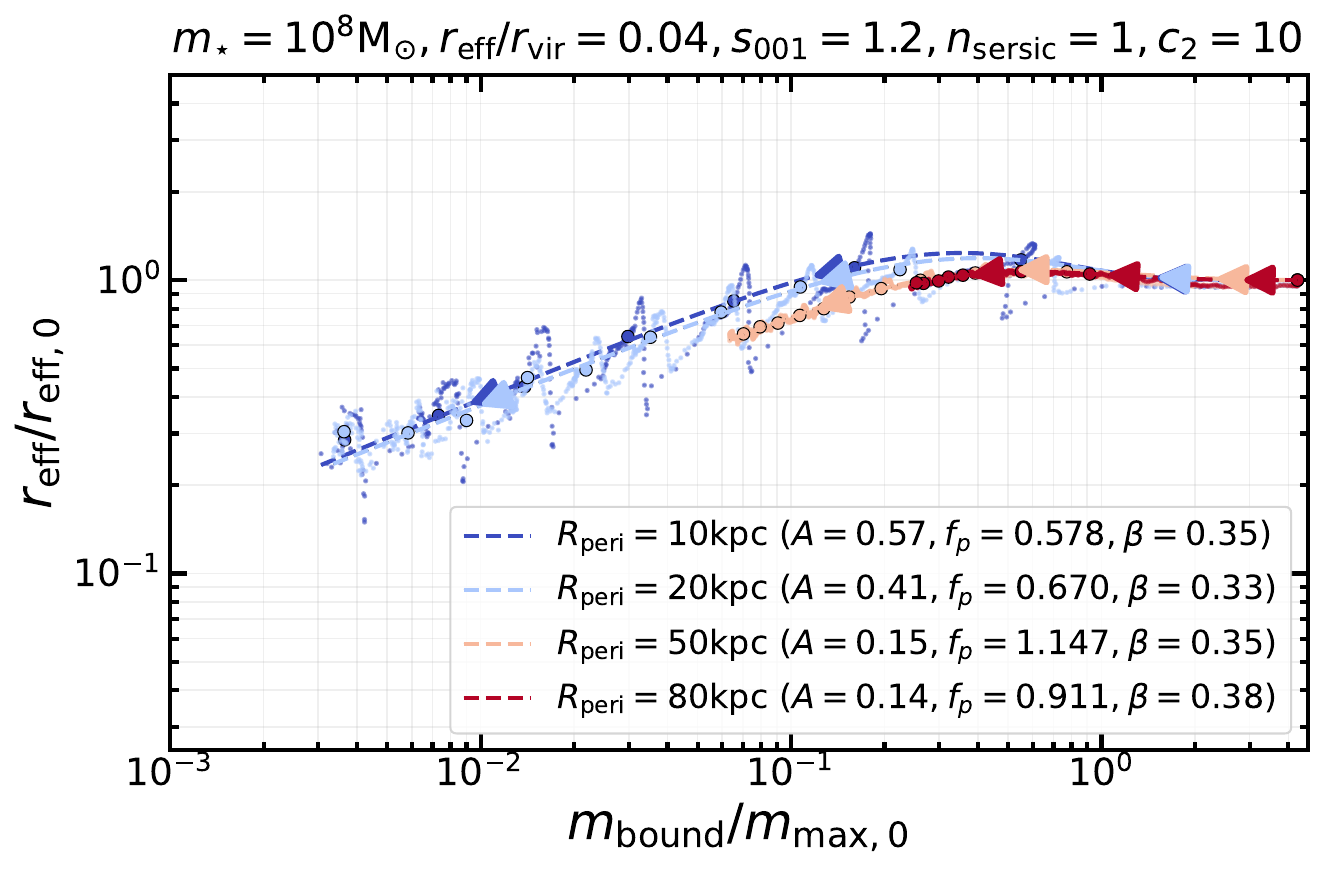}
    \end{subfigure}
    \centering
    \hfill
    \begin{subfigure}
        \centering
        \includegraphics[width=0.45\linewidth]{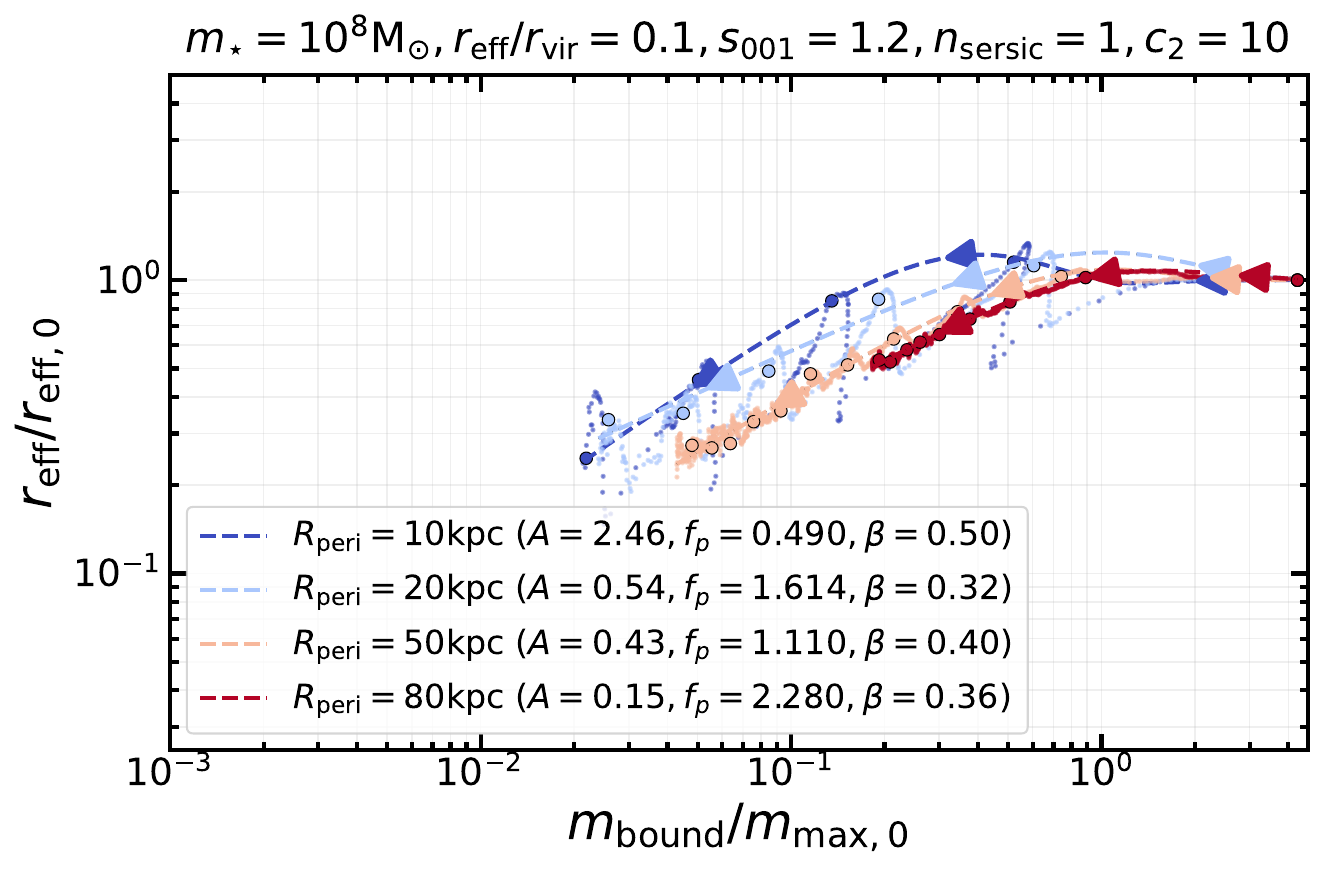}
    \end{subfigure}
    \caption{Orbit dependence of the size tidal track for different stellar compactness. Each panel corresponds to one value of $r_{\rm eff}/r_{\rm vir}$, increasing from $0.01$ to $0.1$, at fixed $m_{\star}=10^{8}\,\rm{M}_{\odot}$, $c_2=10$, $s_{001}=1.2$, and $n_{\rm sersic}=1$. Within each panel, colours denote $R_{\rm peri}=10$, $20$, $50$, and $80\,{\rm kpc}$ at fixed $R_{\rm apo}=100\,{\rm kpc}$. Faint dotted lines show every snapshot, open circles the apocentre snapshots, and dashed curves the fits of Eq.~\ref{eq:freff}, with the fitted $(A,f_{\rm p},\beta)$ listed in the legends. Arrows indicate the direction of evolution. The apocentre-sampled tracks agree between orbits for compact stellar bodies, whereas for diffuse ones the orbital-phase oscillations grow strongly on the plunging orbits, so that the tracks agree only in the envelope traced by these oscillations.}
    \label{fig:orbit_size_reff}
\end{figure*}

\begin{figure*}
    \centering
    \begin{subfigure}
        \centering
        \includegraphics[width=0.45\linewidth]{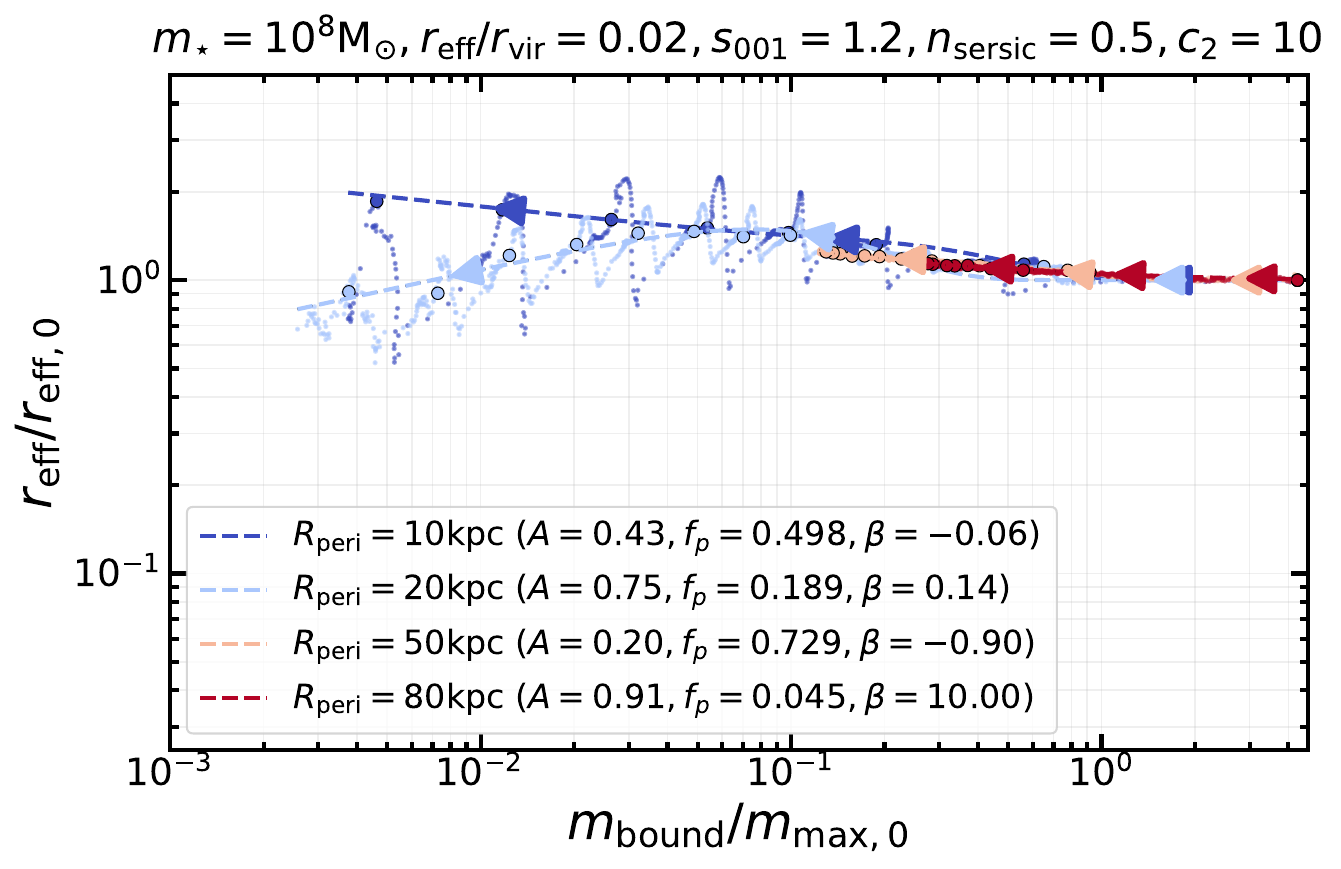}
    \end{subfigure}
    \centering
    \hfill
    \begin{subfigure}
        \centering
        \includegraphics[width=0.45\linewidth]{orbit_convergence_size_reff002_v3.pdf}
    \end{subfigure}
    \vfill
    \centering
    \begin{subfigure}
        \centering
        \includegraphics[width=0.45\linewidth]{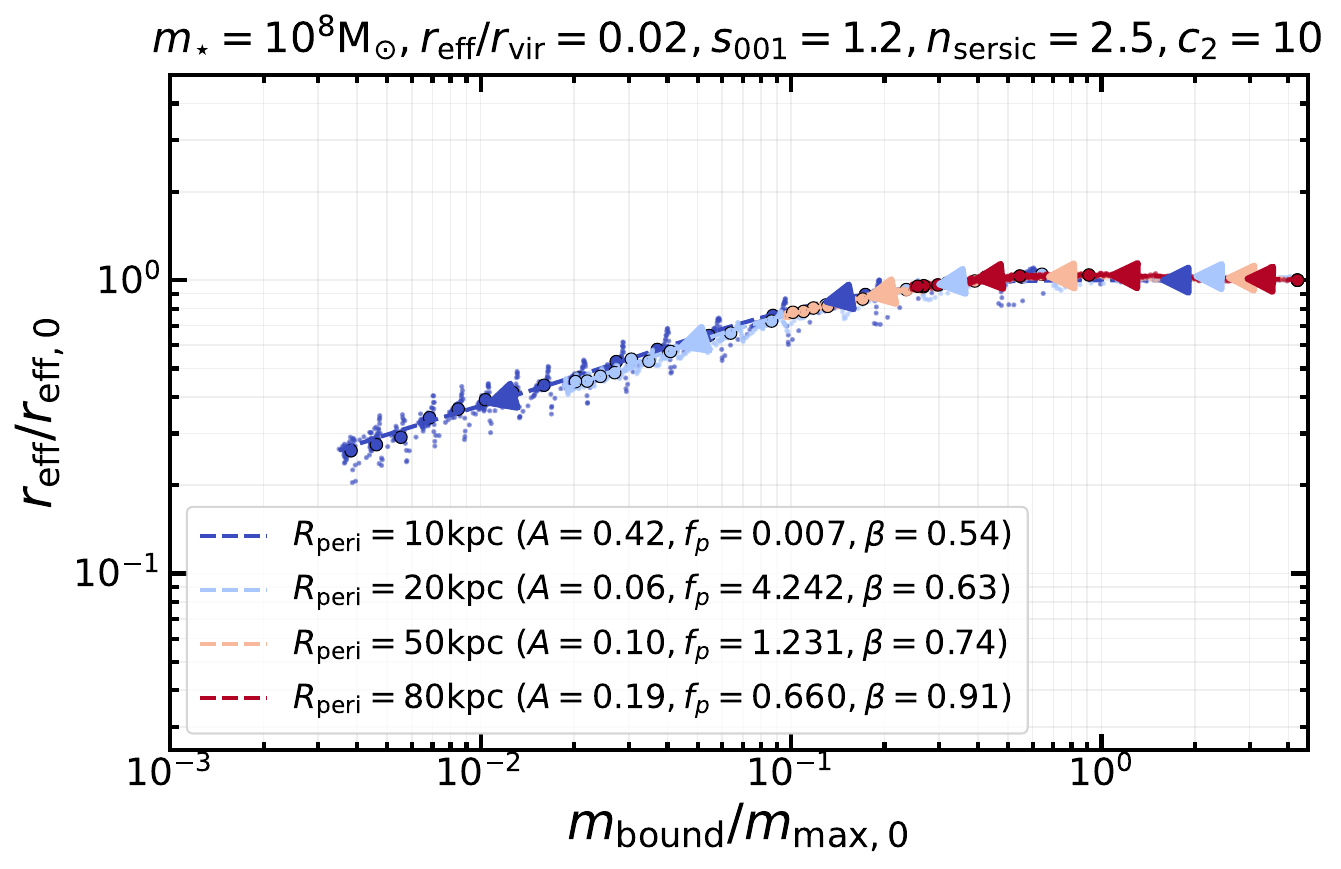}
    \end{subfigure}
    \centering
    \hfill
    \begin{subfigure}
        \centering
        \includegraphics[width=0.45\linewidth]{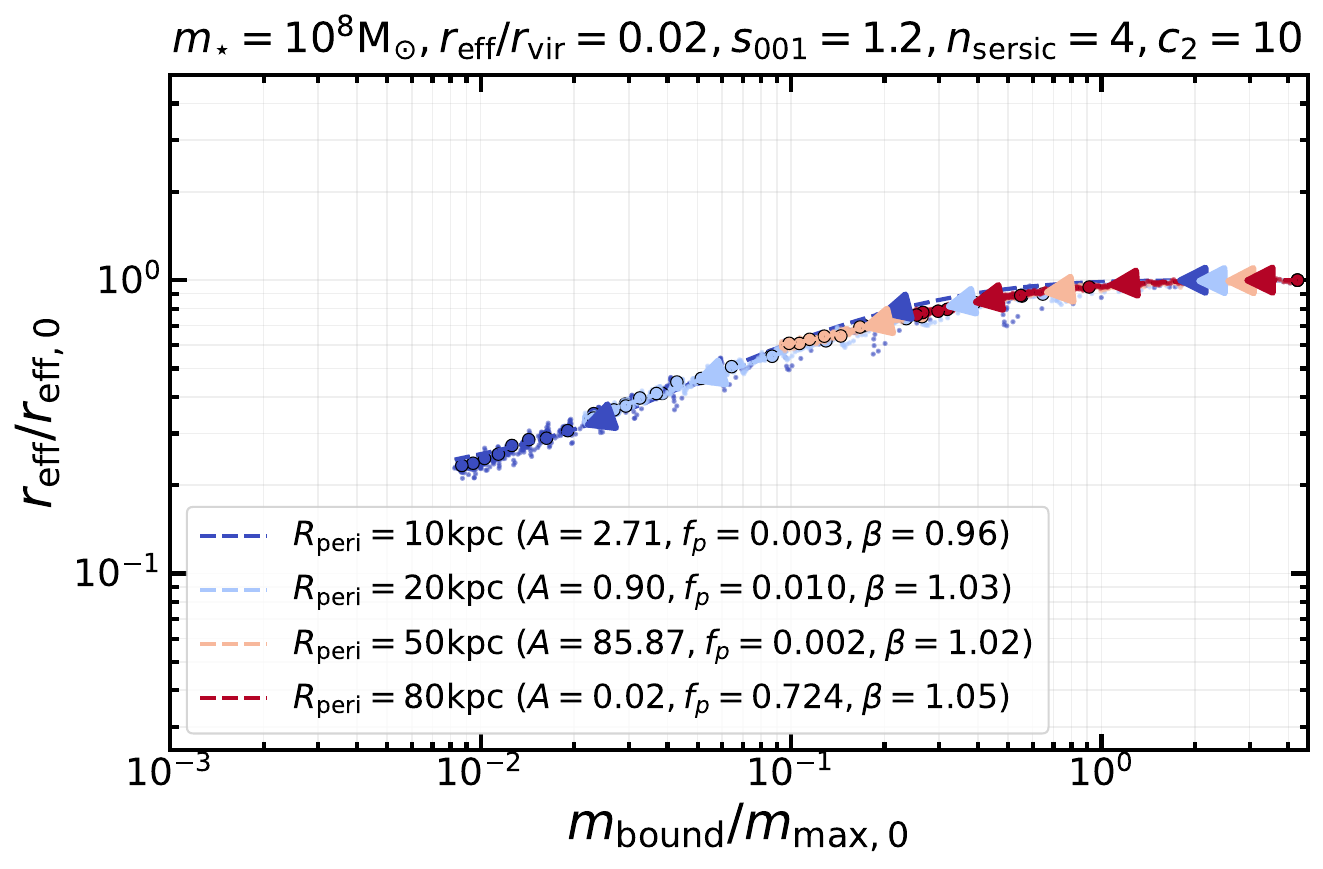}
    \end{subfigure}
    \caption{Orbit dependence of the size tidal track for different S\'ersic indices. Each panel corresponds to one value of $n_{\rm sersic}$, increasing from $0.5$ to $4$, at fixed $m_{\star}=10^{8}\,\rm{M}_{\odot}$, $c_2=10$, $s_{001}=1.2$, and
    $r_{\rm eff}/r_{\rm vir}=0.02$. Colours, symbols, curves, and fitted parameters follow the convention of Fig.~\ref{fig:orbit_size_reff}. Systems with large $n_{\rm sersic}$ follow a single size track on all orbits, whereas stellar bodies with shallow central profiles develop strong orbit-dependent oscillations whose envelope, rather than the apocentre points themselves, is the orbit-independent quantity.}
    \label{fig:orbit_size_nsersic}
\end{figure*}

The tidal tracks of the inner structural properties of subhaloes are generally regarded as being set by the initial internal structure and only weakly dependent on the orbit. This behaviour was established for DM-only quantities such as $v_{\rm max}$ and $r_{\rm max}$, which follow a common track once expressed as a function of $f_{\rm bound}$, largely independently of the pericentre, the orbital circularity, and the number of pericentric passages over which the mass was removed \citep{Penarrubia2008, Penarrubia2010, Green2019, Errani2021}. It is this property that makes tidal tracks a practical tool, and it is what allows us to scan the structural parameter space along a single calibrated orbit for each $(s_{001},c_2)$ combination (Section~\ref{sec:setup}). The same insensitivity is expected for the stellar mass and size tracks, but it has not been tested explicitly. We do so here.

In the fiducial suite, all dwarfs are released at $R_{\rm apo}=100\,{\rm kpc}$ on the orbit calibrated for their halo structure, which for the models considered below corresponds to $R_{\rm peri}=10\,{\rm kpc}$. We rerun the same dwarfs on orbits of higher energy by setting $R_{\rm peri}=20$, $50$, and $80\,{\rm kpc}$ at fixed $R_{\rm apo}=100\,{\rm kpc}$. The initial $N$-body realisations are identical in all cases, and only the tangential velocity at release is changed, so that any difference between the runs is due to the orbit alone.

Fig.~\ref{fig:orbit_mass} shows the stellar mass tracks, and is the counterpart of Fig.~\ref{fig:mass_track} for different orbits. The tracks obtained on different orbits are indistinguishable. The pericentre determines only how far along a given track the satellite proceeds within $20\,{\rm Gyr}$: satellites on wider orbits are stripped more weakly and therefore terminate at larger $x_{\rm bound}$. For $R_{\rm peri}=80\,{\rm kpc}$ the tracks barely reach the second, coupled phase of the stellar mass loss at all. This is precisely why the fiducial suite adopts orbits of as low an energy as the survival of the subhalo permits, so that the full range of the tidal track is sampled.

The size tracks behave in a more complicated way. Fig.~\ref{fig:orbit_size_reff} shows the orbital dependence for the models of the top panel of Fig.~\ref{fig:size_track}, in which the stellar compactness is varied. For the compact systems ($r_{\rm eff}/r_{\rm vir}=0.01$ and $0.02$) the apocentre-sampled size tracks converge across all four orbits. For the most diffuse system ($r_{\rm eff}/r_{\rm vir}=0.1$), however, they do not. The origin of this behaviour is the orbital-phase modulation of $r_{\rm eff}$ between pericentre and apocentre already identified in Section~\ref{sec:size_track}. The amplitude of this modulation grows both with $r_{\rm eff}/r_{\rm vir}$ and with the strength of the tidal shock at pericentre: it is weak for compact stellar bodies at any $R_{\rm peri}$, and for a given compactness it is weak on the wide orbits and strong on the plunging ones. The tracks measured on the wide orbits are therefore close to the smooth secular behaviour, whereas those measured on the $R_{\rm peri}=10\,{\rm kpc}$ orbit sample a strongly oscillating $r_{\rm eff}$ at a particular orbital phase. Considering the envelope traced by these oscillations rather than the apocentre points alone, the size evolution is consistent between orbits even for $r_{\rm eff}/r_{\rm vir}=0.1$, since the wide-orbit tracks lie within the envelope defined by the plunging-orbit runs.

Fig.~\ref{fig:orbit_size_nsersic} shows the corresponding test for the models of the bottom panel of Fig.~\ref{fig:size_track}, in which the S\'ersic index is varied. For large $n_{\rm sersic}$ the size tracks are essentially identical on all four orbits. As $n_{\rm sersic}$ decreases the stellar structure becomes more cored and more readily puffed up, and the periodic modulation on the low-energy orbits becomes pronounced. For $n_{\rm sersic}=0.5$ the apocentre-sampled tracks again fail to overlap, while their envelope remains consistent, in the same manner as for the diffuse models of Fig.~\ref{fig:orbit_size_reff}.

These tests support the use of a single calibrated orbit per halo structure, and at the same time reinforce the caveat raised in Section~\ref{sec:param_dependence}: the size track parameters are least stable precisely for diffuse, low-$n_{\rm sersic}$ dwarfs, whose response to the tidal shock at each pericentre is large enough that a fit to apocentre snapshots retains a residual dependence on the orbit. A quantitative description of the periodic puffing up and recontraction of $r_{\rm eff}$ within an orbital period, which is deliberately not modelled by Eq.~\ref{eq:freff}, is left to future work.

\subsection{\textsc{hbt+}\xspace convergence}\label{sec:hbt_convergence}

\begin{figure}
    \centering
    \includegraphics[width=\linewidth]{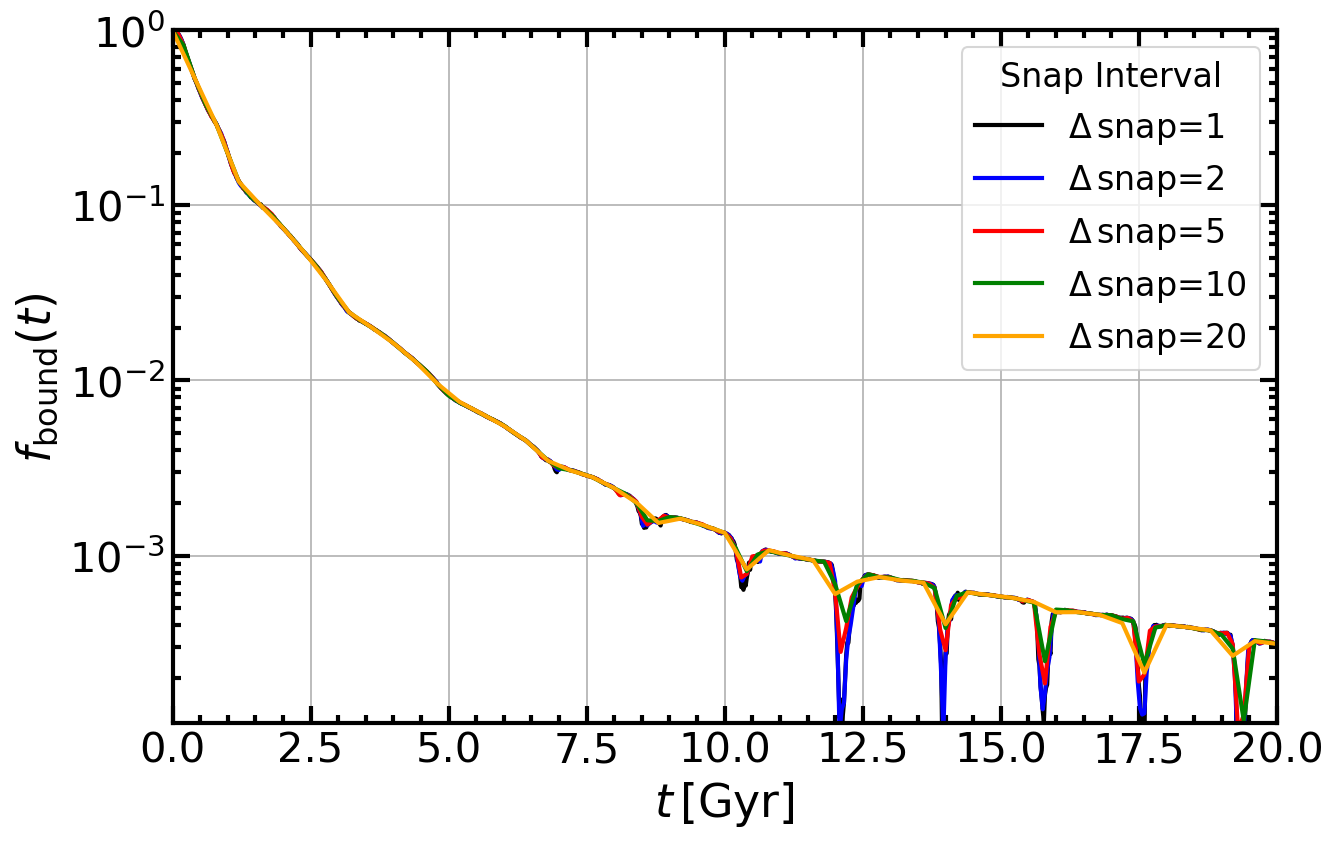}
    \caption{Effect of the snapshot cadence used by \textsc{hbt+}\xspace on the recovered bound-mass history. The bound mass fraction $f_{\rm bound}(t)$ of a representative deeply stripped run is shown as a function of time, with colours denoting the interval $\Delta_{\rm snap}$ between the snapshots passed to \textsc{hbt+}\xspace. The underlying simulation output is identical in all cases. The narrow dips at pericentric passages, which arise from the accumulation of unbinding fluctuations between successive snapshots, weaken as the cadence is coarsened, leaving the saw-tooth mass-loss behaviour characteristic of the lower snapshot cadence of cosmological simulations.}
    \label{fig:hbt_conv}
\end{figure}

The mass-loss histories in Figs.~\ref{fig:preliminary_dmo} and~\ref{fig:masslosscurve} develop narrow, periodic dips in the strongly stripped regime, $f_{\rm bound}\lesssim2\times10^{-3}$, each of which coincides with a pericentric passage. These features are in part an artefact of the unbinding procedure of \textsc{hbt+}\xspace. The bound particle set at a given snapshot is determined iteratively starting from the set identified at the previous snapshot, so that fluctuations introduced near pericentre, where the remnant is transiently compressed and the instantaneous binding energies change rapidly, can be propagated and accumulated from one snapshot to the next. The effect is amplified in our setup because each run is written with $1000$ snapshots over $20\,{\rm Gyr}$, so that a single orbital period is sampled very densely.

To quantify this, we repeat the \textsc{hbt+}\xspace analysis on unchanged simulation output while skipping snapshots, that is, tracking the remnant with intervals $\Delta_{\rm snap}=1$, $2$, $5$, $10$, and $20$. Fig.~\ref{fig:hbt_conv} shows the resulting bound-mass histories. Increasing $\Delta_{\rm snap}$ to $10$ or $20$ substantially reduces the depth of the dips at pericentre and leaves a saw-tooth-like decline. We note that cosmological simulations typically store of order $100$ snapshots in total, so that fewer than $100$ of them cover the evolution of an individual subhalo, a cadence comparable to our $\Delta_{\rm snap}=10$--$20$ analysis. The jagged mass-loss histories commonly seen for subhaloes in cosmological simulations are therefore obtained in a regime in which the pericentric passages are not temporally resolved.






\clearpage
\bibliography{sample701}{}
\bibliographystyle{aasjournalv7}



\end{document}